\documentclass[a4paper,10pt]{article}
\usepackage{jheppub} 
\usepackage{mathrsfs}
\usepackage{comment}
\usepackage[dvipsnames]{xcolor}
\usepackage{float}

\definecolor{darkpastelgreen}{rgb}{0.01, 0.75, 0.24}
\definecolor{pigmentgreen}{rgb}{0.0, 0.65, 0.31}

\arxivnumber{1234.56789} 

\title{Semiclassical Black Hole-White Hole transitions:\\ an analytical treatment}

\author{Julio Arrechea, Stefano Liberati, Massimiliano Spadafora}
\affiliation{SISSA, Via Bonomea 265, 34136 Trieste, Italy}
\affiliation{INFN Sezione di Trieste, Via Valerio 2, 34127 Trieste, Italy}
\affiliation{IFPU—Institute for Fundamental Physics of the Universe, Via Beirut 2, 34014 Trieste, Italy}

\emailAdd{jarreche@sissa.it, liberati@sissa.it, mspadafo@sissa.it}

\abstract{
Recent numerical studies of semiclassical gravity suggest that, in spherically symmetric black holes with both outer and inner horizons, the semiclassical instability of the inner horizon can drive the complete evaporation of the trapped region on timescales shorter than the standard Hawking evaporation time. Even more remarkably, independent simulations indicate that the disappearance of the trapped region is not the end of the evolution: it is followed by the formation of an anti-trapped region, i.e.~a dynamical white hole. In this work, we develop a novel analytic treatment of quantum effects in trapped and anti-trapped regions, showing how these numerical results can be understood within simplified two-dimensional models. We consider collapse models describing the formation of charged and regular black holes and compute the renormalized stress-energy tensor of the $|\textit{in}\rangle$ vacuum state. We show that, within this framework, the emergence of an anti-trapped region is a generic consequence of the amplification of negative energy fluxes propagating along the outgoing direction inside the initial trapped region. This mechanism provides an analytic explanation for the black-hole-to-white-hole transition observed in numerical simulations. Our analysis further suggests that the fluxes generated by the subsequent anti-trapped region, now propagating along the ingoing direction, may in turn trigger the formation of a new trapped region. This opens the possibility of a cascade of black-to-white-hole transitions, potentially ending in a horizon-free spacetime, i.e.~a bouncing scenario, without the need to invoke additional genuinely quantum-gravitational dynamics. Although establishing the complete evolution requires a self-consistent treatment of semiclassical backreaction, the analytic framework developed here identifies which aspects of the underlying mechanism are universal and which depend on the geometry, laying the groundwork for a systematic investigation of semiclassical black-hole-to-white-hole transitions.
}

\begin{document}
\maketitle
\flushbottom

\section{Introduction}

Semiclassical gravity provides a framework for describing the interaction between quantum matter fields and a classical spacetime geometry. In this approximation, the metric is sourced not only by classical matter, but also by the gravitational contribution of the quantum vacuum, leading to the semiclassical Einstein equations
\begin{equation}\label{Eq:SemiEinstein}
G_{\mu\nu} = \frac{8\pi G}{c^4} \left(T_{\mu\nu}^{~\rm{eff}}+\hbar \langle \hat{T}_{\mu\nu} \rangle^{\rm{ren}}\right) ,
\end{equation}
where $T_{\mu\nu}^{~\rm{eff}}$ denotes the effective classical stress-energy tensor, while $\langle \hat{T}_{\mu\nu} \rangle^{\rm{ren}}$ is the vacuum expectation value of the renormalized stress-energy tensor (RSET) of the quantum fields.\footnote{For simplicity, in Eq.~\eqref{Eq:SemiEinstein} we have omitted the terms quadratic in curvature that arise in the regularization procedure used to define $\langle \hat{T}_{\mu\nu} \rangle ^{\rm{ren}}$.}
This framework has led to the remarkable predictions of cosmological particle creation~\cite{Parker:1968mv} and black hole evaporation~\cite{Hawking:1975vcx}, two cornerstones of modern theoretical physics. Indeed, because the construction of a Fock space in curved spacetime is intrinsically ambiguous~\cite{Parker:2009uva,Birrell:1982ix,Wald:1995yp}, regions that are devoid of particles at early times may later become filled with energy fluxes, stimulated by spacetime curvature or dynamics. In suitable regimes, these fluxes can also admit an interpretation in terms of particle creation~\cite{Fulling:1977jm}.

In the special case of a Schwarzschild black hole formed through gravitational collapse, these fluxes will quickly settle down to the stationary values characteristic of the Unruh state,\footnote{The relaxation towards stationarity of the state is exponentially fast (in ingoing Eddington-Finkelstein time $v$) in two-dimensional models~\cite{Barcelo:2007yk,Juarez-Aubry:2018ofz}, but the corresponding four-dimensional physics might display complex transient effects~\cite{Siahmazgi:2025zys} and an overall slower decay rate~\cite{Wilson:2026xxo} throughout this relaxation.} leading to a slow evaporation of the outer horizon scaling as $M_{\textsc{bh}}^3$ in natural units. Such a state corresponds to the vacuum for free-falling observers at horizon crossing, but entails a steady flux of thermal radiation at the Hawking temperature at future null infinity.
Black holes thus cannot exist as isolated systems in semiclassical gravity, as they will end up immersed in a nearly-thermal atmosphere of Hawking radiation~\cite{Hawking:1975vcx,Wald:1975kc,Davies:1976ei,Juarez-Aubry:2018ofz,Levi:2016exv} (see~\cite{Anderson:2025wzp} and~\cite{Jiang:2026qnn} for recent analyses of 2D and 4D RSETs in the region beyond the future horizon). 

The presence of electromagnetic charge and/or angular momentum further modifies this picture by introducing a second, inner horizon, which exposes the black hole interior to enhanced classical and semiclassical effects. In eternal black hole geometries, inner Cauchy horizons are known to be unstable both classically~\cite{Simpson:1973ua, Poisson:1990eh, Ori:1991zz} --- through the so-called mass-inflation instability --- and quantum mechanically~\cite{Hiscock:1977qe,Birrell:1978th,Hiscock1981, Hollands:2019whz, McMaken:2023tft,Balbinot:2023vcm,McMaken:2024fvq, McMaken:2024tpc}, through blueshift instabilities (see~\cite{flanagan1997quantummechanicalinstabilitiescauchy} for an enlightening monograph on their respective physical origins). These instabilities generically lead to 
divergent energy densities at Cauchy horizons, thereby indicating the need for a full nonlinear treatment.

In this article we focus on dynamical trapped and anti-trapped regions, namely transient black-hole and white-hole regions, respectively. Such solutions are characterized by past/future and inner/outer trapping horizons, and do not possess the Cauchy or event horizons normally associated with eternal geometries. This implies that, while the divergent behaviour at inner horizons is generically avoided, some form of exponential build-up is still expected at such boundaries. This was shown explicitly for the dynamical analogue of the mass-inflation instability~\cite{Carballo-Rubio:2024dca}, as well as for its semiclassical counterpart~\cite{Carballo-Rubio:2026gwg}. Remarkably, preliminary studies seem to suggest that the latter may dominate over its classical counterpart~\cite{Barcelo:2022gii}. For this reason, and for the sake of simplicity, in this work we focus on semiclassical effects, while acknowledging that a complete treatment of the evolution of dynamical trapped/anti-trapped regions must also include the role of classical instabilities.

Inner horizons also arise in the agnostic modifications of general relativistic black hole solutions known as regular black holes~\cite{Bardeen1968,Dymnikova:1992ux,Hayward:2005gi,Ansoldi:2008jw}. As their name suggests, these metrics are free of curvature singularities and can be used as effective models for possible quantum-gravity-inspired modifications of General Relativity (GR). Inner horizons are therefore generic features of black holes formed by gravitational collapse whenever some form of repulsive effect is introduced --- whether due to charge, angular momentum, or effective quantum-gravitational corrections --- all while keeping the geometry well defined everywhere. Since the instabilities discussed above are tied to the local blueshift at the inner horizon, they are also expected to occur at the inner horizons of regular black hole solutions~\cite{Carballo-Rubio:2018pmi,Carballo-Rubio:2021bpr,DiFilippo:2022qkl}.

In simple terms, the atmosphere of vacuum polarization and spontaneously created fluxes in which these black holes are immersed is accreted and blueshifted towards the inner horizon, backreacting on the background spacetime and displacing the inner horizon away from its original position. Since vacuum effects violate the general-relativistic pointwise energy conditions~\cite{Ford:1995gb,Visser:1997sd}, they can trigger a timelike expansion of the inner horizon. In combination with the much slower inward contraction of the outer horizon, this results in an inside-out evaporation of the trapped region on timescales shorter than the Hawking time, with numerical simulations typically suggesting timescales of order {$M_{\textsc{bh}}^n$, with $n\gtrsim 2$} in natural units~\cite{barenboim2024drama2dblackhole,boyanov2025semiclassicalevolutiondynamicallyformed}.

These timescales are much shorter than the standard Hawking radiation timescale, $O(M_{\textsc{bh}}^3)$, normally assumed for the evaporation of the trapped region. In combination with the absence of curvature singularities, this has evident implications for the black hole information paradox~\cite{DiFilippo:2025kzh} and calls into question the robustness of the standard Hawking evaporation paradigm when extended beyond the simplest Schwarzschild case. Furthermore, {timescales of order $M_{\textsc{bh}}^2$} are sufficiently long to be compatible with the observed longevity of astrophysical black holes, while still allowing for potentially testable phenomenological predictions; see, e.g.,~\cite{Barrau:2014yka}.

Computing the RSET in black hole interiors is a challenging endeavour (that has seen remarkable progress in recent years~\cite{Levi:2016exv,Zilberman:2019buh, Hollands:2020qpe,McMaken:2023uue}), even more so when incorporating its backreaction~\cite{McMaken:2024fvq,Klein:2024sdd}. To get a grasp on the evolution of black hole interiors incorporating semiclassical effects, it is customary to model the renormalized stress-energy tensor by its two-dimensional counterpart~\cite{Davies:1977pvx,Polyakov:1981rd,Fabbri:2005mw}. This captures the main energetic contribution associated to $s$-wave modes of massless, minimally coupled fields when backscattering is ignored~\cite{Fabbri:2005mw}. While one should exert caution when drawing conclusions from these two-dimensional models (see e.g.~\cite{Frolov:1999an}), we will not dive into discussions on their validity here, leaving them for future investigations.

Through a two-dimensional modelling of semiclassical effects, recent works have solved the semiclassical equations~\eqref{Eq:SemiEinstein} numerically in double-null coordinates in two different, yet related, contexts: a regular black hole model in two-dimensional dilaton gravity~\cite{barenboim2024drama2dblackhole,barenboim2025evaporationregularblackholes}, and a four-dimensional Reissner-Nordstr\"om black hole~\cite{boyanov2025semiclassicalevolutiondynamicallyformed}. In both works, an initially Minkowskian background metric is matched with the respective black hole metric along a collapsing null shell. This transition is perceived by the (initially empty) $|in\rangle$ vacuum state, which reacts to the sudden change in the background by generating stress-energy fluxes along both the ingoing and outgoing null directions. 

In this paper, we study the properties of the $|in\rangle$-vacuum RSET for generic, fixed background spacetimes through analytic techniques. This
exploration allows us to propose an analytical interpretation of the results found in the above mentioned numerical backreaction analyses. 
In what follows, we summarize succinctly our main findings.

In the first place, the RSET flux component along the ingoing direction drives the dynamics of the inner and outer horizons, which merge in finite time as seen by an outside observer. This alone might appear as unsurprising, but the reactive vacuum has one more ace up its sleeve. The fluxes along the outgoing directions become amplified by the blueshifting effects of the inner horizon. In this way, they can cause light cones to tilt outwards. This happens in an especially aggressive way in the untrapped region inside the inner horizon, leading to the formation of an anti-trapped region that grows with time. The anti-trapped region can appear bounded by inner and outer horizons, as in~\cite{barenboim2024drama2dblackhole,barenboim2025evaporationregularblackholes}, or unbounded, as in~\cite{boyanov2025semiclassicalevolutiondynamicallyformed}. Our findings indicate that, to some extent, the lifetime of the anti-trapped region depends strongly on the properties of the background spacetime. 

White holes are the time-reversed version of black holes: regions of spacetime inaccessible to infalling matter. While they naturally appear as vacuum solutions in GR, no simple physical mechanism can generate them dynamically. Analogously to gravitational collapse forming black holes, a white hole could be thought of as originating from an explosive dispersal of matter. 
Formation of white holes in gravitational collapse has been studied in~\cite{hergott2025dynamicalmodelblackhole,Zhang:2025iyq, Barcel__2016,Martin_Dussaud_2019, Han_2023}. With the exception of loop quantum gravity-inspired models (see e.g.~\cite{Haggard:2015iya,Bianchi:2018mml}), usually it is hard to pinpoint a concrete mechanism responsible for these transitions. Our analysis shows that semiclassical gravity turns out to be a robust candidate since the anti-trapped region forms generically and on a timescale much shorter than the one predicted by quantum gravity-inspired calculations. 

If we ignore the well-known classical instabilities that plague anti-trapped regions~\cite{PhysRevLett.33.442,Barcel__2016}, they are nonetheless expected to be subject to a Hawking-like evaporation~\cite{PhysRevD.21.2736}. The RSET fluxes generated can, on the one hand, set the lifetime of the anti-trapped region and, on the other hand, stimulate the creation of a subsequent trapped region. Once this process ignites, it can result in a chain of self-propelled black-to-white hole transitions. By emitting radiation to infinity, this process is accompanied by a natural damping mechanism. One possible end-state of this chain effect could be reached when the energy amplification produced by the last black (white) hole is not large enough to ensure the formation of the next white (black) hole.

To get a hold on the roots of the aforementioned effects, in this paper we study the properties of the $|in\rangle$ vacuum in spacetimes describing the formation and evaporation of black holes (white holes) through collapsing (expanding) null shells. The techniques used follow the same spirit as the works~\cite{Hiscock:1980ze,Hiscock:1981xb,Balbinot:2023vcm, Carballo-Rubio:2026gwg}, but our results significantly expand on them while clarifying many physical characteristics of the results found in the numerical works~\cite{barenboim2024drama2dblackhole,barenboim2025evaporationregularblackholes,boyanov2025semiclassicalevolutiondynamicallyformed}.
In this way, we identify which elements of black-to-white hole transitions are universal, and which are model-dependent, allowing us to formulate qualitative predictions that can be tested in self-consistent numerical evolutions.

This paper is organized as follows. Section~\ref{Sec:preliminaries} introduces some useful preliminary concepts on trapped regions and 2D RSETs. Section~\ref{section: evanescentBH} calculates the RSET generated by an evanescent black hole formed by the collapse of a pair of positive and negative mass null shells. Section~\ref{section: evanescentWH} does the same computation for an evanescent white hole. Section~\ref{sec:BH-WH_transition} argues how black holes and white holes are connected dynamically via the backreaction effects of the RSET, and shows how our analytic treatment is in agreement with previous numerical results in the literature. We conclude with some discussion in Section~\ref{Sec:Discussion}.

\section{Preliminaries}
\label{Sec:preliminaries}

In the following sections we present a brief review of several notions that are well established in the literature and are relevant for the purposes of this work. Throughout the paper, we will work in $\hbar=c=G=1$ units.

\subsection{Trapped Regions and Horizons}
\label{sec: Trapped and Anti-trapped regions}
Black holes (and white holes) are solutions of GR characterized by the presence of an event horizon, defined as the boundary of the region of spacetime from which no future-directed outgoing (past-directed ingoing) null geodesic can reach future (past) null infinity $\mathscr{I}^+$ ($\mathscr{I}^-$) \cite{ashtekar2004,huber2025dynamicalhorizonsblackhole}.
In dynamical spacetimes, the standard notion of an event horizon is often not suitable, as it is a global concept defined in terms of the entire causal structure of the spacetime, demanding knowledge of its full future evolution.
To analyze gravitational collapse or evaporating black holes, it is useful to introduce quasi-local notions of horizons, defined in terms of the behaviour of null geodesic congruences in a finite region of spacetime. The dynamics of the latter is governed by the Raychaudhuri equation. For a congruence of null geodesics with tangent vector $n^\mu$, the expansion scalar $\theta$ evolves according to
\begin{equation}
\frac{d\theta_n}{d\lambda_n}
=
-\frac{1}{2}\theta_n^2 + \kappa  \theta_n
-\sigma_{\mu\nu}\sigma^{\mu\nu}
+\omega_{\mu\nu}\omega^{\mu\nu}
- R_{\mu\nu}n^\mu n^\nu ,
\label{Raychaudhuri}
\end{equation}
where $\sigma_{\mu\nu}$ is the shear tensor,  $\omega_{\mu\nu}$ is the vorticity and $\kappa$  is the surface gravity such that $n^\nu\nabla_\nu n^\mu = \kappa n^\mu$, which vanishes if $\lambda_n$ is an affine parameter. For hypersurface-orthogonal null congruences one has $\omega_{\mu\nu}=0$, and using Einstein's equations the last term can be related to the stress--energy tensor $T_{\mu\nu}$.

Consider now a closed space-like two-surface $\mathcal{S}$ and the two independent future-directed null vectors orthogonal to it, denoted by $k^\mu_{+}$ (outgoing) and $k^\mu_{-}$ (ingoing). The associated expansions $\theta_{+}$ and $\theta_{-}$ measure the variation of the transverse area along the two null directions  (i.e. $\theta_\pm = \frac{1}{A_\mathcal{S}}\frac{\delta A_\mathcal{S}}{\delta \lambda_\pm}$). The sign of the expansion $\theta_i$ indicates if the transverse area defined above is expanding ($\theta_i>0$) or contracting ($\theta_i<0$) evolving along the parameter $\lambda_i$.
The 2D space-like surface $\mathcal{S}$ can thus be classified according to the signs of the null expansions $\theta_{+}$ and $\theta_{-}$. The different cases are summarized in Table~\ref{tab:surface_classification}.
\begin{table}[h]
\centering
\begin{tabular}{|c|c|c|}
\hline
\textbf{Surface} & $\theta_{+}$ & $\theta_{-}$  \\
\hline
Trapped & $<0$ & $<0$\\
\hline
Anti-trapped & $>0$ & $>0$ \\
\hline
Untrapped & $>0$ & $<0$\\
\hline
\end{tabular}
\caption{Classification of space-like surfaces according to the sign of their null expansions.}
\label{tab:surface_classification}
\end{table}

A trapped (anti-trapped) region on a 3D hypersurface $\Sigma$ is the set of all the points in $\Sigma$ through which a trapped (anti-trapped) surface passes~\cite{Wald:1984rg,Hawking:1973uf}, otherwise the region is untrapped.
A \textit{marginal surface} is defined by the condition that one of the expansions vanishes (they can be defined as the boundary of a trapped region).  A 3D hypersurface foliated by such marginal surfaces is called a \emph{trapping horizon}.   The classification of trapping horizons depends on the sign of the non-vanishing expansion and on the variation of the vanishing one along the other direction.
The four possible types of trapping horizons are reported in the Table \ref{tab:trapping_horizons}.
\begin{table}[h]
\centering
\begin{tabular}{|c|c|c|c|}
\hline
\textbf{} & $\theta_{+}$ & $\theta_{-}$ & \textbf{$\mathcal{L}_{k_i}\theta_{k_j}$} \\
\hline
Future outer trapping horizon or FOTH & $0$ & $<0$ & $\dfrac{d \theta_{+}}{d \lambda_-} < 0$ \\
\hline
Future inner trapping horizon or FITH & $0$ & $<0$ & $\dfrac{d \theta_{+}}{d \lambda_-} > 0$ \\
\hline
Past outer trapping horizon or POTH & $>0$ & $0$ & $\dfrac{d \theta_{-}}{d \lambda_+} < 0$ \\
\hline
Past inner trapping horizon or PITH & $>0$ & $0$ & $\dfrac{d \theta_{-}}{d \lambda_+} > 0$ \\
\hline
\end{tabular}
\caption{Classification of trapping horizons according to the null expansions and their derivatives.}
\label{tab:trapping_horizons}
\end{table}

Future trapping horizons are typically associated with black hole horizons, while past  trapping horizons characterize white hole or cosmological horizons (see e.g.~\cite{Helou:2015zma}). We note that, in general, a trapping horizon is not necessarily a null hypersurface as the event horizon, but can also be spacelike or timelike depending on the sign of the flux component of the stress--energy tensor that pierces it, in particular it is spacelike (timelike) if this component is positive (negative). Further detail can be found in~\cite{Helou:2015zma,Hayward:1994bu,Ashtekar:2004cn,Helou:2016xyu,Vertogradov:2024fbd}.

Let us consider a spherically symmetric and static spacetime. The metric can be written as
\begin{equation}
ds^2 = -f(r)dx_+dx_- + r^2(x_+,x_-)d\Omega^2 ,
\end{equation}
where $x_+,x_-$ are null coordinates, $r$ is the areal radius
and $d\Omega^2$ is the metric on the unit two-sphere.
For the two families of null congruences $k^\mu_{\pm}$ with affine parameters $\lambda_{\pm}$,
the expansions take the simple form \cite{ashtekar2004,huber2025dynamicalhorizonsblackhole}
\begin{equation}
\theta_{\pm} = \frac{2}{r}\,\frac{dr}{d\lambda_{\pm}} .
\label{spherical_expansions}
\end{equation}

Let us now give concrete examples of the type of spacetimes that we will consider throughout this manuscript. 

\subsection{Charged and regular black holes}
Let us now briefly discuss the black hole spacetimes under consideration. The first example will be given by the Reissner--Nordstr\"om spacetime~\cite{1916AnP...355..106R, 1918KNAB...20.1238N}. Choosing a particular pair of $x_{+},x_{-}$ null coordinates, the line element can be written as
\begin{equation}
ds^2 = -f(r)\, du\, dv + r^2 d\Omega^2,
\label{double-null_sperical_metric}
\end{equation}
with
\begin{equation}
    u=t-r_*,\qquad v=t+r_*,
\end{equation}
the outgoing and ingoing Eddington-Finkelstein coordinates, and $r_*$ is the radial tortoise coordinate
\begin{equation}
    r_*=\int \frac{dr}{f(r)}.
\end{equation}

\subsubsection{Reissner-Nordstr\"om black hole}
The Reissner-Nordstr\"om metric is given by
\begin{equation}
    f(r) = 1 - \frac{2M}{r} + \frac{Q^2}{r^2},
\end{equation}
Some useful quantities that characterize this metric are the radial position of the outer and inner horizons, and their surface gravity, i.e.~respectively,
\begin{equation}
r_\pm = M \pm \sqrt{M^2 - Q^2}, \qquad \kappa_\pm= \pm \frac{\sqrt{M^2-Q^2}}{r_\pm^2}.
\end{equation}
Finally, the tortoise coordinate can be explicitly integrated to yield
\begin{equation}
r_* = \int \frac{dr}{f(r)}=r + \frac{1}{2\kappa_+} \ln|\kappa_+(r - r_+)|
- \frac{1}{2|\kappa_-|} \ln|\kappa_-(r - r_-)|.
\end{equation}

The Reissner-Nordstr\"om metric contains a pair of horizons if $Q<M$, a single, extremal horizon if $Q=M$ and no horizons if $Q>M$. Since these characteristics are also shared by the rotating Kerr solution, the Reissner-Nordstr\"om metric is broadly used as a proxy for studying the effects of having inner horizons while retaining the benignity of spherical symmetry. However, this spacetime contains a curvature singularity at $r=0$, which can even be naked for $Q>M$. Note also that such singularity is generically timelike and hence associated inherently with a Cauchy horizon, i.e.~a breakdown of classical predictability. So, to further study the properties of spacetimes displaying inner horizons without these nuisances, we shall also consider in what follows regular black holes.

\subsubsection{Regular black holes}\label{subsubsec:RBHs}
Regular black holes are a class of solutions to the gravitational field equations characterized by the absence of spacetime singularities in their interior. Unlike classical black hole solutions of GR, where curvature invariants typically diverge at the center, regular geometries are constructed such that all invariants remain finite everywhere. This is usually achieved by modifying the high-curvature region of the metric, replacing the singular core with a regular one, which can be of de Sitter \cite{Bardeen1968,Dymnikova:1992ux,Hayward:2005gi,Ansoldi:2008jw}, anti-de Sitter \cite{Arrechea:2025nlq}, or Minkowski type \cite{Simpson:2019mud, Panassiti:2025diw}.

These solutions are of considerable interest for at least two reasons. On the one hand, they provide an effective framework to probe possible quantum gravity effects or modified theories of gravity, where singularity resolution is expected. On the other hand, in the collapse models we will consider in the next Sections, regular black holes can avoid the formation of pathological structures such as Cauchy horizons, leading to a better-defined and more physically consistent global spacetime structure.

As an example of regular black hole metric we select the Bardeen solution~\cite{Bardeen1968}
\begin{equation}
f(r) = 1 - \frac{2M}{r}\left(\frac{r}{\sqrt{r^2+\ell ^2}}\right)^{3},
\label{Bardeen_metric}
\end{equation}
which has a pair of horizons as long as $\ell<4M/(3\sqrt{3})$, a single extremal horizon if $\ell=4M/(3\sqrt{3})$ and no horizons if $\ell>4M/(3\sqrt{3})$. This metric does not have analytic expressions for the position of the horizons and the tortoise coordinate. Fortunately, this does not significantly complicate our upcoming analysis.

Some other regular black hole metrics we are going to use are:
\begin{enumerate}


    \item {Minkowski core~\cite{Simpson:2019mud}:}
    \begin{equation}
    \label{eq:Mink}
         f(r) = 1 - \frac{\textstyle 2M}{\textstyle r} \, e^{\textstyle  -(\ell/r)^n}; \qquad n\in \mathbb{N^+}.
    \end{equation}

    \item {Anti-de Sitter core Bardeen~\cite{Arrechea:2025nlq}:  } 
    \begin{equation}
    \label{eq:AdS}
    f(r) = 1 - \frac{\textstyle 2M}{\textstyle r}\left(\frac{\textstyle r}{\textstyle \sqrt{r^2+\ell ^2}}\right)^{3}\left(\frac{\textstyle r^2-\ell^2}{\textstyle r^2+\ell^2}\right).    
    \end{equation}

\end{enumerate}
All these metrics are characterized by the same properties of the Bardeen black hole in \eqref{Bardeen_metric}, with some critical value of $\ell$ which determines if the spacetime has two, one, or zero horizons. Their main difference is the behaviour of these metrics near $r=0$. The Bardeen metric looks like the static patch of de Sitter spacetime locally around the origin. Modifying this condition can produce regular black hole spacetimes with Minkowski~\cite{Simpson:2019mud,Panassiti:2025diw} and anti-de Sitter cores~\cite{Arrechea:2025nlq}.

\subsection{Quantum fields in curved spacetimes and Polyakov approximation}
As a proxy to model the bare bones of semiclassical physics, in this work we consider a massless scalar field minimally coupled to gravity. Its dynamics is governed by the action
\begin{equation}
S[\phi,g] =
-\frac{1}{2}\int d^4x \sqrt{-g}\,
g^{\mu\nu}\nabla_\mu \phi \nabla_\nu \phi ,
\end{equation}
where $g_{\mu\nu}$ is the spacetime metric and $\nabla_\mu$ denotes the covariant derivative associated with it. Varying the action with respect to the field yields the covariant Klein--Gordon equation
\begin{equation}
\Box \phi = 0 ,
\end{equation}
where $\Box = g^{\mu\nu}\nabla_\mu\nabla_\nu$.
The stress--energy tensor of the scalar field is obtained by varying the action with respect to the metric,
\begin{equation}
T_{\mu\nu} =
-\frac{2}{\sqrt{-g}}
\frac{\delta S}{\delta g^{\mu\nu}} ,
\end{equation}
which adopts the form
\begin{equation}
T_{\mu\nu}
=
\nabla_\mu \phi \, \nabla_\nu \phi
-
\frac{1}{2} g_{\mu\nu}
\left(
\nabla^\alpha \phi \, \nabla_\alpha \phi
\right).
\label{scalar_SET}
\end{equation}

The quantization procedure we follow is the usual canonical quantization described in the literature~\cite{Birrell:1982ix, Parker:2009uva}. Once the field is quantized, its stress-energy tensor (SET) is promoted to be an operator. Selecting a quantum state of the field, the mean value of the SET is, once covariantly renormalized~\cite{Christensen:1977jc, Davies:1977pvx}, the observable that enters in the semiclassical Einstein equations~\eqref{Eq:SemiEinstein} and backreacts on the metric. 

As discussed in the introduction, it appears impossible to obtain a general analytic expression for the RSET in 4D, let alone to estimate its backreaction. Nonetheless, in spherically symmetric geometries the semiclassical backreaction problem can be simplified through dimensional reduction.

\subsubsection{Polyakov Approximation}
Consider a four-dimensional spherically symmetric spacetime with line element
\begin{equation}
ds^2 =
g_{ab}(x^c) dx^a dx^b
+
r^2(x^c)d\Omega^2 ,
\label{4D-metric}
\end{equation}
where $a,b=0,1$ denote coordinates on the radial--temporal submanifold. The scalar field can be decomposed into spherical harmonics,
\begin{equation}
\phi(t,r,\vartheta,\varphi)
=
\sum_{\ell m}
\frac{\psi_{\ell m}(t,r)}{r}
Y_{\ell m}(\vartheta,\varphi) .
\end{equation}
Substituting this decomposition into the Klein--Gordon equation leads to a set of effective 2D equations for the radial modes,
\begin{equation}
\left[
-\nabla_a\nabla^a
+
V_\ell(r)
\right]\psi_{\ell m}=0 ,
\end{equation}
where $V_\ell(r)$ is an effective potential containing contributions from the angular momentum barrier and curvature.

The Polyakov approximation consists in retaining only the dominant contribution coming from the $s$-wave sector ($\ell=0$) and neglecting the effective potential. In practice, this eliminates backscattering effects that should nonetheless be negligible near horizons, since the effective potential vanishes there.\footnote{This does not necessarily imply that the Polyakov approximation will always approximate the 4D RSET near any horizon. Take as an example the mismatch between the Polyakov RSET and the 4D RSET at Cauchy horizons~\cite{Arrechea:2024ajt}.} Under these assumptions the dynamics reduces to that of a massless scalar field propagating in a two-dimensional spacetime with metric $g_{ab}$.

The RSET of the four-dimensional theory can then be approximated by the RSET of the corresponding two-dimensional theory~\cite{Polyakov:1981rd,Fabbri:2005mw} via
\begin{equation}
\langle \hat{T}_{\mu\nu} \rangle^{\rm (4D)}
=
\frac{1}{4\pi r^2}
\langle \hat{T}_{ab} \rangle^{\rm (2D)} \delta^{a}_{\mu}\delta^{b}_{\nu} ,
\label{Polyakov_approx}
\end{equation}
where $\langle \hat{T}_{ab} \rangle^{\rm (2D)}$ is the renormalized stress--energy tensor of a massless scalar field propagating in the two-dimensional geometry $(\mathcal{M}_2,g_{ab})$ \cite{Davies:1977pvx,Fabbri:2005mw}. The multiplicative radial factor is introduced to ensure the covariant conservation of the 4D RSET. In the above expression, Greek letters take four spacetime values and Latin ones two. The $s$-wave Polyakov approximation succeeds in reproducing the main features of the Boulware, Unruh, Hartle-Hawking, and $|\textit{in}\rangle$ states~\cite{Fabbri:2005mw, Arrechea:2023fas}. It has also been proven to reproduce Hawking evaporation qualitatively~\cite{Parentani:1994ij}.

Within this approximation, the problem of determining the semiclassical stress tensor reduces to the evaluation of the much simpler 2D RSET. In the following we summarize the properties of the renormalized stress tensor in 2D that will be used in our analysis.

\subsection{Renormalized stress-energy tensor in 2D}
\label{sec:2D_RSET}
Let us now review the general treatment for the RSET of a scalar field on a curved background in 2D as developed over the last four decades (see e.g.~\cite{Balbinot:2023vcm,Hiscock:1980ze,Hiscock:1981xb}). The starting point is that any metric in 2D is locally conformally flat, so that one can always choose a set of null coordinates $(x_+,x_-)$ --- where $(x_+,x_-)$ label respectively outgoing and ingoing null rays --- for which the line element takes the form
\begin{equation}
    ds^2 = -e^{2 \sigma(x_+,x_-)}dx_+ dx_-\,.
    \label{general_null_metric}
\end{equation}

Finally, note that both the usual double null and Kruskal coordinates, typically used in describing black hole spacetimes, are of the $(x_+,x_-)$ kind.

In this frame, the scalar field obeying the 2D Klein-Gordon equation can be expanded in Fourier modes as
\begin{equation}
\psi(x_+,x_-)
=
\int_0^\infty \frac{d\omega}{\sqrt{4\pi\omega}}
\left[
a_\omega^{(+)} e^{-i\omega x_+}
+
a_\omega^{(-)} e^{-i\omega x_-}
+
\text{h.c.}
\right].
\label{mode_expansion}
\end{equation}
After canonically quantizing this field, the operators $a_\omega^{(+)},a_\omega^{(-)}$ satisfy the usual algebra of the quantum harmonic oscillators. Let us notice that in general, the concept of ``vacuum" as the state devoid of particle content is not globally defined. Indeed, if we choose a particular spacelike foliation $(\partial_\tau,\Sigma_{\tau})$, and we set the quantum state as the vacuum on one slice $\Sigma_{\tau_0}$, i.e. $a^{(-)}_\omega (\tau_0)|x_\pm\rangle = 0 $, it is not assured that for a general leaf $\Sigma_\tau$ the state will remain empty of particles, i.e. $a^{(-)}_\omega (\tau)|x_\pm\rangle \neq 0 $, unless $\partial_\tau$ is a time-like Killing vector of the spacetime~\cite{Birrell:1982ix,Parker:2009uva}. From here onwards we only consider spacetimes where such a time-like Killing vector $\partial_\tau$ exists (at least on some compact support).

Once the quantum state is defined, directly computing the mean value of~\eqref{scalar_SET} produces a divergent result: a regularization procedure is necessary. One route to derive the RSET is through a covariant point-splitting procedure~\cite{Davies:1977pvx}. However, the simplicity of the two-dimensional theory allows to fully determine it via covariant conservation $\nabla^a\langle \hat{T}_{ab}\rangle =0$ and the 2D Weyl anomaly $\langle \hat{T}_{~a}^a\rangle=-c R/24\pi $ (we follow here the procedure and notation from~\cite{Fabbri:2005mw}); where $R=-8e^{-2\sigma}\partial_+\partial_-\sigma$ is the 2D Ricci scalar,  the conformal charge for the scalar field is $c=1$, and the operator is evaluated on a generic quantum state $|\Psi\rangle$. 

The RSET assumes then the particular form

\begin{subequations}
\label{eq:RSET2D}
\begin{align}
       \langle \hat{T}_{++} \rangle &= -\frac{1}{12 \pi } \left( (\partial_+ \sigma)^2 -  \partial^2_+ \sigma\right)   -\frac{1}{12 \pi }t_+(x_+), \label{eq:RSET2Da}\\
       \langle \hat{T}_{--} \rangle &= -\frac{1}{12 \pi } \left( (\partial_- \sigma)^2  - \partial^2_- \sigma\right)  -\frac{1}{12 \pi }t_-(x_-), \label{eq:RSET2Db}\\
      \langle \hat{T}_{+-} \rangle &= -\frac{1}{12 \pi } \partial_+ \partial_- \sigma.\label{eq:RSET2Dc}
\end{align}
\end{subequations}
The functions $t_+(x_+)$ and $t_-(x_-)$ are  arbitrary and arise upon integration, they encode entirely the RSET dependence on the choice of quantum state. These two functions correspond to the normal-ordered RSET~\cite{Fabbri:2005mw}
\begin{equation}
    \langle \Psi|:T_{ab}: |\Psi\rangle = -\frac{1}{12\pi}t_a(x_a)\delta_{ab}.
\end{equation}
If the RSET has the form \eqref{eq:RSET2D} with $t_+(x_+)=t_-(x_-)=0$ then the vacuum state $| \Psi \rangle$ selected by the modes~\eqref{mode_expansion} is the so called Boulware state. In this state the RSET is zero at both null infinities and it is only characterized by the geometric part, determined by the derivatives of the conformal factor $\sigma$ independently on the quantum state. This term is often called the \textit{zero-temperature} or Boulware term, and we shall denote it here by $\Theta_{ab}(\sigma)$ for conciseness.
So, in the end, for a generic state we can write {in a given base}
\begin{equation}
    \langle \Psi|\hat{T}_{ab} |\Psi\rangle = \Theta_{ab}(\sigma) + \Delta_{ab}(x_+,x_-),
    \label{2D_RSET_psi}
\end{equation}
where we have defined the function $\Delta_{ab} (x_+,x_-) \equiv  \langle \Psi|:T_{ab}: |\Psi\rangle$. 

Now, let us perform the change of coordinates $(x_+,x_-)\rightarrow(\tilde{x}_+, \tilde{x}_-)$ such that the metric reads 
\begin{equation}
        ds^2 = -e^{2 \tilde{\sigma}(\tilde{x}_+, \tilde{x}_-)}d\tilde{x}_+ d\tilde{x}_-.
    \label{general_null_metric_changed}
\end{equation}
The normal-ordered stress energy tensor is not a primary operator, i.e.~it is not invariant under the above diffeomorphisms. Its anomalous transformation law is given by the Virasoro anomaly~\cite{Fabbri:2003vy,Fabbri:2005mw} 
\begin{equation}
    \Delta_{\tilde{a}\tilde{b}}(x_{+}, x_-)= \left(\frac{dx_a}{d\tilde{x}_{\tilde{a}}}\right)\left(\frac{dx_b}{d\tilde{x}_{\tilde{b}}}\right)\Delta_{ab}(x_+,x_-) - \frac{1}{24 \pi} \{x_a,\tilde{x}_{\tilde{a}} \} \delta_{ab}\delta_{\tilde{a}\tilde{b}},
    \label{Delta_transform}
\end{equation}
where $\{x_a,\tilde{x}_{\tilde{a}}\}$ denotes the Schwarzian derivative between the coordinates $x_a$ and $\tilde{x}_a$, reviewed in~\eqref{schwarzian}.
The transformation property \eqref{Delta_transform} ensures that the RSET is covariant under changes of coordinates. Notice that the transformation $(x_+,x_-)\rightarrow(\tilde{x}_+, \tilde{x}_-)$ does not imply a change in the vacuum state, which is still defined with respect to the modes~\eqref{mode_expansion}.  
As we have said before, choosing the quantum state $|\Psi \rangle\equiv |x_\pm \rangle$ to be vacuum with respect to the mode expansion \eqref{mode_expansion} is equivalent to setting $\Delta_{ab}(x_+,x_-)=0$; the covariance of the RSET and \eqref{Delta_transform} allows us then to write the equation \eqref{2D_RSET_psi} as
\begin{equation}
    \langle x_\pm|T_{\tilde{a}\tilde{b}} |x_\pm\rangle = \Theta_{\tilde{a}\tilde{b}}(\tilde{\sigma})   - \frac{1}{24 \pi} \{x_a,\tilde{x}_{\tilde{a}} \} \delta_{ab}\delta_{\tilde{a}\tilde{b}}.
    \label{2D_RSET_pm_vacuum}
\end{equation}
In 2D spacetimes describing the collapse (or expansion) of null shells, the line elements at the past and future of the shells are related by null coordinate transformations. In the upcoming sections, we will exploit this property and relation~\eqref{2D_RSET_pm_vacuum} to  easily calculate the RSET in these spacetimes.

Consider now a stationary spacetime of the form 
\begin{equation}
    ds^2 =-f(r) \,du_i\,dv_i = -f(r)dt_i^2 + \frac{dr^2}{f(r)},
\end{equation}
where we have introduced a generic set of double null coordinates $(u_i,v_i)$ for which a time-like Killing vector $\partial_{t_i}$ exists. 

In this setup, the zero-temperature term becomes\footnote{The vacuum state $|\rm B \rangle$ defined by the mode decomposition with $(x_+,x_-)=(u,v)$ in \eqref{mode_expansion} sets $\Delta_{ab}(u,v)=0$ in \eqref{2D_RSET_psi}. With this choice the RSET is fully determined by the zero-temperature term $\langle {\rm{B}}|T_{ab} | {\rm{B}}\rangle = \Theta_{ab}(\sigma)$. This defines the Boulware state, the vacuum state compatible with staticity and asymptotic flatness. That is where the nomenclature ``zero-temperature'' comes from.}
\begin{subequations}
\label{eq:vacuum_polarization}
\begin{align}
       \langle B|T_{u_iu_i} |B\rangle &=\Theta_{u_iu_i}(f)= -\frac{1}{192 \pi }B(r),
       \label{vpol_Tuu}\\
       \langle B|T_{v_iv_i} |B\rangle &=\Theta_{v_iv_i}(f)= -\frac{1}{192 \pi } B(r), \label{vpol_Tvv}\\
      \langle B|T_{u_iv_i} |B\rangle &=\Theta_{u_iv_i}(f)= \frac{1}{96 \pi } f(r) f(r)'',\label{vpol_Tuv}
\end{align}
\end{subequations}
where the $'$ denotes derivatives with respect to the $r$ coordinate and where the function $B(r)$ is defined as 
\begin{equation}
    B(r) = f'(r)^2 - 2f(r) f''(r).
    \label{B_function}
\end{equation} 

In this frame and for the vacuum state $|x_\pm \rangle$\footnote{Let us notice that the difference between RSETs evaluated in the vacua $|B \rangle$ and $|x_\pm \rangle$ amounts to a Schwarzian derivative (see  also \cite{Fabbri:2005mw}), but the origin of this Schwarzian is different from the change of coordinates which happens in~\eqref{Delta_transform}.  We showed though that both transformations are algebraically identical.}, using~ \eqref{2D_RSET_pm_vacuum}, the full expression of the RSET is
\begin{subequations}
\label{eq:general_vacuum}
\begin{align}
       \langle x_\pm|T_{u_iu_i} |x_\pm\rangle &= \langle {\rm{B}}|T_{u_iu_i} | {\rm{B}}\rangle - \frac{1}{24 \pi} \{x_+,u_i \},
       \label{xp_Tuu}\\
       \langle x_\pm|T_{v_iv_i} |x_\pm\rangle &= \langle {\rm{B}}|T_{v_iv_i} |{\rm{B}}\rangle - \frac{1}{24 \pi} \{x_-,v_i \},
       \label{xm_Tvv}\\
      \langle x_\pm|T_{u_iv_i} |x_\pm\rangle &= \langle {\rm{B}}|T_{u_iv_i} |{\rm{B}}\rangle.
      \label{xm_Tuv}
\end{align}
\end{subequations}

Usually, if $f(r)$ represents the metric of a black hole, there is a broader set of variables to choose in~\eqref{mode_expansion} in place of $(x_+,x_-)$ that result in non-equivalent vacua states~\cite{Balbinot:2023vcm, Birrell:1982ix,Parker:2009uva}. Note that this RSET has only up to second derivatives of the metric, which dramatically simplifies backreaction analyses.

\subsubsection{Regularity of the RSET}
 Let us now identify the mode decomposition~\eqref{mode_expansion} with the one associated to the null coordinates $(u,v)$. Such coordinates diverge at both the past/future horizons, making the modes oscillate infinitely there.\footnote{In principle, an infinitely-oscillatory mode basis need not correspond to a singular RSET, the clearest example provided by the Boulware state RSET in four-dimensional extremal black holes~\cite{Anderson:1995fw,Arrechea:2024cnv}.}
To construct states regular at the past or future horizons, we can instead use ingoing $(\bar{v},r)$ or outgoing $(\bar{u},r)$ horizon-penetrating Eddington--Finkelstein coordinates (which we shall abbreviate as IEF and OEF, respectively). In what follows, we shall discuss explicitly the case of a future horizon and IEF and simply deduce by symmetry the results for a past horizon and OEF.

So, let us consider IEF coordinates, $(\bar{v},r)$, defined w.r.t. to the double null system as $\bar{v}=v$ and $2r^\ast(r)=v-u$, and use them to evaluate the regularity of the RSET. In this system the metric takes the form\footnote{Note that this metric is explicitly regular at every future trapping horizon $r=r_h$ such that $f(r_h)=0.$ }
\begin{equation}
    ds^{2}=-f(r)\,d\bar{v}^{2}+2\,d\bar{v}\,dr\,,
\end{equation}
and the components of the RSET are
\begin{subequations}
\label{eq:TIEF}
\begin{align}
T_{\bar{v}\bar{v}}^{\mathrm{IEF}}
    &= T_{uu} + 2\,T_{uv} + T_{vv},
    \label{eq:TIEFa} \\[3pt]
T_{\bar{v}r}^{\mathrm{IEF}}
    &= -\frac{2}{f}\!\left(T_{uu} + T_{uv}\right),
    \label{eq:TIEFb} \\[3pt]
T_{rr}^{\mathrm{IEF}}
    &= \frac{4}{f^{2}}\,T_{uu}.
    \label{eq:TIEFc}
\end{align}
\end{subequations}

Now, a vacuum state is compatible with a stationary black hole if the associated RSET remains finite at the horizon. So, we require the above RSET to be regular in the IEF frame. This will be true if all its component of the RSET are finite, yielding the conditions on the RSET components in the $(u,v)$ system
\begin{subequations}
\label{regularBH_RSET}
\begin{align}
\quad
&\langle T_{uu} \rangle^{\textsc{bh}}_{\rm phys}
    = \frac{\left\langle \Psi \right| \hat{T}_{uu} \left| \Psi \right\rangle}{f^2}
    < \infty,
\label{regular_Tuu}
\\[3pt]
\quad
&\langle T_{uv} \rangle^{\textsc{bh}}_{\rm phys}
    = \frac{\left\langle \Psi \right| \hat{T}_{uv} \left| \Psi \right\rangle}{f}
    < \infty,
\label{regular_Tuv}
\\[3pt]
\quad
&\langle T_{vv} \rangle^{\textsc{bh}}_{\rm phys}
    = \left\langle \Psi \right| \hat{T}_{vv} \left| \Psi \right\rangle
    < \infty.
\label{regular_Tvv}
\end{align}
\end{subequations}
In the above expressions, the ``$\textrm{phys}$" suffix indicates the suitably rescaled components of the RSET of the $(u,v)$ frame, whose behaviour characterizes the regularity of the RSET in the well-behaved IEF frame. We will frequently make reference to the above components throughout the manuscript, but the reader can also find expressions for the energy fluxes directly perceived by freely-falling and static observers in Appendix \ref{app:physRSET}. 

The equivalent relations for past horizons and OEF, are identical but exchanging $T_{uu}$ and $T_{vv}$, i.e.,
\begin{subequations}\label{regularWH_RSET}
\begin{align}
\quad
&\langle T_{uu} \rangle^{\textsc{wh}}_{\rm phys}
    = \left\langle \Psi \right| \hat{T}_{uu} \left| \Psi \right\rangle
    < \infty,
\label{regular_Tuu_wh}
\\[3pt]
\quad
&\langle T_{uv} \rangle^{\textsc{wh}}_{\rm phys}
    = \frac{\left\langle \Psi \right| \hat{T}_{uv} \left| \Psi \right\rangle}{f}
    < \infty,
\label{regular_Tuv_wh}
\\[3pt]
\quad
&\langle T_{vv} \rangle^{\textsc{wh}}_{\rm phys}
    = \frac{\langle \Psi | \hat{T}_{vv} | \Psi\rangle}{f^2}
    < \infty.
\label{regular_Tvv_wh}
\end{align}
\end{subequations}

It follows from the above expressions that, in order to be regular on the future (past) horizon, the $T_{uu}$ ($T_{vv}$) component of the RSET must go to zero as $ \sim (r-r_h)^2$ near a non-degenerate horizon $r_h$ as $f^2$ does.\footnote{By non-degenerate, we mean a horizon with non-vanishing surface gravity.}
In order to check this point let us then study the near-horizon behaviour of the relevant RSET component~\eqref{xp_Tuu}, this is determined by \eqref{vpol_Tuu} which in turn depends on $B(r)$.

Let us start, by recalling that for any analytical metric function we can write \mbox{$f(r)=\sum_0^\infty \frac{f^{(n)}(\bar{r})}{n!}(r-\bar{r})^n$}, where by $f^{(n)}$ we denote the $n$-th derivative. Consequently, the Taylor series expansion of $B(r)$ around a general point $\bar{r}$ is
\begin{equation}
        B(r)= B(\bar{r})- 2\left(f(\bar{r})f^{(3)}(\bar{r})\right)(r - \bar{r}) - \left(f'(\bar{r})f^{(3)}(\bar{r}) + f(\bar{r})f^{(4)}(\bar{r})\right) (r - \bar{r})^2 +\mathcal{O}(r - \bar{r})^3.
    \label{B_expansion}
\end{equation}
which on a non-degenerate horizon $r_h$ such that $f(r_h)=\mathcal{O}\left(r-r_{h}\right)$ simplifies to\footnote{In what follows we use primes for spatial derivatives up to second order and the $f^{(n)}$ notation for $n>2$.}
\begin{equation}
    B(r) = f'(r_h)^2 - f'(r_h)f^{(3)}(r_h)(r - r_h)^2 + \mathcal{O}\left(r-r_h
    \right)^3.
    \label{B_horizon}
\end{equation}
It is now possible to see that the Boulware term in~\eqref{xp_Tuu} does not go ``per se" to zero at the horizon. Thus, in order to fulfill the regularity condition~\eqref{regular_Tuu}, one has to require its contribution to be cancelled by an equal and opposite contribution from the Schwarzian $\{u,u_i\}$, i.e.
\begin{equation}
\lim_{r \rightarrow r_h} \left(-\frac{1}{24\pi}\{x_+,u\}\right) = \frac{1}{192\pi} f'(r_h)^2 + \mathcal{O}(r-r_h)^2.
\end{equation} 

Having introduced this preliminary material, we are now ready to analyze the properties of 2D RSETs in spacetimes describing the formation and evaporation of black and white holes. 

\section{Evanescent Black Hole}
\label{section: evanescentBH}
\begin{figure}[h]
    \centering
    \includegraphics[width=0.8\linewidth]{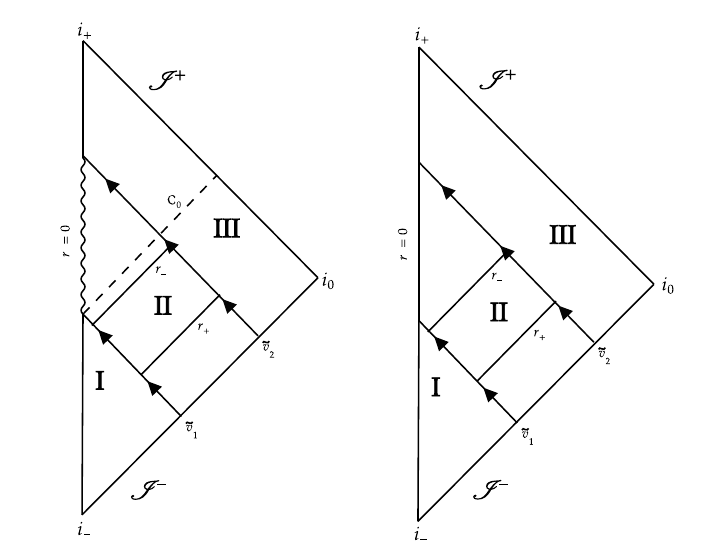}
    \caption{Penrose diagram representing formation and evaporation of a Reissner--Nordstr\"om black hole (on the left) and of a regular black hole (on the right). These spacetimes are built such that they are Minkowski in regions I and III, while region II is a black hole spacetime, which contains a trapped region. Two null shells of matter are placed in $\tilde{v}_1$ and $\tilde{v}_2$. $r_-$ and $r_+$ are respectively the inner and the outer trapping horizons (FITH and FOTH). Finally, $\mathcal{C}_0$ is the Cauchy horizon, given by the first null ray emanating from the timelike singularity; the regular BH will be used mostly due the absence of the latter.  
    }
    \label{fig:spherical_penrose}
\end{figure}

The aim of this section is first to construct simple spacetimes simulating the formation and the evaporation of a black hole.
The simplest way to model said situation consists in introducing a pair of ingoing null-shells at the times $\tilde{v}_1,\tilde{v}_2 $. The first shell, which creates the black hole, describes the collapse of classical matter and thus carries a positive mass. The second shell, which destroys the black hole by carrying a negative mass, concentrates along a constant $v$ line the entire evolution of the evaporation process. This is a simple extension of the models originally considered by Hiscock in the seminal works~\cite{Hiscock:1980ze,Hiscock:1981xb} or more recently e.g.~in~\cite{Cardoso:2023guh}. 
We present in  Fig.~\ref{fig:spherical_penrose} the Penrose diagrams of two exemplary scenarios, namely a Reissner--Nordstr\"om black hole and of a regular black hole.

With this setup, the spacetime naturally splits into three regions, described by the metrics and null coordinates below:
\begin{table}[htbp]
\centering
\renewcommand{\arraystretch}{1.6}
\begin{tabular}{|l|c|c|c|}
\hline
\textbf{Region} & \textbf{Metric} & \textbf{Null coordinates} & \textbf{Areal radius} \\
\hline
$I):\ v < \tilde{v}_1$ &
$ds_1^2 = -\,du_1\,dv$ &
$u_1 = t_1 - r,\quad v = t_1 + r$ &
$r = \dfrac{v - u_1}{2}$ \\
\hline
$II):\ \tilde{v}_1 < v < \tilde{v}_2$ &
$ds_2^2 = -\,f(r)\,du_2\,dv$ &
$u_2 = t_2 - r^{\ast},\quad v = t_2 + r^{\ast}$ &
$\displaystyle r^{\ast} = \int_{r_0}^{r}\frac{dx}{f(x)}
= \dfrac{v - u_2}{2}$ \\
\hline
$III):\ v > \tilde{v}_2$ &
$ds_3^2 = -\,du_3\,dv$ &
$u_3 = t_3 - r,\quad v = t_3 + r$ &
$r = \dfrac{v - u_3}{2}$ \\
\hline
\end{tabular}
\caption{Summary of the three spacetime regions, their corresponding metrics, null coordinates, and areal radius relations.}
\label{table:regions}
\end{table}

\subsection{Matching conditions}
In order to be consistent with this construction, it is necessary to guarantee the validity of the Israel junction conditions for null shells \cite{Balbinot:2023vcm}. Indeed, continuity of the induced metric at the shell implies that the radial coordinate (and, by extension, the $u$ coordinate) must be continuous. In what follows we shall consider an outgoing  null ray entering the trapped region at some radius $x_1$ and  advanced time $\tilde{v}_1$ and leaving it at some radius $x_2$ and advanced time $\tilde{v}_2$. Due to the peeling/anti-peeling behaviour of null rays close to the horizons of the trapped region if the ray had initial retarded time $u_1$ in region I, it will have a different retarded time $u_2$ once leaving region II.
\begin{enumerate}
\item Matching at $\tilde{v}_1$: $r(u_1,\tilde{v}_1) = r(u_2, \tilde{v}_1)\equiv x_1$
        \begin{equation}
            \frac{\tilde{v}_1-u_2}{2} =   \int_{r_0}^{x_1}\frac{dx}{f(x)}; \hspace{5mm} x_1 = \frac{\tilde{v}_1-u_1}{2} \hspace{0.5 cm} \Rightarrow \hspace{0.5 cm}
            \frac{du_2}{du_1} = \frac{1}{f(x_1)}.
            \label{v1_match}
        \end{equation}
    \item Matching at $\tilde{v}_2$: $r(u_2,\tilde{v}_2) = r(u_3, \tilde{v}_2)\equiv x_2$
        \begin{equation}
         \frac{\tilde{v}_2-u_2}{2} =\int_{r_0}^{x_2}\frac{dx}{f(x)} ;\hspace{5mm} x_2 = \frac{\tilde{v}_2-u_3}{2}\hspace{0.5 cm} \Rightarrow \hspace{0.5 cm}
            \frac{du_2}{du_3} = \frac{1}{f(x_2)}.
            \label{v2_match}    
    \end{equation}
\end{enumerate}
where $r_0$ is a lower integration limit used to regularize $r^*$ when the spacetime is singular at $r=0$. When the latter is not present (as for a regular BH), we set $r_0 =0$. Spacetimes constructed from the collapse of null shells also have the advantage that the $\Delta_{ab}$ terms in the RSET, which only involve the Schwarzian derivative between null coordinates in different patches, can be easily calculated. 

\subsection{Null geodesics behaviour}
\label{sec: null_geodesic}
Let us analyze in more detail the behaviour of the radial coordinate during the black hole phase (region II). Let us denote the lifetime of the black hole in the $v$ coordinate as $\Delta v=\tilde{v}_2 - \tilde{v}_1$.
From the matching conditions \eqref{v1_match} and \eqref{v2_match} we obtain a relation between $x_1(u_1)$ and $x_2(u_2)$. The two functions are related through the integral equation:
\begin{equation}
     \int_{x_1}^{r(x_1,v)}\frac{dx}{f(x)} = \frac{v - \tilde{v}_1}{2}, \qquad \int_{x_1}^{x_2}\frac{dx}{f(x)} = \frac{\Delta v}{2}.
    \label{r1_r2_relation}
\end{equation}   
In physical terms, the relation \eqref{r1_r2_relation}, defines $x_2$ as the value of the radial coordinate $r$ acquired by an outgoing null ray after a time $\Delta v$, starting from $x_1$, that is, $x_2=r(u_2,\Delta v)$. Equivalently, one can write, from
the definition of $r^\ast$ in Table \ref{table:regions},
\begin{equation}
    \begin{cases}
\frac{\textstyle dr(u_2,v)}{\textstyle dv}= \frac{\textstyle f(r)}{\textstyle 2}\,,\\
r(u_2,\tilde{v}_1)=x_1\,.
\end{cases}
\label{out_ray_eq}
\end{equation}

\begin{figure}
    \centering
    \includegraphics[width=0.9\linewidth]{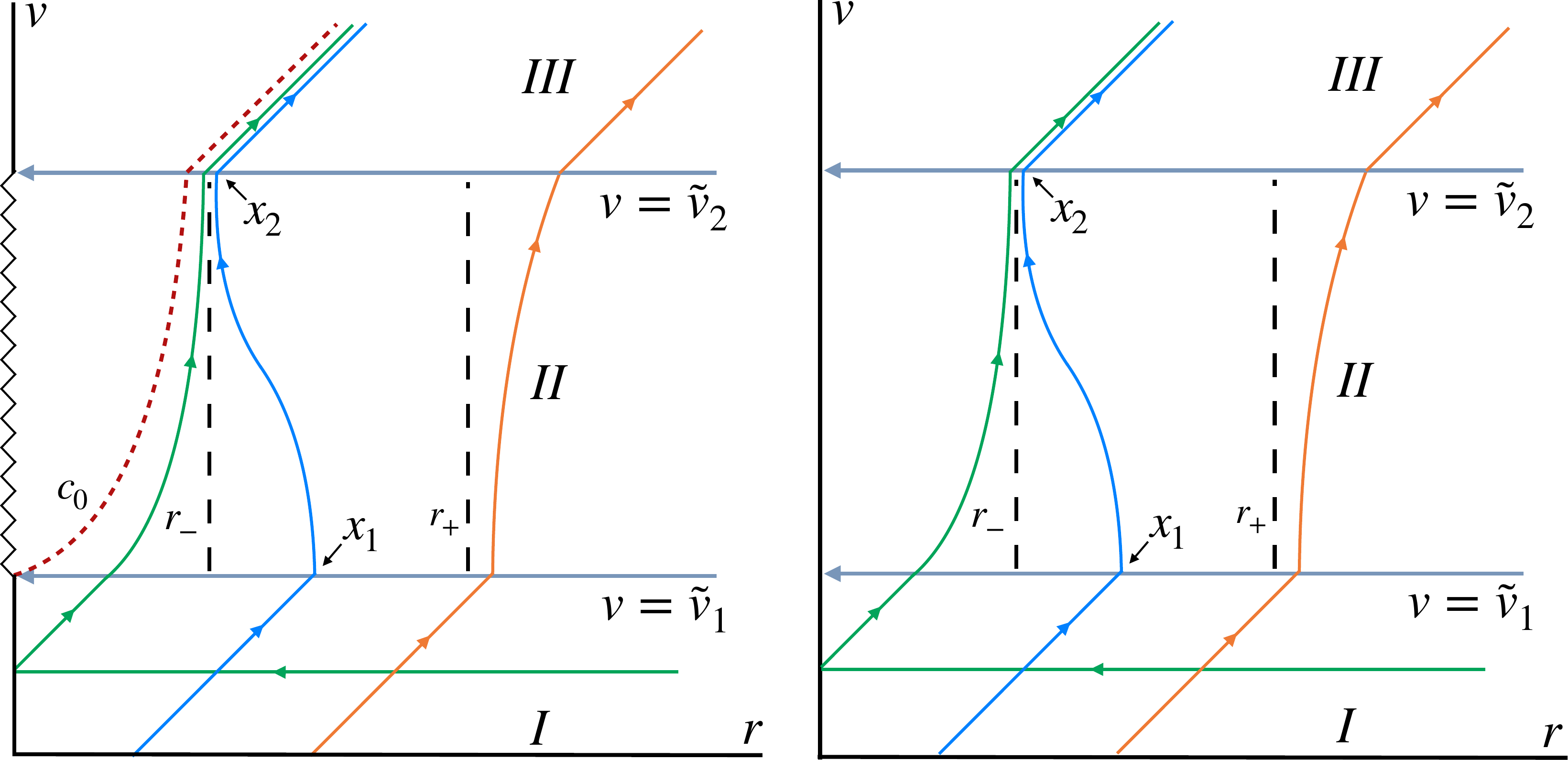}
    \caption{Graphical representation of outgoing null geodesics for Reissner--Nordström (left) and regular spacetimes (right). The main difference between the two cases lies in formation of the Cauchy horizon $\mathcal{C}_0$. One can see that null rays exhibit different behaviours depending on the radial position $x_1$ at which they cross from region I to region II.
    For $x_1 \in [0, r_-)$ (green), the rays increase in radius without surpassing $r_-$ but approaching it. For $x_1 \in (r_-, r_+)$ (blue), they decrease in radius, again approaching $r_-$. Light rays that originate closer to the outer horizon require a larger $\Delta v$ to approach the inner horizon; hence, the value of $\Delta v$ for which the approximation $x_2 \sim r_-$ holds depends on $r$.
    For $x_1 \in (r_+, \infty)$, the null rays grow indefinitely in radius, reaching $\mathscr{I}^+$ for large time.}
    \label{fig:vrdiagBH}
\end{figure}

Let us assume that in region II the spacetime is characterized by a pair of non-degenerate inner and outer horizons $r_-$ (FITH) and $r_+$ (FOTH), as depicted in Fig~\ref{fig:spherical_penrose}. We see that the behaviour of the radial coordinate acquired by the null ray depends on the form of $f(r)$. In particular the sign of $f(r)$ determines if the region is trapped ($f(r)<0$) or untrapped ($f(r)>0$).\footnote{If $f(r)<0$, from \eqref{out_ray_eq}, the radial position of the outgoing null ray decreases with time, according to Sec. \ref{sec: Trapped and Anti-trapped regions}, this defines a negative expansion and a trapped region. } For $f(r)=0$, the function $r(u_2,v)$ is constant, the constant-$r$ surfaces associated to the horizons $r_+,r_-$ is an outgoing null curve.\footnote{According to \cite{Vertogradov:2024fbd}, this is because in our model the BH is static in region II. In a dynamical situation in which we allow the BH to accrete or loose mass, the trapping horizon can be respectively space-like or time-like. } Simply put, the radial position of an outgoing null ray that enters region II at $v=\tilde{v}_{1}$ will grow with $v$ if it finds itself inside the inner horizon (green in Fig. \ref{fig:vrdiagBH}) or outside the outer horizon (orange in Fig. \ref{fig:vrdiagBH}), and decrease with $v$ if it lives in the region between horizons (blue in Fig. \ref{fig:vrdiagBH}). 

\subsection{RSET in the $|in\rangle$ state}
We choose  $(\bar{u}, \bar{v}) = (u_1,v)$ as canonical quantization variables in \eqref{mode_expansion}, which define the quantum in-state $|in\rangle$ as the vacuum in the $\mathscr{I}^-$ of region I; essentially, we suppose that the state is in the Minkowski vacuum (vanishing RSET) in the far past. 

The spacetime we built does not admit a globally defined time-like Killing vector, but by construction, all the regions I, II and III separately have one ($\partial_{t_1},\partial_{t_2},\partial_{t_3}$ respectively). This allows to apply the construction we defined in Sec.~\ref{sec:2D_RSET} separately to each region. 

To be more precise, we can rewrite the metric in Table~\ref{table:regions} in the canonical coordinates $(\bar{u},\bar{v})$. By the definition of the $|in\rangle$ state, we have $\Delta_{ab}(\bar{u},\bar{v})=0$. The term that changes among regions is the zero-temperature term $\langle {\rm B}|T_{ab} (\bar{u},\bar{v})| {\rm B}\rangle$ due to the global $v$-dependence of the conformal factor when expressed in terms of the coordinates $(u_{1},v)$.

In order to compute the RSET using the same frame as in Table~\ref{table:regions}, we apply the coordinates transformation derived from~\eqref{Delta_transform}. From~\eqref{eq:vacuum_polarization}, we have
\begin{equation}
    \langle in| T_{ab} | in \rangle = \langle {\rm B}|T_{ab} |{\rm B}\rangle - \frac{1}{24 \pi} \{ u_1,u_i \} \delta_{a u_i}\delta_{b u_i}. 
\end{equation}
where $i=1,2,3$ labels the region (I, II, III), while $a,b = u_i,v$. We obtain
\begin{enumerate}
    \item Region I:
        \begin{equation}
            \langle T_{ab} \rangle_{\rm phys}^{\textsc{bh}}= \langle in| T_{ab} | in \rangle = 0.
        \end{equation}
    \item Region II:
        \begin{subequations}
        \label{eq:RSET-regionII}
        \begin{align}
            &\langle T_{u_2u_2} \rangle_{\rm phys}^{\textsc{bh}}=\frac{\langle in|T_{u_2u_2} |in\rangle}{f(r)^2} = \frac{1}{192 \pi} \frac{\left[B(x_1) - B(r) \right]}{f(r)^2}, 
            \label{Tuu2}\\
            &\langle T_{vv} \rangle_{\rm phys}^{\textsc{bh}}=\langle in|T_{vv} |in\rangle = -\frac{1}{192 \pi}B(r),
            \label{Tvv2}\\
            &\langle T_{u_2v} \rangle_{\rm phys}^{\textsc{bh}}=\frac{\langle in|T_{u_2v} |in\rangle}{f(r)} =  \frac{1}{96 \pi} f''(r).
            \label{Tuv2}
        \end{align}
        \end{subequations}

    \item Region III:
        \begin{subequations}
        \label{eq:RSET-regionIII}
        \begin{align}
             &\langle T_{u_3u_3} \rangle_{\rm phys}^{\textsc{bh}}=\langle in|T_{u_3u_3} |in\rangle =\frac{1}{192 \pi}\frac{\left[ B(x_1) - B(x_2)\right]}{f(x_2)^2},   \label{Tuu3} \\
            &\langle T_{vv} \rangle_{\rm phys}^{\textsc{bh}}=\langle in|T_{vv} |in\rangle = 0, \label{Tvv3}\\
            &\langle T_{u_3v} \rangle_{\rm phys}^{\textsc{bh}}=\langle in|T_{u_3v} |in\rangle = 0. \label{Tuv3}
        \end{align}
        \end{subequations}
\end{enumerate}


Let us check the regularity of the RSET in the different regions. 
Evaluating the RSET along a generic null line corresponding to a horizon ($x_1 \rightarrow r_h \implies x_2 \rightarrow r_h $), We obtain (for the explicit computation check Appendix \ref{app:reg_check}):
\begin{enumerate}
    \item Region I:
        \begin{equation}
            \langle T_{ab} \rangle_{\rm phys}^{\textsc{bh}}|_{r_h} = 0.
        \end{equation}
    \item Region II:
        \begin{subequations}
        \label{eq:RSET-horizonII}
        \begin{align}
            &\langle T_{u_2u_2} \rangle_{\rm phys}^{\textsc{bh}}|_{r_h} = \frac{1}{384 \pi}\frac{f^{(3)}(r_h)}{\kappa_h} \left[1 -e^{-2\kappa_h(v-\tilde{v}_1)} \right], 
            \label{Tuu2_horizon}\\
            &\langle T_{vv} \rangle_{\rm phys}^{\textsc{bh}}|_{r_h} = -\frac{1}{48 \pi}\kappa_h^2,
            \label{Tvv2_horizon}\\
            &\langle T_{u_2v} \rangle_{\rm phys}^{\textsc{bh}}|_{r_h} = \frac{1}{96 \pi} f''(r_h).
            \label{Tuv2_horizon}
        \end{align}
        \end{subequations}

    \item Region III:
        \begin{subequations}
        \label{eq:RSET-horizonIII}
        \begin{align}
             &\langle T_{u_3u_3} \rangle_{\rm phys}^{\textsc{bh}}|_{r_h} = \frac{1}{384\pi}\frac{f^{(3)}(r_h)}{ \kappa_h} \left[ 1 - e^{-2 \kappa_h \Delta v}\right],
             \label{Tuu3_horizon}\\
            &\langle T_{vv} \rangle_{\rm phys}^{\textsc{bh}}|_{r_h} = 0,\\
            &\langle T_{u_3v} \rangle_{\rm phys}^{\textsc{bh}}|_{r_h} = 0.
        \end{align}
        \end{subequations}
\end{enumerate}
Where $\kappa_h =f'(r_h)/2$ is the surface gravity at the horizon $r_h$. Recall that $\kappa_{h}>0$ for a FOTH and $\kappa_{h}<0$ for a FITH. This determines the sign of the exponent in the above exponentials, and thus the growth or decrease of the RSET with $v$. It is also clear from the above expressions that the RSET satisfies the regularity conditions~\eqref{regularBH_RSET} on both the horizons $r_-$ and $r_+$.

These results illustrate the marked difference between the ingoing and outgoing RSET components.
The ingoing component $\langle T_{vv}\rangle $, Eq.~\eqref{Tvv2_horizon}, which is transverse to trapping horizons, is always negative on them and proportional to their respective surface gravities squared. This energy flux, which jumps from $0$ to a constant value along the incoming shell, is being accreted into the black hole at the outer horizon, and it depends solely on the horizon surface gravity; indeed, after evaporation (region III), this RSET component vanishes. The universal, negative sign of this flux is of fundamental importance, as it provides the mechanism responsible for black hole evaporation. In a self-consistent picture, it will drive the FITH and the FOTH closer together, eventually making the trapped region disappear in a finite $v$ time~\cite{barenboim2024drama2dblackhole, boyanov2025semiclassicalevolutiondynamicallyformed}. 

On the contrary, the outgoing component $T_{uu}$ is not completely determined by the surface gravity of the horizon, but it is related to the derivative of the spacetime curvature instead (recall that the 2D Ricci scalar is just $R\propto f''(r)$). Indeed, the sign of the outgoing energy fluxes that are emitted after the evaporation time gets entirely determined by $f^{(3)}(r_h)$. This quantity is sensitive to the particular background that we are considering, and there is no universal sign that we can assign to it. 

The nature of the horizon (whether a FITH or a FOTH) determines the asymptotic behaviour of the outgoing flux at late times:
\begin{enumerate}
    \item FOTH:
        $ \kappa_h>0 \implies \lim_{\Delta v \rightarrow \infty}|\langle T_{u_3u_3} \rangle_{\text{phys}}^{\textsc{bh}}|_{r_h}\approx\frac{1}{384\pi}\frac{|f^{(3)}(r_h)|}{|\kappa_h|} < \infty. $
    \item FITH:
        $ \kappa_h<0 \implies \lim_{\Delta v \rightarrow \infty}|\langle T_{u_3u_3} \rangle_{\text{phys}}^{\textsc{bh}}|_{r_h} \approx \frac{1}{384\pi}\frac{|f^{(3)}(r_h)|}{|\kappa_h|} e^{2 |\kappa_h| \Delta v}\to\infty. $
\end{enumerate}
We see that along the FITH energy accumulates exponentially fast during the lifetime of the evanescent black hole. Such energy is in the end (once the evaporation finishes) released as a highly collimated outgoing pulse centred at $u_3 = \tilde{v}_2 - 2 r_h$.

The importance of the $T_{uu}$ component lies not in driving the evaporation of the trapping horizons, but in changing the sign of the ingoing expansion $\theta_{-}$ throughout the interior of the black hole. Through its backreaction effects, this component is thus responsible for the creation of an anti-trapped region during the black hole phase. This phenomenon has been observed in numerical simulations of evaporating charged and regular black holes~\cite{barenboim2024drama2dblackhole,boyanov2025semiclassicalevolutiondynamicallyformed,barenboim2025evaporationregularblackholes}. Our analysis suggests that the formation of this anti-trapped region is a generic consequence of the evolution of the $|\textit{in}\rangle$-state RSET inside a trapped region. However, the specific properties of this anti-trapped region (lifetime and size) 
will depend on the particular choice of background spacetime characterizing the original black hole.

Let us now extract more information from the above expressions by considering two limits amenable to analytic treatment: $\Delta v\to 0$ and $\Delta v \to \infty$.

\subsection{Early times limit $(v - \tilde{v}_1\rightarrow 0)$}
\label{subsec:smalldv_BH}
Let us focus on the region II, we want to compute the value of the outgoing RSET $\langle T_{u_2u_2} \rangle_{\text{phys}}^{\textsc{bh}}$ in a specific point $r(x_1,v)$ with $v \in [\tilde{v}_1,\tilde{v}_2]$, in the limit $\,dv = v-\tilde{v}_1 \rightarrow 0$ (early times evolution). We define the function $\delta(x_1)$ such that $r(x_1,v) = x_1 + \delta$, from \eqref{r1_r2_relation} we can write 
\begin{equation}
    \int_{x_1}^{x_1 + \delta}\frac{dx}{f(x)} = \frac{v-\tilde{v}_1}{2},
    \label{delta_relation}
\end{equation}
being obvious that $\lim_{\,d v \rightarrow 0^+} \delta (x_1) = 0$. In this limit, we can approximately solve the above integral through the Taylor expansion  $f(x) \sim f(x_1) + f'(x_1)(x-x_1)+\cdots$. In this regime it is straightforward to find
\begin{equation}
    \delta(x_1) = \frac{f(x_1)}{f'(x_1)} \left(e^{f'(x_1) \frac{\,d v}{2}} -1\right).
\end{equation}
Let us note that $\delta(r_-) = \delta(r_+)=0$, verifying that null rays sitting at the horizons stay at constant $r$ values, as seen from Fig.~\ref{fig:spherical_penrose}.

If $\delta$ is small, it is possible to obtain a compact expression for the RSET in region II. Expanding Eq.~\eqref{Tuu2} around $\delta=0$ to first order, and then expanding for $dv\to0$, we obtain
\begin{equation}
    \langle T_{u_2u_2} \rangle_{\text{phys}}^{\textsc{bh}} \simeq\frac{1}{96\pi}
    \frac{f^{(3)}(x_1)}{ f'(x_1)}\left( 1 - e^{-f'(x_1)\frac{\,d v}{2}}\right) 
    \simeq \frac{f^{(3)}(x_1)}{192\pi}\,d v
    \simeq \frac{f^{(3)}(x_2)}{192\pi}\,d v.
    \label{Tuu2_smalldv}
\end{equation}
In the limit $\Delta v \rightarrow 0 $, the same component in region III is simply the expression \eqref{Tuu2_smalldv} computed in $v=\tilde{v}_2$ ( $x_2 = r(x_1,\tilde{v}_2)$), namely
\begin{equation}
    \langle T_{u_3u_3} \rangle_{\text{phys}}^{\textsc{bh}}
    \simeq \frac{f^{(3)}(x_1)}{192\pi}\Delta v.
    \label{Tuu3_smalldv}
\end{equation}
The flux does not change in $v$ in region III, the flux accumulated at $\tilde{v}_2$ is frozen and released from $r_-$ to increasing radial values.
Eq.~\eqref{Tuu2_smalldv} shows us that the outgoing flux in this regime is generated by the variation of the 2D curvature (or, equivalently, the trace anomaly) that a null outgoing light-ray ``detects" along their way through region II. Indeed, this result is more illuminating if we look at the ``accumulation" that happens in region II; with the same approximation~\eqref{delta_relation}, the RSET in region II, from \eqref{Tuu2_smalldv} and \eqref{regular_Tuu} reads
\begin{equation}
        \langle in|T_{u_2u_2} |in\rangle 
    \simeq \frac{f^2(x_1) f^{(3)}(x_1)}{192\pi}\,d v .
    \label{RTuu2_smalldv}
\end{equation}

The total outgoing flux accumulated in an interval $v-v_{1}$ is naturally computed as 
\begin{equation}
    \langle in|T_{u_2u_2} |in\rangle 
    = \frac{1}{192\pi}\int_{\tilde{v}_1}^{v} f^2[r(u_2, v')] f^{(3)}[r(u_2, v')]d v' .
    \label{Tuu2_integral}
\end{equation}
This can be thought of as the total outgoing flux along a null ray at $u_2=$const.~from its entry into region II at $(x_1,\tilde{v}_1)$ up to some time $v$.

After using $dr=f(r)dv/2$, the general result~\eqref{Tuu2} is obtained again. More explicitly,
\begin{align}
    \langle in|T_{u_2u_2} |in\rangle 
    &= \frac{1}{96\pi}\int_{x_1}^{r(u_2,v)} f(r') f^{(3)}(r')d r' \\
    &= -\frac{1}{192\pi}B(r')\big|^{r}_{x_1}\notag\\
    &= \frac{B(x_1)-B(r)}{192\pi}\notag,
    \label{Tuu2_integral_radius}
\end{align}
Hence, moving along an outgoing null ray for an infinitesimal lapse of advanced time $\,dv$,  
the energy accumulated is proportional to ``how quickly" the 2D curvature --- the Ricci scalar which is in turn proportional to $f''\left(r(x_1,v)\right)$ --- is changing. This is surprising because the nature of the outgoing component of the RSET seems to be strongly model-dependent and, as said in the previous section, different w.r.t the ingoing one, which is constant in time and depends mainly on the surface gravity of the horizon.\footnote{Note however, that such behavior is most probably an artifact of the Polyakov approximation: in 2D the trace anomaly is simply proportional to the scalar curvature, while this would not be generally the case in $(3+1)D$} When the BH is evaporated leaving only a Minkowski spacetime, i.e.~once one gets in region III, the outgoing flux stored until the instant $\tilde{v}_2$ stops growing and is released towards $\mathscr{I}^+$. Note however, that for sufficiently large $\Delta v$ the backreaction of these exponentially-increasing outgoing fluxes is expected to be non-negligible long before the trapped region fully evaporates. 

\subsection{Late times limit $(v - \tilde{v}_1\rightarrow \infty)$}

\label{bigdv_BH}
Another limit of remarkable interest is $v-\tilde{v}_{1}\to\infty$ (with $\Delta v \rightarrow\infty$ as well), which describes a stationary black hole. Although it does not need to describe the actual end-point of the trapped region --- which cannot be eternal due to the horizons evaporation --- this limit allows us to verify if the $|\textit{in}\rangle$-vacuum approaches the Unruh-vacuum one at late-times. Surprisingly, we shall see that this is not exactly true close to the inner horizon.

Let us detail the behaviour of $x_2(x_1)$ in region $\rm{II}$ for the black hole interior, defined by $x_1 \in [0,r_+]$. For long living black holes, Eq.~\eqref{r1_r2_relation} reduces approximately to $x_2 \sim r_-$  $\forall x_1 \in [0,r_+)$.
Under this assumption, and combining expressions~\eqref{B_expansion},~\eqref{r1_r2_relation}, and~\eqref{Tuu3}, we obtain
\begin{equation}
         \langle T_{u_2 u_2} \rangle_{\text{phys}} \simeq
         \frac{f^{(3)}(r_-)}{384\pi\kappa_-}
         + \frac{1}{192\pi r_-^2}\left(\frac{B(x_1)}{4 \kappa_-^2}-1\right)
         e^{4 |\kappa_-|\left[ R_{\textsc{bh}}(x_1) + \left(v-\tilde{v}_{1}\right)/2\right]}.
         \label{Tuu2_bigdv}
      \end{equation}
Where we have defined 
    \begin{equation}
      R_{\textsc{bh}}(x_1) = \lim_{|\epsilon| \rightarrow 0}\int_{r_\epsilon}^{x_1}\frac{dx}{f(x)} + \frac{1}{2\kappa_-}\ln(|\epsilon|),
      \label{RBH}
    \end{equation}
with $r_\epsilon= r_-(1+\epsilon)$ and $\epsilon= |\epsilon|$sign$(x_1-r_-)$.
As before, the outgoing flux in region III is\footnote{The corresponding large-$\Delta v$ coefficient is
\[
\lim_{\Delta v\to\infty}\frac{\langle T_{u_3u_3}\rangle_{\rm phys}}{e^{2|\kappa_-|\Delta v}}
=\frac{e^{4|\kappa_-|R_{\textsc{bh}}(x_1)}}{192\pi r_-^2}
\left(\frac{B(x_1)}{4\kappa_-^2}-1\right).
\]
This is the quantity represented in Figs.~\ref{fig:RNTuulong} and~\ref{fig:BardeenTuuLong}.}
\begin{equation}
         \langle T_{u_3 u_3} \rangle_{\text{phys}} \simeq
         \frac{f^{(3)}(r_-)}{384\pi\kappa_-}
         + \frac{1}{192\pi r_-^2}\left(\frac{B(x_1)}{4 \kappa_-^2}-1\right)
         e^{4 |\kappa_-|\left[ R_{\textsc{bh}}(x_1) + \Delta v/2\right]}.
         \label{Tuu_bigdv}
\end{equation}
For sufficiently large $ v$, the outgoing flux perceived along outgoing null geodesics close to the inner horizon  exhibits an exponential growth in the advanced coordinate $v$, with exponent proportional to the multiplicative factor \(\eqref{RBH}\), which depends on the parameter $x_1$. The ingoing flux stays negative and constant. In the exterior region, the exponent in~\eqref{Tuu_bigdv} changes sign (being determined by the positive outer horizon surface gravity $\kappa_+$) and the outgoing flux approaches a constant value. 

When the evolution of these fluxes in $v$ is measured at constant-radius surfaces we expect the following qualitative behaviour.
In the region $r>r_{-}$, this state approaches the Unruh-vacuum RSET at late times, in consistency with the results derived in~\cite{Anderson:2025wzp} for the Schwarzschild spacetime. 

In the limit $\Delta v\to \infty$ the ray at $\tilde{v}_2=$const.~in region II, becomes a Cauchy horizon at $r=r_{-}$. In this case, there is no meaningful sense in which the $|\textit{in}\rangle$-state RSET can be said to approach the Unruh one there, since outgoing fluxes blow up exponentially in physical coordinates for the former, while being always divergent with a different behaviour for the latter~\cite{Balbinot:2023vcm, Carballo-Rubio:2026gwg}. Hence, at $r=r_{-}$ (contrarily to $r=r_{+}$), the $|\textit{in}\rangle$-state RSET does not approach any known stationary state at late times. This calls into question how well the 4D Unruh RSET describes the late time limit of gravitational collapse at the inner horizon. These subjects deserve a standalone investigation that goes beyond the current scope of this work.

\subsection{Numerical results for charged and regular black holes}
For the rest of the manuscript, when numerical graphs are shown, all quantities are normalized with respect to the ADM mass of the black hole, i.e.~$\tilde{T}_{ab}=M^2T_{ab}$, $\tilde{r}_h=r_h/M$, and $\tilde{\kappa}_h=M\kappa_h$. For regular black holes, the parameter that determine the geometry of the spacetime is $L=\ell/M$; for each of them there is a critical value $L_c$ corresponding to the extremal geometry; e.g.~for the Bardeen black hole the critical value is $L_c=4/(3\sqrt{3})$. Let us now present some exemplary cases.

\begin{figure}
    \centering
    \includegraphics[width=0.8\linewidth]{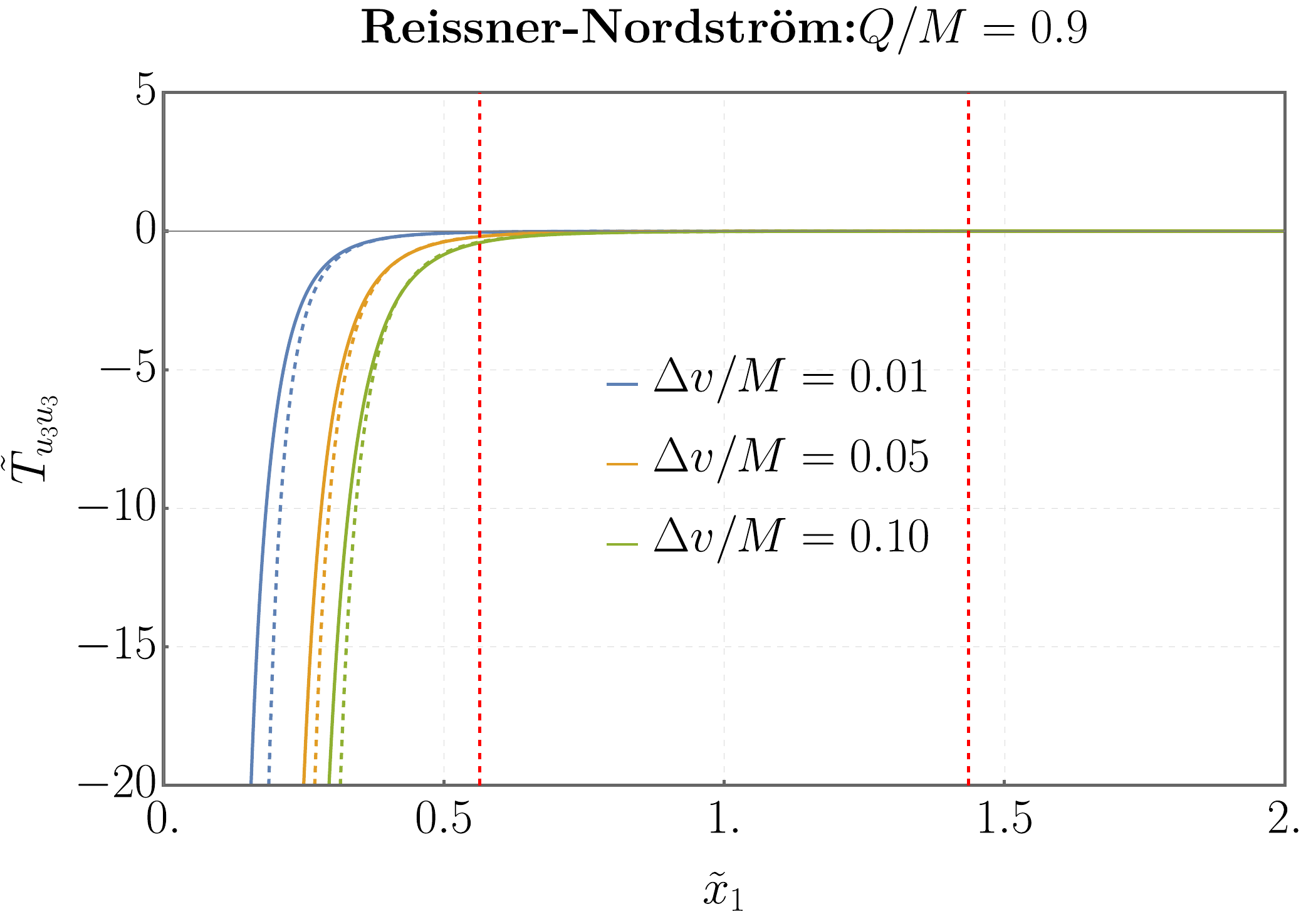}

    \caption{ Outgoing flux produced by the formation and evaporation of a Reissner-Nordstr{\"o}m black hole with $Q/M=0.9$. The red dashed lines are the locations of the inner and outer horizons, and the continuous blue, yellow, and green curves correspond to the exact fluxes generated black holes of different lifetimes $\Delta v$. Their dashed counterparts represent the $\Delta v\to 0$ analytic approximation. These fluxes are always negative in the interior region, and diverge negatively as the Cauchy horizon $\tilde{x}_{1}=0$ is approached.}
    \label{fig:RNTuu}
\end{figure}

\begin{figure}
    \centering
    \includegraphics[width=0.8\linewidth]{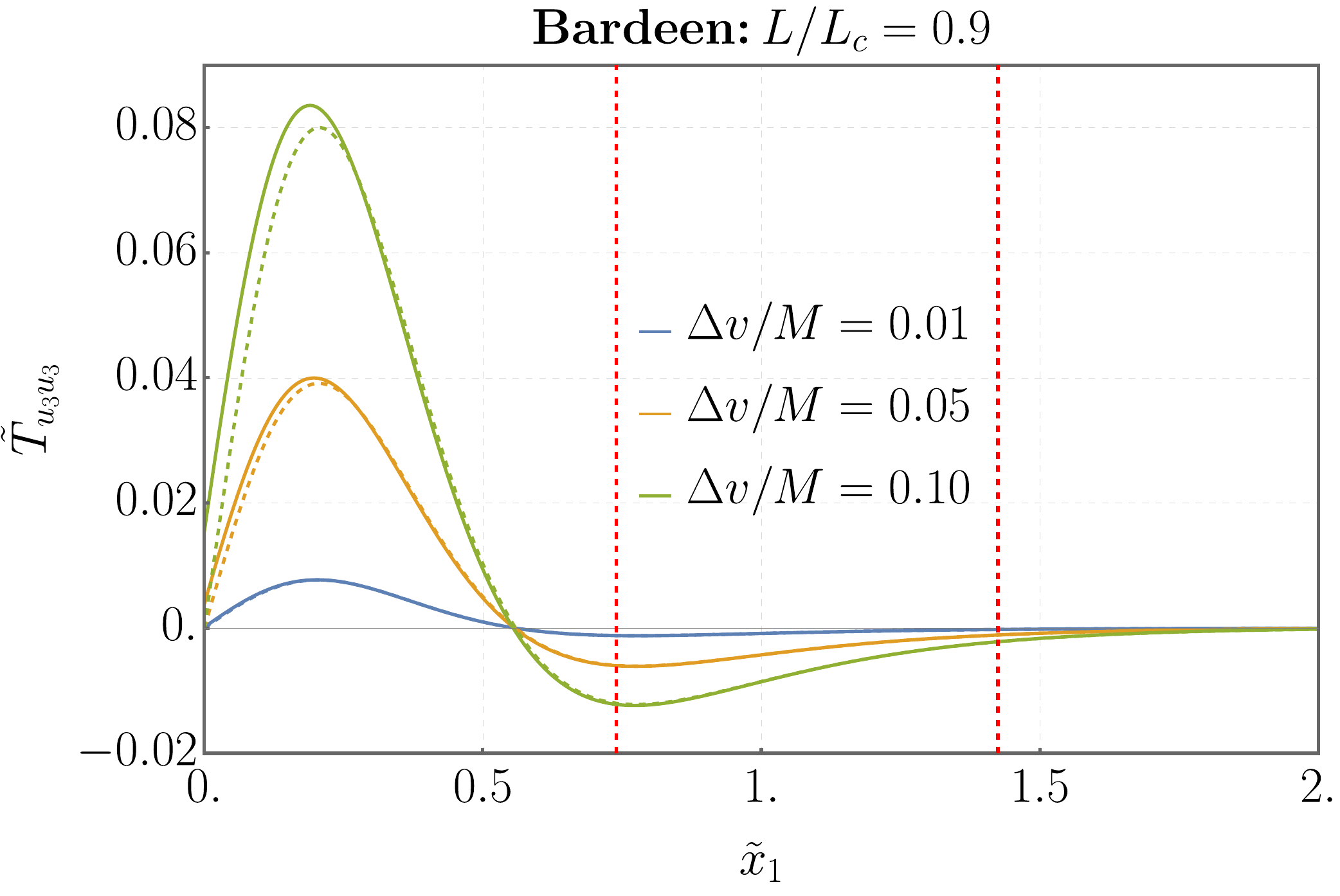}
    \caption{ Outgoing flux produced by the formation and evaporation of a Bardeen regular black hole with $L/L_c=0.9$. The red dashed lines are the locations of the inner and outer horizons, and the continuous blue, yellow, and green curves correspond to the exact fluxes generated black holes of different lifetimes $\Delta v$. Their dashed counterparts represent the $\Delta v\to 0$ analytic approximation. Regions of negative and positive values are found for the outgoing flux.}
    \label{fig:BardeenTuu}
\end{figure}

We have computed numerically the outgoing fluxes for different background geometries and lifetimes, finding exact agreement with our analytic approximations. Figures~\ref{fig:RNTuu} and~\ref{fig:BardeenTuu} show the~$\langle T_{u_3 u_3}\rangle^{\text{BH}}_{\text{phys}}$ component (i.e., the outgoing flux in region III) produced by the formation and evaporation of Reissner-Nordstr\"om and Bardeen black holes with different lifetime $\Delta v$. The dashed lines represent the $\Delta v\to0$ analytic approximation derived in Subsec.~\ref{subsec:smalldv_BH}, while the continuous lines are exact numerical results. 
We have considered black holes close to extremality for illustrative purposes, but the qualitative shape of the RSET does not change significantly far from extremality. 

The main difference between Reissner-Nordstr\"om and Bardeen is that, in the first case, $\langle T_{u_3 u_3}\rangle_{\text{phys}}$ is mostly negative, whereas in the second case it displays regions of negative and positive values. This is directly linked to the respectively singular or regular behaviour of the geometry at $r=0$. Different models of regular black holes can exhibit more complicated shapes, but a region with negative flux values is present in every model we have analyzed. This is quite remarkable, as it is evidence that the presence or absence of a singularity behind the inner horizon could influence fluxes eventually reaching infinity at late times.

\begin{figure}
    \centering
    \includegraphics[width=0.8\linewidth]{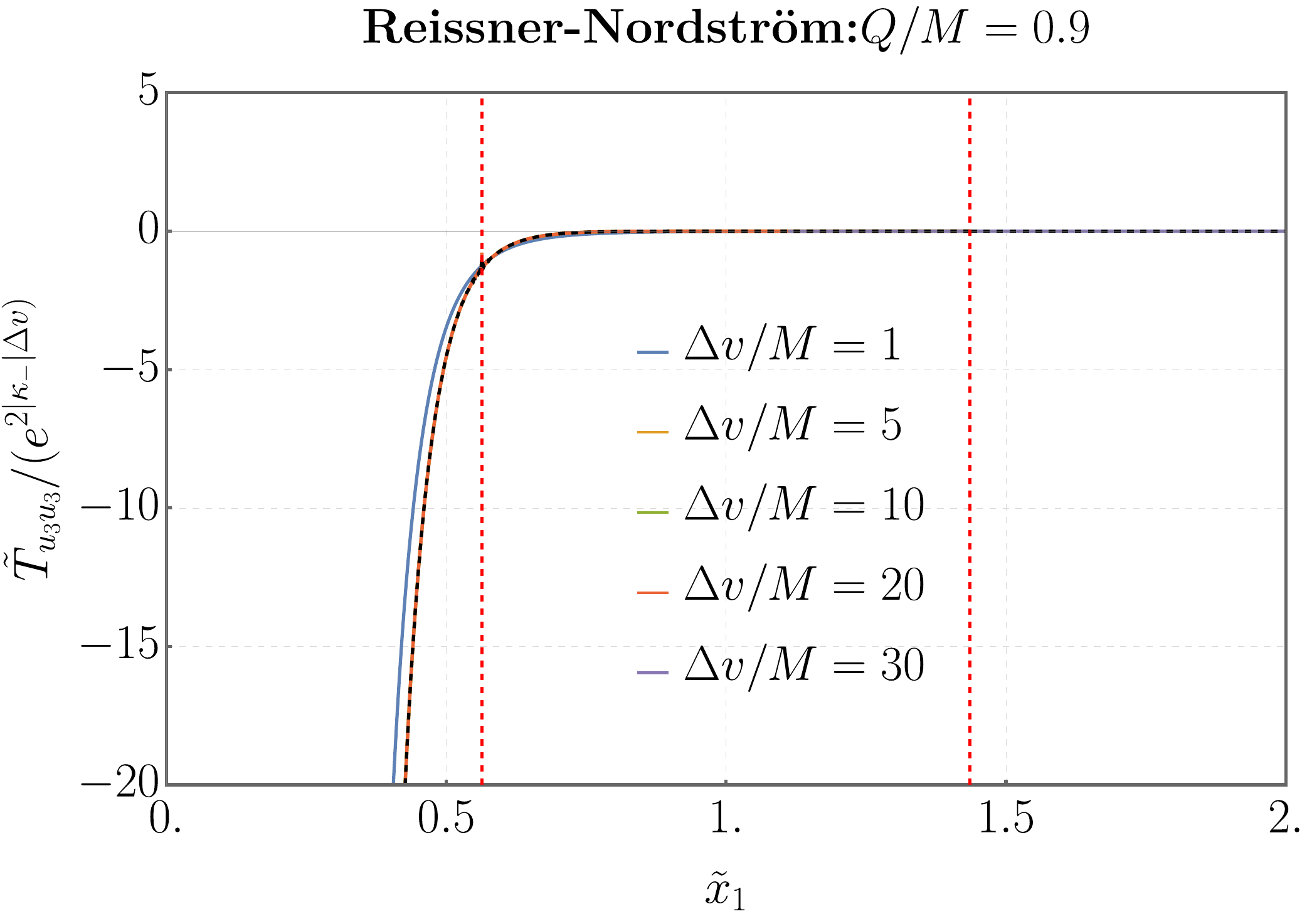}

    \caption{Outgoing flux produced by the formation and evaporation of a Reissner-Nordstr{\"o}m black hole with $Q/M=0.9$ and large lifetime $\Delta v$. The red dashed lines are the locations of the inner and outer horizons, and the continuous blue, yellow, green, and red curves correspond to the exact fluxes. The exponential growth of this quantity has been subtracted, showing that for sufficiently large $\Delta v$, the coefficient of this growth approaches a fixed function of $\tilde{x}_{1}$ (in dashed black; see Eq.~\eqref{Tuu_bigdv}).}
    \label{fig:RNTuulong}
\end{figure}

\begin{figure} 
\centering
    \includegraphics[width=0.8\linewidth]{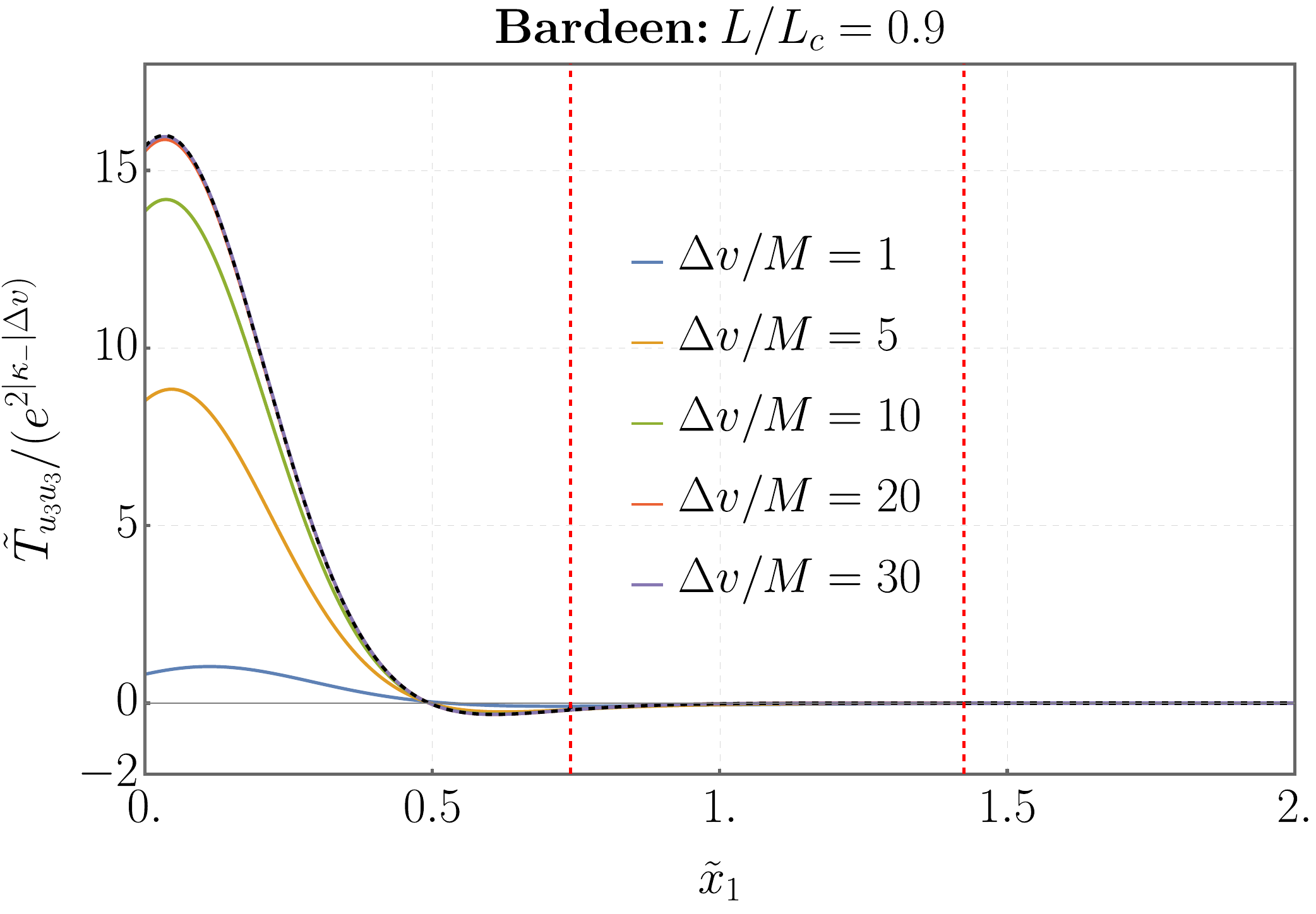}
    \caption{ Outgoing flux produced by the formation and evaporation of a Bardeen regular black hole with $L/L_c=0.9$ and large lifetime $\Delta v$. The red dashed lines are the locations of the inner and outer horizons, and the continuous blue, yellow, green, and red curves correspond to the exact fluxes. The exponential growth of this quantity has been subtracted, showing that for sufficiently large $\Delta v$, the coefficient of this growth approaches a fixed function of $\tilde{x}_{1}$ (in dashed black; see Eq.~\eqref{Tuu_bigdv}).}
    \label{fig:BardeenTuuLong}
\end{figure}

Indeed, the background initial shape of $\langle T_{u_3 u_3}\rangle_{\text{phys}}$ acts as a ``seed" that is then exponentially amplified in $v$, as can be seen in Figs.~\ref{fig:RNTuulong} and~\ref{fig:BardeenTuuLong}. Values for large $\Delta v$ are in excellent agreement with the analytic approximation~\eqref{Tuu_bigdv} (for region III, we just replace $v-\tilde{v}_{1}$ by $\Delta v$).
 
As we will discuss in Section~\ref{sec:BH-WH_transition}, when these outgoing fluxes backreact on the spacetime, they can generate an anti-trapped region. The size this anti-trapped region will depend on $\Delta v$, and its lifetime will be tightly related to the initial $\langle T_{{u_3 u_3}}\rangle_{\text{phys}}$ ``seed". In particular, a compact (finite-lived) anti-trapped region requires the presence of regions with negative and positive outgoing fluxes. One can already infer this is the case by looking at the formation and evaporation of white holes, which will be the topic of next Section. 

\section{Evanescent White Hole}
\label{section: evanescentWH}
\begin{figure}
    \centering
    \includegraphics[width=0.4\linewidth]{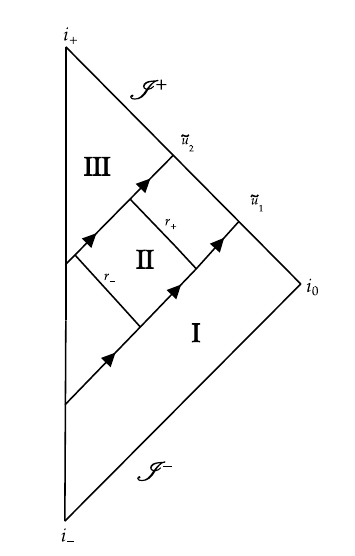}
    \caption{Penrose diagram representing Formation and evaporation of a Regular White hole. This spacetime is built such that they are Minkowski in regions I and III, in region $II$ an anti-trapped region is present. Two null outgoing shells of matter are placed in $\tilde{u}_1$ and $\tilde{u}_2$. $r_-$ and $r_+$ are respectively the inner and the outer apparent horizons (PITH and POTH). }
    \label{fig:WH_penrose}
\end{figure}
Observing the spacetimes obtained in numerical simulations \cite{barenboim2024drama2dblackhole,boyanov2025semiclassicalevolutiondynamicallyformed}, after the evaporation  of the trapped region one sees the formation and eventually the subsequent evaporation (only for regular black hole spacetimes) of an anti-trapped region. After that, for some collection of parameters, the formation of a second trapped region can be observed and the process continues in cascade. It is therefore worthwhile to reformulate the model presented in the Section~\ref{section: evanescentBH} in order to investigate the semiclassical effects associated with an anti-trapped region.
In this Section, we build a toy model for an evanescent anti-trapped region (i.e. a white hole). The construction is exactly the same as for the black hole, but using outgoing shells in place of ingoing, and swapping the signs of their masses (the first and second shells carry negative and positive mass, respectively).\footnote{By a classical point of view, the spacetime constructed in Fig. \ref{fig:WH_penrose} is the time reverse ( $v\rightarrow - u$) of the regular Penrose diagram in Fig. \ref{fig:spherical_penrose}.} Being the time reversed solution of a gravitational collapse, a white hole is created, classically, by an ``explosion" of matter (which is why in the model two outgoing rather than ingoing shells of matter $\tilde{u}_1$ and $\tilde{u}_2$ are present). 
The various spacetime regions can be labelled as
\begin{center}
\renewcommand{\arraystretch}{1.6}
\begin{tabular}{|l|c|c|c|}
\hline
\textbf{Region} & \textbf{Metric} & \textbf{Null coordinates} & \textbf{Radius} \\
\hline
$I): u < \tilde{u}_1$ &
$ds_1^2 = -\,du\,dv_1$ &
$  u = t_1 - r, \quad v_1 = t_1 + r$ &
$ r = \dfrac{v_1 - u}{2}$ \\
\hline
$ II): \tilde{u}_1 < u < \tilde{u}_2$ &
$ds_2^2 = -\,f(r)\,du\,dv_2$ &
$\ u = t_2 - r^{\ast}, \quad v_2 = t_2 + r^{\ast}$ &
$ r^{\ast} = \int_{r_0}^{r}\frac{dx}{f(x)}
= \dfrac{v_2 - u}{2}$ \\
\hline
$ III): u > \tilde{u}_2$ &
$ds_3^2 = -\,du\,dv_3$ &
$ u = t_3 - r, \quad v_3 = t_3 + r$ &
$ r = \dfrac{v_3 - u}{2}$ 

\label{Table_regions}\\
\hline

\end{tabular}
\end{center}
By applying the matching conditions we get:  
\begin{enumerate}
\item Matching at $\tilde{u}_1$: $r(\tilde{u}_1,v_1) = r(\tilde{u}_1,v_2)\equiv y_1$
        \begin{equation}
            \frac{v_2-\tilde{u}_1}{2} =   \int_{r_0}^{y_1}\frac{dx}{f(x)}; \hspace{5mm} y_1 = \frac{v_1-\tilde{u}_1}{2} \hspace{0.5 cm} \Rightarrow \hspace{0.5 cm}
            \frac{dv_2}{dv_1} = \frac{1}{f(y_1)}.
            \label{u1_match}
        \end{equation}
    \item Matching at $\tilde{u}_2$: $r(\tilde{u}_2,v_2) = r(\tilde{u}_2, v_3)\equiv y_2$
        \begin{equation}
            \frac{v_2-\tilde{u}_2}{2} =   \int_{r_0}^{y_2}\frac{dx}{f(x)}; \hspace{5mm} y_2 = \frac{v_3-\tilde{u}_2}{2} \hspace{0.5 cm} \Rightarrow \hspace{0.5 cm}
            \frac{dv_2}{dv_3} = \frac{1}{f(y_2)}.
            \label{u2_match}    
    \end{equation}
\end{enumerate}

\subsection{Null geodesics behavior}

Let us call $\Delta u=\tilde{u}_2-\tilde{u}_1$. The radial position of the ingoing null rays in region II follows the equations
\begin{equation}
    \int_{y_1}^{r(u,y_1)}\frac{dx}{f(x)} = -\frac{ u - \tilde{u}_1}{2},
    \qquad \int_{y_1}^{y_2}\frac{dx}{f(x)} = -\frac{\Delta u}{2},
    \label{y1_y2_relation}
\end{equation}
It is identical to the BH case \eqref{r1_r2_relation} but $v \rightarrow -u$ ( or $\Delta v \rightarrow -\Delta u$). This change of sign is crucial since it inverts the behavior of the two horizons and flips the sign of the RSET. 

As before, the problem can be rewritten as 
\begin{equation}
    \begin{cases}
\frac{\textstyle dr(u,v_2)}{\textstyle du}=-\frac{\textstyle f(r)}{\textstyle 2}\,,\\
r(\tilde{u}_1,v_2)=y_1\,.
\label{in_ray_eq}
\end{cases}
\end{equation}
Again, $y_2(y_1)$ really depends on the behavior of $f(r)$, but in the opposite way of before.

\begin{figure}
    \centering
    \includegraphics[width=0.48\linewidth]{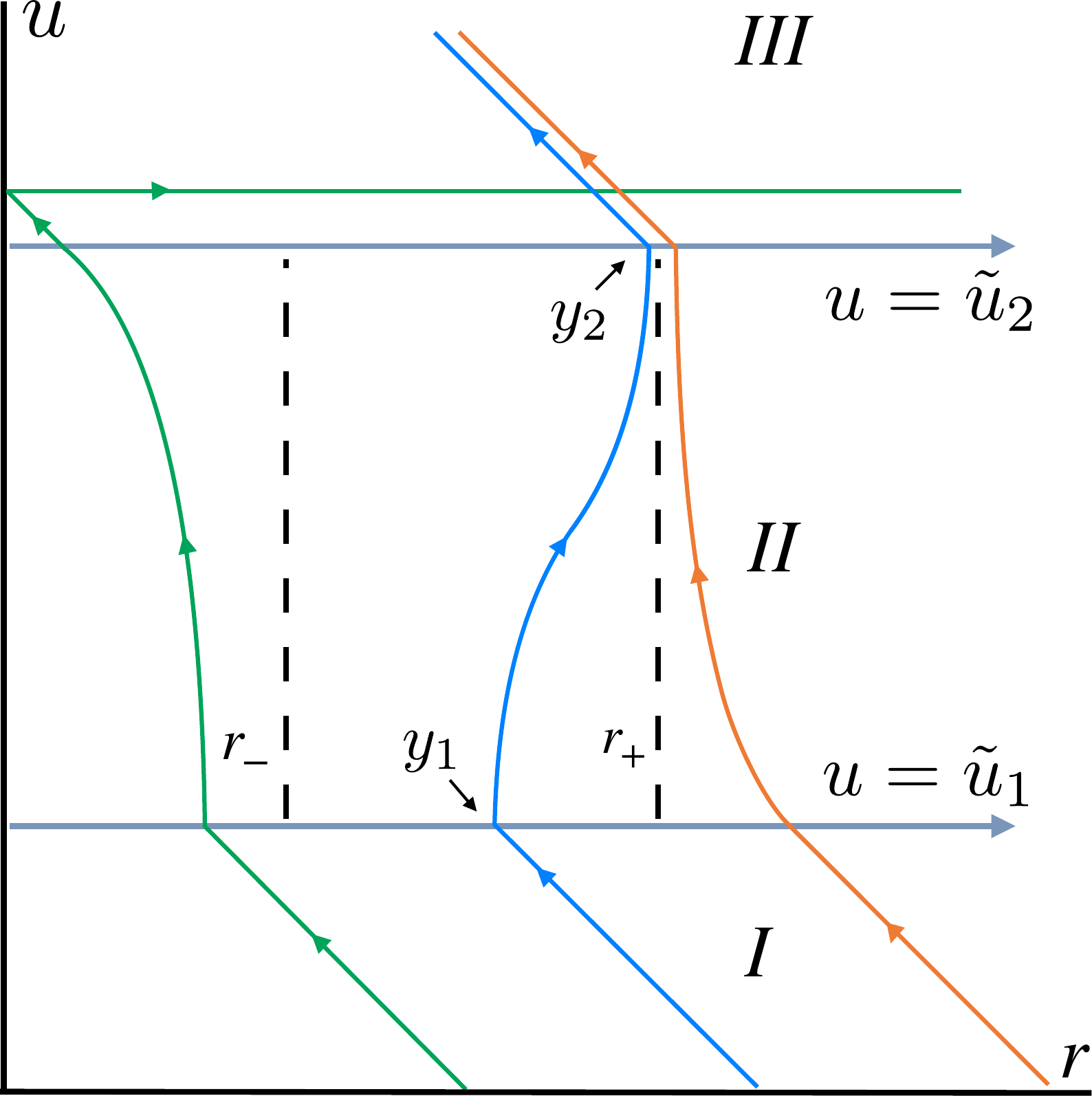}
    \caption{
    Graphical representation of ingoing null geodesics for regular spacetimes.  One can see that null rays exhibit different behaviours depending on the radial position $y_1$ at which they cross from region I to region II.
    For $y_1 \in [0, r_-)$ (green), the rays decrease in radius until they approach $r=0$. For $y_1 \in (r_-, r_+)$ (blue), they increase in radius without crossing $r_+$. Light rays that originate closer to the inner horizon require a larger $\Delta u$ to approach the outer horizon; hence, the value of $\Delta u$ for which this approximation holds depends on $r$.
    For $y_1 \in (r_+, \infty)$, the null rays decrease in radius, approaching $r_+$ exponentially fast.}
    \label{fig:urdiagBH}
\end{figure}

Let us assume that in region II the spacetime is characterized by a pair of non-degenerate inner and outer horizons $r_-$ (PITH) and $r_+$ (POTH), as depicted in Fig~\ref{fig:WH_penrose}. The sign of $f(r)$ determines if the region is anti-trapped ($f(r)<0$) or untrapped ($f(r)>0$).\footnote{If $f(r)<0$, from \eqref{in_ray_eq}, the radial position of the ingoing null ray increases with time, according to Sec. \ref{sec: Trapped and Anti-trapped regions}, this defines a positive expansion and an anti-trapped region. } The description (as shown in Fig. \ref{fig:urdiagBH}) is exactly the time reversed of the BH scenario depicted in Fig. \ref{fig:vrdiagBH}.

We notice that not all ingoing rays propagating through region II subsequently enter region III, but only those for which $y_1 > y_i$, where $y_2(y_i)=0$. All rays with $y_1 < y_i$ reach $r=0$ and then become outgoing.
Moreover, $y_i$ is determined by
\begin{equation}
    \int_{0}^{y_i}\frac{dx}{f(x)} = \frac{\Delta u}{2},
    \label{yi_expression}
\end{equation}
and therefore depends on $\Delta u$. Equation \eqref{yi_expression} is formally identical to \eqref{r1_r2_relation} with $x_1=0$, $x_2=y_i$, and $\Delta v = \Delta u$. This should not be surprising, since, as discussed earlier, this model is kinematically the time reverse of the evanescent black hole considered in Sec.~\ref{section: evanescentBH}. 

One can interpret $y_i$ as the radial coordinate reached in a time $\Delta u$ by an outgoing null ray originating from $r=0$; from all the considerations in \ref{sec: null_geodesic}
\begin{subequations}
\label{eq:y_i-limits}
\begin{align}
    &\lim_{\Delta u \rightarrow \infty} y_i =r_-,\\
    &\lim_{\Delta u \rightarrow 0} y_i =0.
\end{align}
 \end{subequations}

\subsection{RSET in the $|in\rangle$ state}

In this case we set the in state $|in\rangle$  choosing $(\bar{u}, \bar{v}) = (u,v_1)$ as canonical quantization variables, which, again, corresponds to the Minkowski vacuum in the far past. We have:

\begin{enumerate}
    \item Region I:
        \begin{equation}
            \langle T_{ab} \rangle_{\rm phys}^{\textsc{wh}}= \langle in| T_{ab} | in \rangle = 0.
        \end{equation}
    \item Region II:
        \begin{subequations}
        \label{eq:RSET-WHregionII}
        \begin{align}
            &\langle T_{uu} \rangle_{\rm phys}^{\textsc{wh}}=\langle in|T_{uu} |in\rangle = -\frac{1}{192 \pi}B(r),
            \label{Tuu2_wh}\\
            &\langle T_{v_2v_2} \rangle_{\rm phys}^{\textsc{wh}}=\frac{\langle in|T_{v_2v_2} |in\rangle}{f(r)^2} = \frac{1}{192 \pi} \frac{\left[B(y_1) - B(r) \right]}{f(r)^2}, 
            \label{Tvv2_wh}\\
            &\langle T_{uv_2} \rangle_{\rm phys}^{\textsc{wh}}=\frac{\langle in|T_{uv_2} |in\rangle}{f(r)} =  \frac{1}{96 \pi} f''(r).
            \label{Tuv2_wh}
        \end{align}
        \end{subequations}

    \item Region III:
        \begin{subequations}
        \label{eq:RSET-WHregionIII}
        \begin{align}
            &\langle T_{uu} \rangle_{\rm phys}^{\textsc{wh}}=\langle in|T_{uu} |in\rangle = 0, \label{Tuu3_wh}\\
             &\langle T_{v_3v_3} \rangle_{\rm phys}^{\textsc{wh}}=\langle in|T_{v_3v_3} |in\rangle =\frac{1}{192 \pi}\frac{\left[ B(y_1) - B(y_2)\right]}{f(y_2)^2},   \label{Tvv3_wh} \\
            &\langle T_{uv_3} \rangle_{\rm phys}^{\textsc{wh}}=\langle in|T_{uv_3} |in\rangle = 0. \label{Tuv3_wh}
        \end{align}
        \end{subequations}
\end{enumerate}

It is easy to see that the RSET is regular if on the past horizon we impose the conditions \eqref{regularWH_RSET}.
Computing the RSET on a generic non-degenerate horizon $r_h$ (in the same way we did in Sec. \ref{section: evanescentBH}) we obtain
\begin{enumerate}
    \item Region I:
        \begin{equation}
            \langle T_{ab} \rangle_{\rm phys}^{\textsc{wh}}|_{r_h} = 0.
        \end{equation}
    \item Region II:
        \begin{subequations}
        \label{eq:RSET-WHhorizonII}
        \begin{align}
            &\langle T_{uu} \rangle_{\rm phys}^{\textsc{wh}}|_{r_h} =  -\frac{1}{48 \pi}\kappa_h^2, \\
            &\langle T_{v_2v_2} \rangle_{\rm phys}^{\textsc{wh}}|_{r_h} =\frac{1}{384 \pi}\frac{f^{(3)}(r_h)}{\kappa_h} \left[1 -e^{2\kappa_h(u-\tilde{u}_1)} \right], \\
            &\langle T_{uv_2} \rangle_{\rm phys}^{\textsc{wh}}|_{r_h} = \frac{1}{96 \pi} f''(r_h). 
        \end{align}
        \end{subequations}

    \item Region III:
        \begin{subequations}
        \label{eq:RSET-WHhorizonIII}
        \begin{align}
             &\langle T_{uu} \rangle_{\rm phys}^{\textsc{wh}}|_{r_h} = 0,\\
            &\langle T_{v_3v_3} \rangle_{\rm phys}^{\textsc{wh}}|_{r_h} = \frac{1}{384\pi}\frac{f^{(3)}(r_h)}{ \kappa_h} \left[ 1 - e^{2 \kappa_h \Delta u}\right],
            \label{Tvv3_horizon_wh}\\
            &\langle T_{uv_3} \rangle_{\rm phys}^{\textsc{wh}}|_{r_h} = 0.
        \end{align}
        \end{subequations}
\end{enumerate}
Where $\kappa_h = \frac{f'(r_h)}{2}$ is again the surface gravity at the horizon $r_h$.
Differently from before, the RSET is regular on both horizons $r_-,r_+$ with the prescription \eqref{regularWH_RSET}; this tells us that the trapping horizons now bound an anti-trapped region.
Moreover, the role of the two horizons is flipped with respect to Sec. \ref{section: evanescentBH}. Explicitly
\begin{enumerate}
    \item POTH:
        $ \kappa_h>0 \implies \lim_{\Delta u \rightarrow \infty}|\langle T_{v_3v_3} \rangle_{\rm phys}^{\textsc{wh}}|_{r_h} = \frac{\textstyle 1}{\textstyle  384\pi}\frac{\textstyle  |f^{(3)}(r_h)|}{\textstyle |\kappa_h|} e^{\textstyle  2 |\kappa_h| \Delta u} \rightarrow\infty. $
    \item PITH:
        $ \kappa_h<0 \implies \lim_{\Delta u \rightarrow \infty}|\langle T_{v_3v_3} \rangle_{\rm phys}^{\textsc{wh}}|_{r_h} =\frac{\textstyle  1}{\textstyle  384\pi}\frac{\textstyle  |f^{(3)}(r_h)|}{\textstyle  |\kappa_h|}  < \infty. $
\end{enumerate}
So there is in this case an exponentially fast accumulation of energy along the outer horizon of the anti-trapped region, which will eventually be released in a single bolt at the drastic opening in our toy model. Let us also notice that the RSET is regular everywhere and we do not have Cauchy horizons. We are considering regular spacetimes only, which allows us to circumvent some of the problems described in~\cite{Wald:1975kc}.
\newline

\subsection{Early times limit $( u - \tilde{u}_1\rightarrow 0)$}
Let us define the function $\delta(y_1)$ such that $r(u,y_1) = y_1 + \delta(u,y_1)$ (eventually $y_2(y_1) = r(\tilde{u}_2,y_1)$), from \eqref{y1_y2_relation} we can write: 
\begin{equation}
    \int_{y_1}^{y_1 + \delta}\frac{dx}{f(x)} = -\frac{u - \tilde{u}_1}{2},
    \label{deltay_relation}
\end{equation}
it's obvious that $\lim_{\Delta u \rightarrow 0^+} \delta (y_1) = 0$. At this point, noting that \eqref{deltay_relation} and \eqref{delta_relation} are related by $v \rightarrow -u$ ($\Delta v \rightarrow -\Delta u$), we can directly write the expression analogous to \eqref{Tuu3_smalldv}.    
\begin{equation}
    \langle T_{v_2v_2} \rangle_{\rm phys}^{\textsc{wh}} \simeq -\frac{f^{(3)}(y_1)}{192\pi}\,du, \qquad
    \langle T_{v_3v_3} \rangle_{\rm phys}^{\textsc{wh}} \simeq -\frac{f^{(3)}(y_1)}{192\pi}\,\Delta u.
    \label{Tuu_smalldu}
\end{equation}
The minus sign is physically related to what was discussed for the black hole. The interpretation given there was that, locally, null rays accumulate energy according to the rate of variation of the trace anomaly $\propto R$. Here the interpretation is exactly the same; however, since the rays are ingoing rather than outgoing, the variation in $r$ has the opposite sign (this information is encoded in the time-reversal transformation $v \rightarrow -u$, and, since ingoing, the null rays are moving along $v \sim $const.).

\subsection{Late times limit $(u - \tilde{u}_1\rightarrow \infty)$}

It is more interesting to study the limit $ \,du = u - \tilde{u}_1 \to \infty$, since it is in this regime that the qualitative difference between black holes and white holes emerges. As in the previous analysis, the behavior of the RSET is controlled by the relation $y_2(y_1)$, whose form depends on the region considered.

\begin{enumerate}
\item \textbf{White hole interior}, $y_1 \in [y_i,r_+]$.

The behavior of $r(u,y_1)$ (or $y_2$ at $u=\tilde{u}_2$) is not homogeneous on the interior of the WH (as it was for the BH), but depends on whether it lies in untrapped or  anti-trapped regions. In the interval $y_1 \in [y_i,r_-)$, one has $\lim_{u \to \infty} r(u,y_1) \to 0$, and the corresponding contribution to the RSET remains finite (we assume no singularity at $r=0$). Let us notice that for $ \Delta u \rightarrow \infty$ we have $y_i\rightarrow r_-$ so, for large $\Delta u$, this region becomes exponentially smaller.
Here the RSET reads:
\begin{equation}
    \langle T_{v_2v_2} \rangle_{\rm phys}^{\textsc{wh}} =\frac{1}{192 \pi}{\left[ B(y_1) - B(0)\right]}.
\end{equation}

On the contrary, for $y_1 \in [r_-,r_+)$ the relation $y_2(y_1)$ approaches $y_2 \sim r_+$ in the large $\Delta u$ limit. As a consequence, combining Eqs.~\eqref{B_expansion}, \eqref{y1_y2_relation} and \eqref{Tvv3_horizon_wh}, one obtains  
\begin{equation}
\langle T_{v_2v_2} \rangle_{\rm phys}^{\textsc{wh}} \simeq
\frac{f^{(3)}(r_+)}{384\pi\kappa_+}
+ \frac{1}{192\pi r_+^2}\left(\frac{B(y_1)}{4 \kappa_+^2}-1\right)
e^{4 \kappa_+\left[(u - \tilde{u}_1)/2-R_\textsc{wh}(y_1) \right]}.
\label{Tvv_bigdu}
\end{equation}
Where we have defined 
    \begin{equation}
      R_\textsc{wh}(y_1) = \lim_{|\epsilon| \rightarrow 0}\int_{r_\epsilon}^{y_1}\frac{dx}{f(x)} + \frac{1}{2\kappa_+}\ln(|\epsilon|),
      \label{RWH}
    \end{equation}
    with $r_\epsilon= r_+(1+\epsilon)$ and $\epsilon= |\epsilon|$sign$(y_1-r_+)$.
Thus, in the anti-trapped region, the RSET exhibits exponential growth in $u$. When $u=\tilde{u}_2$, the growth is frozen (constant along $u$), so the RSET in region III is obtained by substituting \mbox{$u-\tilde{u}_1 \rightarrow \Delta u$}. The behavior found in this section is analogous to that found in the trapped region, with the important difference that the relevant horizon (the one responsible for the focusing, and hence for the exponential growth) is now $r_+$ instead of $r_-$.

\item \textbf{White hole exterior}.

The main qualitative difference with respect to the black hole scenario arises in the exterior region. Due to the accumulation of ingoing rays near the outer horizon, there exists $y_{ext}$ such that $r(u,y_1) \sim r_+$ remains valid for $y_1\in(r_+, y_{ext}]$. In this region, the RSET retains the same exponentially growing form as in Eq.~\eqref{Tvv_bigdu}.

For $y_1 \gg y_{ext}$, the relation $y_2(y_1)$ departs from this regime and the behavior of the RSET smoothly connects to the one found in the corresponding black hole exterior.
The focusing on the outer horizon $r_+$ is the main origin of the Eardley instability \cite{PhysRevLett.33.442}, which we do not consider in this model, as we neglect standard, classical matter. The exponential growth we observe here is the semiclassical version of this classical instability.
\end{enumerate}

\begin{figure}
    \centering
    \includegraphics[width=0.7\linewidth]{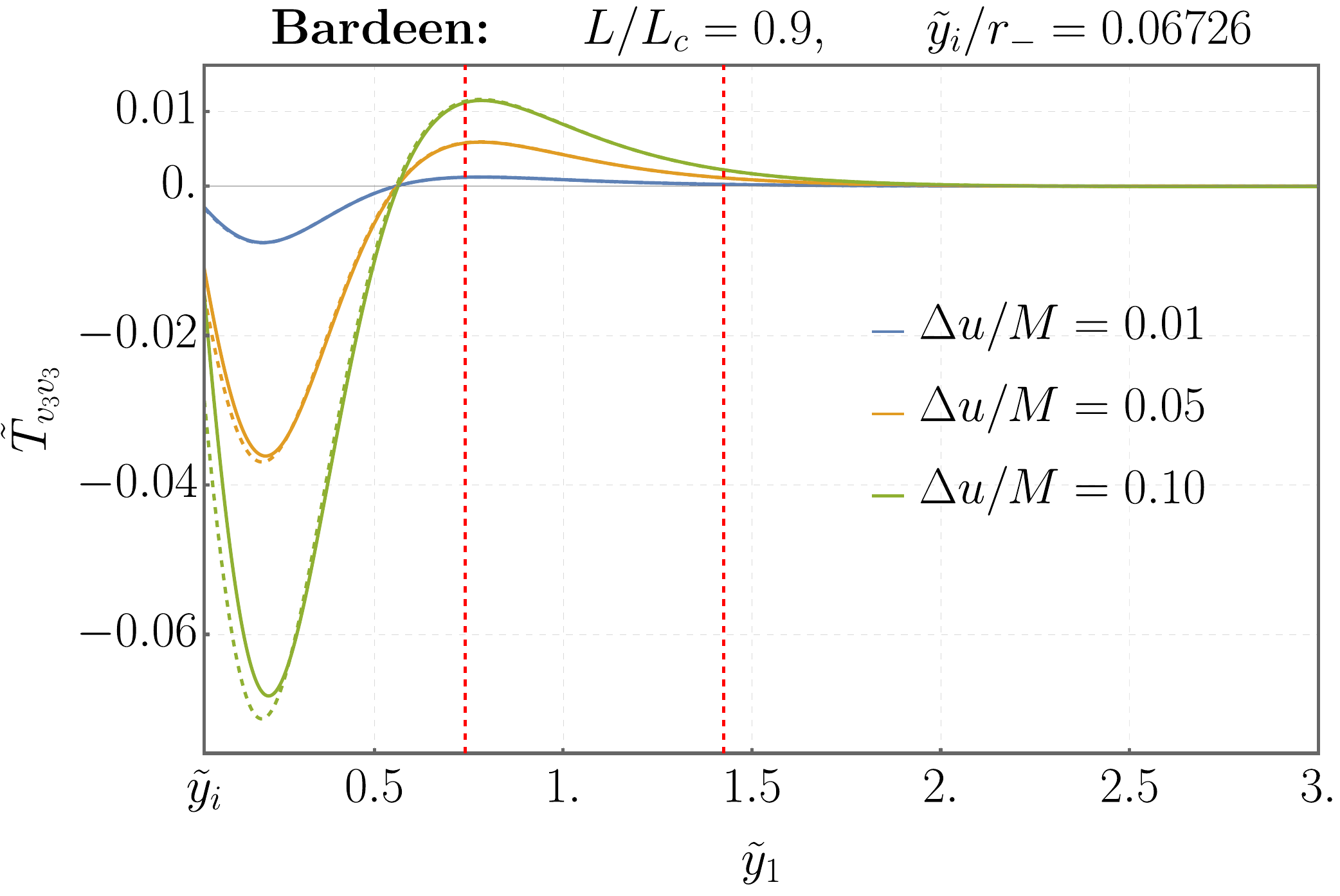}
    \caption{ Ingoing flux produced by the formation and evaporation of a Bardeen regular white hole with $L/L_c=0.9$. The red dashed lines are the locations of the inner and outer horizons (PITH and POTH), and the continuous blue, yellow, and green curves correspond to the exact fluxes generated by white holes of different lifetimes $\Delta u$. Their dashed counterparts represent the $\Delta u\to 0$ analytic approximation. Regions of negative and positive values are found for the ingoing flux. The value of $\tilde{y}_i$ is taken with respect to the greatest value of $\Delta u$.}
    \label{fig:BardeenTvv_WH}
\end{figure}

\begin{figure} 
\centering
    \includegraphics[width=0.7\linewidth]{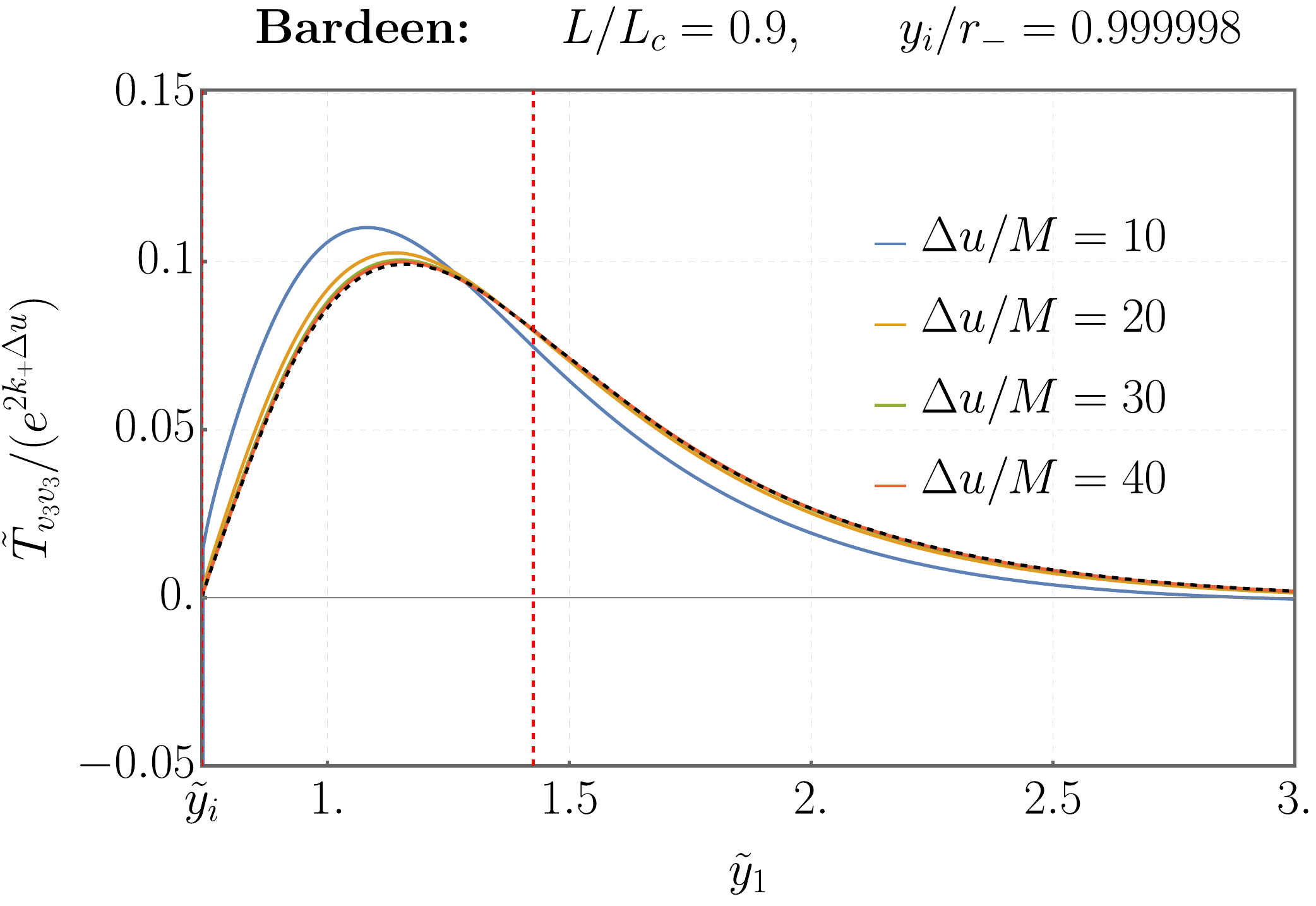}
    \caption{ Ingoing flux produced by the formation and evaporation of a Bardeen regular white hole with $L/L_c=0.9$ and large lifetime $\Delta u$. The red dashed lines are the locations of the inner and outer horizons (the inner horizon overlaps with the vertical axis), and the continuous blue, yellow, green, and red curves correspond to the exact fluxes. The exponential growth of this quantity has been subtracted, showing that for sufficiently large $\Delta u$, the coefficient of this growth approaches a fixed function of $\tilde{y}_{1}$ (in dashed black).
    The value of $\tilde{y}_i$ is taken with respect to the greatest value of $\Delta u$.}
    \label{fig:BardeenTvvLong_WH}
\end{figure}

In summary, the exponential amplification of the RSET is a common feature of both black holes and white holes, but the key difference lies in the region where this amplification takes place: near the inner horizon for black holes, and near the outer horizon for white holes. This reflects the interchange of the roles of trapped and anti-trapped regions in the two geometries. We report some examples of the ingoing fluxes for different $u$ in Fig.~\ref{fig:BardeenTvv_WH} and Fig.~\ref{fig:BardeenTvvLong_WH}. Now that we have studied the fluxes generated by both the trapped and the anti-trapped regions, we can estimate how the ingoing and outgoing expansions are modified in their presence. We will first analyze in detail the effects of the fluxes generated by the trapped region and then provide a qualitative discussion of those generated by the anti-trapped region.


\section{Black Hole to White Hole transition}
\label{sec:BH-WH_transition}
Solving the semiclassical Einstein equations is, in general, an extremely difficult task. A standard approach is instead to compute the renormalized stress--energy tensor (RSET) on fixed backgrounds and then infer the qualitative effects of their backreaction on the geometry. In the spirit of extracting general physical lessons while retaining the simplicity of our construction, we will make use of the toy models discussed in Sections~\ref{section: evanescentBH} and~\ref{section: evanescentWH} to infer how trapped and anti-trapped regions will evolve after their formation. Let us advance that, in presence of regular trapped (anti-trapped) regions, the outgoing (ingoing) component of the RSET 
can undergo exponential amplification in $v$ ($u$) time. This indicates that  their backreaction becomes significant in regions II (and, by extension, in III) and that the spacetime evolution cannot, in general, be approximated by a simple transition from a black/white hole geometry to Minkowski space. Accordingly, the analysis below should be understood as a diagnostic of the backreaction suggested by the fixed-background RSET, rather than as a self-consistent solution of the semiclassical Einstein equations.
Another important disclaimer is that, up to this point, all our results for the RSET have been derived for two-dimensional geometries. From now on, however, the analysis of the expansions will be carried out in four dimensions using the Polyakov approximation introduced in Sec. \ref{sec:2D_RSET}. As can be seen from Eq. \eqref{Polyakov_approx}, this approximation becomes problematic at $r=0$ even in regular spacetimes~\cite{Parentani:1994ij}. This behaviour simply indicates that the validity of the approximation breaks down in the vicinity of the origin.  Nevertheless, this issue does not affect our analysis, since the integration domain is restricted to $x_1/M\in[\epsilon,3]$, with $\epsilon>0$. Under these conditions, $\forall, v\geq\tilde{v}_1$, one has $r(x_1,v)>0$, so the region where the approximation may fail is never probed.

\subsection{Evolution of null expansions}
The characteristic exponential amplification of the $|\textit{in}\rangle$-state RSET is such that its signature backreaction effects can be deduced without the need to solve the semiclassical Einstein equations. For simplicity, we direct our attention to the dynamical evolution of an initially trapped region and its tendency towards stimulating, semiclassically, the formation of an anti-trapped region. The natural way to identify the presence of trapped and anti-trapped horizons is to look at the expansions $\theta_+$ and $\theta_-$. The Raychaudhuri equation~\eqref{Raychaudhuri} links these expansions with the ingoing and outgoing RSET components, respectively. In what follows, we will analyze how these expansions deviate from their classical (stationary) values when the RSET is present as a source. Notice that, in doing so, we are using the 2D RSET components as an approximation of the 4D RSET by means of the Polyakov approximation.

We stress that the approach we are adopting constitutes a substantial approximation to the real semiclassical dynamics. We consider the formation of a charged or regular black hole via an ingoing null shell as in Fig.~\ref{fig:spherical_penrose} and compute the RSET in region II using Eqs.~\eqref{eq:RSET-regionII}. The initial metric and its associated RSET do not satisfy the semiclassical Einstein equations; therefore, the expansions we compute are not the exact ones that would arise from a fully self-consistent solution. Nevertheless, this approach provides useful insight into the possible dynamics and qualitatively reproduces features of the numerical solutions obtained in~\cite{barenboim2025evaporationregularblackholes} and~\cite{boyanov2025semiclassicalevolutiondynamicallyformed}. It also allows us to diagnose the expected tendency of the null expansions in a wide range of background spacetimes without the need for complex numerical simulations.

So we start by looking at the spacetime discussed in Section \ref{section: evanescentBH}. To make the connection between the RSET and the geometry more transparent, it is useful to rewrite the Raychaudhuri equation in terms of the variable $x_1$. For the sake of completeness, let us discuss the behaviour of the outgoing and ingoing expansions in regions II (black hole region) and III (Minkowski region).

\subsubsection{Null expansions in region II}
In region II, all components of the RSET are non-vanishing; therefore, in principle, we must analyse the dynamics of both the outgoing and ingoing expansions, $\theta_{+,i}$ and $\theta_{-,i}$ respectively, where the subscript $i$ specifies to which of the three regions --- I, II, III --- we are referring to.

First, let us discuss the outgoing expansion $\theta_{+,2}$. 
We parametrize the geodesic congruence along which the expansion is computed by $\lambda_+ = v$ with $x^\mu(\lambda_+) = \big(u_{0}, v, \vartheta_0, \phi_0)$, which selects a null vector $n^\mu = \left(0, 1, 0, 0\right)$. Let us notice that $v$ is not an affine parameter, indeed $n^\nu\nabla_\nu n^\mu = f'(r)n^\mu/2$, so the Raychaudhuri Eq.~\eqref{Raychaudhuri} reads
\begin{equation}
\frac{d\theta_{+,2}}{dv}
= -\frac{1}{2}\theta_{+,2}^2 + \frac{f'(r)}{2}\theta_{+,2}
-\frac{2}{r^2} \langle T_{vv}^{\rm (2D)}(r,v)\rangle .
\label{Raych_u2}
\end{equation}
where $\theta_{+,2}$ is computed along a null outgoing worldline, i.e. depends on $x_1$ through $r(x_1,v)$\footnote{We mention that, in principle, the presence of the RSET changes also the function $r(x_1,v)$. The $r$-constant lines present in Figures \ref{fig:dynamical_RN}, \ref{fig:dynamical_bardeen},\ref{fig:Minkowski_dynamics},\ref{fig:AdSBardeen-dynamics} have been obtained solving the differential equation $\frac{dr}{dv}=\frac{r}{2}\theta_{+,i}$ (from \eqref{spherical_expansions}) imposing as initial condition $r(x_1,\tilde{v}_1)=x_1$. Obviously, $\theta_{+,i}$, with $i=1,2$, are the expansions from \eqref{Raych_u2} and \eqref{theta_plus_3}.}
 
In the classical case ($\langle T^{\rm (2D)}_{vv}\rangle=0$), the outgoing expansions assumes, in virtue of~\eqref{spherical_expansions}, the form $\theta_{+,2}=f(r)/r$.
To solve the Raychaudhuri equation we suppose, at the initial time $v=\tilde{v}_1$, that 
\begin{equation}
    \theta_{+,2} (x_1,\tilde{v}_1)=\frac{1}{x_1}f(x_1).
\end{equation}

For the ingoing expansion $\theta_-$, instead, we consider ingoing radial null geodesics defined by $x^\mu(\lambda_-) = \big(u_2(\lambda_-), v_0, \vartheta_0, \phi_0)$.
Choosing the affine parameter $\lambda_- = -2r$ (with $r$ computed at a constant $v$), the null tangent vector reads $l^\mu = dx^\mu/d\lambda_-
= \left(1/f(r), 0, 0, 0\right)$. With the convention $ds^2=-f\,du\,dv$, this choice gives $l^\mu n_\mu=-1/2$.
The Raychaudhuri equation~\eqref{Raychaudhuri} then takes the form
\begin{equation}
\frac{d\theta_{-,2}}{d\lambda_-}
= -\frac{1}{2}\theta_{-,2}^2
-\frac{2}{r^2} \, \frac{\langle T_{u_2u_2}^{(2D)}(r,v)\rangle}{f(r)^2 }.
\end{equation}
To integrate this expression, we now write it in terms of the variable $x_1$. Using the relation $dr/dx_{1} = f(r)/f(x_{1})$,
one obtains
\begin{equation}
\frac{d\theta_{-,2}}{dx_1}
= \frac{f(r)}{f(x_1)}\left[\theta_{-,2}^2
+ \frac{4}{r^2}\, \frac{\langle T_{u_2u_2}^{(2D)}(x_1,v)\rangle}{f(r)^2}
\right].
\label{Raych_v2}
\end{equation}
In the classical case ($\langle T^{(2D)}_{u_2u_2}\rangle=0$), the ingoing expansions assumes, following Eq.~\eqref{spherical_expansions}, the form $\theta_{-,2}=-{1}/{r}$. To find the solution, we integrate from a large $x_{1}$ leveraging the asymptotic decay of $T^{\rm (2D)}_{u_2u_2}$, i.e., $\lim_{x_1 \rightarrow \infty} T^{(2D)}_{u_2u_2} = \mathcal{O}(x_1^{-3})$, which implies $\lim_{x_1 \rightarrow \infty} \theta_{-,2} = -1/r$.

\subsubsection{Null expansions in region III}
In this case, the only non-vanishing component of the RSET is given by Eq.~\eqref{Tuu3}, which simplifies further the analysis. The outgoing expansion $\theta_{+,i}$  is the same we would have in the classical theory, since the ingoing component of the RSET is null in this region. We obtain the standard Minkowski expansion from \eqref{spherical_expansions}
\begin{equation}
    \theta_{{+,3}}=\frac{1}{r}.
    \label{theta_plus_3}
\end{equation}
For the ingoing expansion, We choose $\lambda =u_3$ as affine parameter and the null curve $x^\mu(u_3) = \big(u_3, v_0, \vartheta_0, \phi_0)$. The ingoing expansion $\theta_-$ satisfies
\begin{equation}
\frac{d \theta_{-,3}}{du_3} = - \frac{\theta_{-,3}^2}{2} - 2 \frac{\langle T_{u_3u_3}^{\rm (2D)} (x_1) \rangle}{r^2}.
\end{equation}
Using the chain of relations $u_3 \to u_2 \to u_1 \to x_1$, one finds 
$dx_{1}/du_{3}=-f(x_{1})/[2f(x_{2})]$.
The Raychaudhuri equation becomes\footnote{We note that, by comparing Eqs.~\eqref{Raych_v2} and \eqref{Raych_v3}, we can conclude that $\theta_{-,2}$ and $\theta_{-,3}$ are continuous across $\tilde{v}_2$. Obviously, the same does not hold for $\theta_{+,2}$ and $\theta_{+,3}$.}
\begin{equation}
\frac{d \theta_{-,3}}{dx_1}
= \frac{f(x_2)}{f(x_1)}\left[
\theta_{-,3}^2 +  \frac{4}{r^2}\langle T_{u_3u_3}^{(2D)}(x_1)\rangle
\right].
\label{Raych_v3}
\end{equation}
Also in this case we use $\lim_{x_1 \rightarrow \infty} \theta_{-,3} = -1/r$ as the boundary condition with which to integrate Raychaudhuri's equation.

We are now equipped with the tools allowing us to deduce how the defining characteristic of a black hole formed by gravitational collapse, i.e.~its capacity to trap or anti-trap light rays, will evolve under the influence of the quantum fluxes generated by its own formation.

\subsection{Approximate evolution of black hole interiors through concrete examples }

Let us now illustrate the evolution of black holes formed by gravitational collapse in presence of semiclassical effects. Since there is a rich and complex phenomenology, we start by giving a sufficiently generic qualitative picture, to later dive into concrete examples. In the following, we mostly discuss the behaviour of the expansions for different times $dv = v - \tilde{v}_1$ after black hole formation. 
The roles played by the ingoing and outgoing RSET components will be clarified as well.

We situate ourselves at the start of region II, defined by $dv = 0$ ($v=\tilde{v}_1$). Here, region $r_- < r < r_+$ is trapped, whereas regions $r>r_+$ and $r<r_-$ are untrapped.  We have an initially vanishing outgoing flux (in virtue of~\eqref{Tuu2_smalldv}, which guarantees $\theta_{-,2} = -\frac{1}{r}$ and $\theta_{+,2} = f(r)/r$). The evolution of the outgoing expansion $\theta_{+,{2}}$ is determined by the ingoing flux~\eqref{Tvv2}, which is negative nearly everywhere in the spacetime, and always negative on the horizons for any black hole. Near a trapping horizon $r=r_{h}$, we have $\theta_{+,{2}}\sim 0$, hence
\begin{equation}
    \frac{d\theta_{+,{2}}}{dv}\simeq -\frac{2}{r^2}\langle T_{vv}^{\rm (2D)}(r,v)\rangle\propto\frac{f'(r)^2}{ r^2}>0.
\end{equation}
This implies that, locally around the trapping horizons, the outgoing expansion $\theta_{+,2}$ grows with $v$. At the outer horizon this tends to displace the corresponding marginal surface inward, whereas at the inner horizon it tends to displace it outward, thereby reducing the size of the trapped region. In the approximate evolutions obtained below by integrating the Raychaudhuri equation with the fixed-background RSET, the two marginal surfaces approach one another and, for the examples considered, merge after a finite advanced-time interval at $v=v_{\rm extr}$. We use this merger time as the prescription for the second matching surface, $\tilde v_2\equiv v_{\rm extr}$. Beyond this surface the prescribed background is Minkowskian and $\theta_{+,3}=1/r>0$ everywhere. The finite-time disappearance found in this construction should therefore be understood as a feature of the Raychaudhuri diagnostic, consistent with the self-consistent numerical evolutions of \cite{barenboim2025evaporationregularblackholes,boyanov2025semiclassicalevolutiondynamicallyformed}, rather than as a complete solution of the semiclassical Einstein equations.

It is important to stress that, due to the above mentioned inner horizon dynamics, the trapped region disappears in an advanced time interval generically shorter than the Hawking time $\mathcal{O}(M_\textsc{bh}^3)$. This was also suggested by actual simulations~\cite{boyanov2025semiclassicalevolutiondynamicallyformed} for a transient Reissner--Nordstr\"om black hole estimating something closer to $\mathcal{O}(M_\textsc{bh}^2)$. This result, if confirmed for regular geometries without a central singularity, could imply the opening of the trapped region well before a Page time (which is again $\mathcal{O}(M_\textsc{bh}^3)$) and a possible resolution of the information loss problem~\cite{Gralla:2025gzl,DiFilippo:2025kzh}. 

Let us also stress that numerical simulations do not display an asymptotic approach to an extremal end-state, as suggested in some analyses~\cite{Carballo-Rubio:2018pmi}, but rather reach it in finite time, leaving no extremal remnant behind. This seems to indicate that the adiabatic thermal evolution assumed in~\cite{Carballo-Rubio:2018czn} is grossly violated by the inner horizon dynamics in the last phases of the evaporation.

This is \emph{per se} already striking, but semiclassical physics still leaves margin for the unexpected.
Looking at equation~\eqref{Raych_v2}, we observe the profile of the outgoing flux as a function of decreasing $x_1$ (the boundary connecting region I to region II) determines the  evolution of $\theta_{-,2}$.  
For some fixed $v > \tilde{v}_1$ a negative outgoing flux tends to decrease the derivative of $d\theta_{-,2}/dx_{1}$. 
If large enough, this flux can make $\theta_{-,2}$ reach a local minimum value and start to grow moving towards smaller $x_1$\footnote{ Let us notice that $x_1$ is a decreasing function of $u_1$  ($x_1= (\tilde{v}_1-u_1)/2$) and so also of $u_2$ and $u_3$. This means that going forward in $u$ time is equivalent on moving towards smaller values of $x_1$.}, potentially crossing $0$ and generating an anti-trapping horizon. A positive outgoing flux would have the contrary effect, contributing positively towards $d\theta_{-,2}/dx_{1}$, and thus making the ingoing expansion more negative. If this happens when $\theta_{-,2}>0$, i.e. inside an anti-trapped region, it eventually causes its evaporation.\footnote{Let us stress that the identification of anti-trapping horizons (PITH and POTH ) requires to study the evolution of $\theta_-$ along the complementary null direction, e.g. the derivative with respect to $\lambda_+ = v$ (see the definitions in the Table \ref{tab:trapping_horizons}).} The profile of the outgoing flux, which is determined by the function $f(r)$ as showed from Eqs.~\eqref{Tuu3_smalldv} and \eqref{Tuu_bigdv}, and Figures~\ref{fig:RNTuu} and~\ref{fig:BardeenTuu}, depends on the type of black hole under consideration. In the following, we will see how the shape of these fluxes determines whether the anti-trapped region has a finite lifetime or not.  

Consider now how the evolution of the anti-trapped region extends into region III. The two-dimensional RSET is effectively frozen at the matching null surface $\tilde{v}_2$ separating regions II and III. If an anti-trapped region has been formed in region II, it will be present initially in region III. However, the four-dimensional Polyakov RSET is suppressed by the factor $1/r^2$. As a consequence, it is progressively dissipated (radiated away) as $v$ increases, reducing the size of the anti-trapped region. This argument suggests that the anti-trapped region should have a finite lifetime. We turn now to the discussion of specific examples.

Before turning to discuss specific cases, we want to highlight that white holes generate ingoing fluxes that grow exponentially with $\Delta u$. These ingoing fluxes, which we have obtained in Section~\ref{section: evanescentWH}, if large and positive enough, can trigger the formation of a trapped region with the same mechanism discussed above. The outgoing fluxes generated by this trapped region can, subsequently, trigger the formation of an anti-trapped one.This suggests that quantum vacuum effects may constrain the lifetime of trapped and anti-trapped regions and provide a dynamical link between black-hole and white-hole geometries.

\subsubsection{Reissner-Nordstr\"om black hole}
Let us now analyze the main feature of the Reissner-Nordstr\"{o}m black hole (for reference see Fig. \ref{fig:RNTuu}). It is straightforward to describe the behaviour of the expansions at early times, indeed in the limit $d v \to 0$, the result derived in Sec.~\ref{subsec:smalldv_BH} shows that the RSET in region II reduces to a local expression of the form
\begin{equation}
\frac{\langle T_{u_2u_2}^{(2D)} \rangle }{f^2(r)}
\simeq \frac{1}{192\pi} f^{(3)}(x_1)d v.
\end{equation}
In this regime the behavior of the RSET is entirely determined by the third derivative of the metric function.
One finds
\begin{equation}
f^{(3)}(r) = \frac{12M}{r^4} - \frac{24Q^2}{r^5}
= \frac{12}{r^5}(Mr - 2Q^2).
\end{equation}
The sign of $f^{(3)}(r)$ is therefore determined by the quantity $Mr - 2Q^2$. Defining the critical radius
\begin{equation}
r_c \equiv \frac{2Q^2}{M},
\end{equation}
one obtains
\begin{equation}
\frac{\langle T_{u_2u_2}^{(2D)} \rangle}{f^2(r)}
\begin{cases}
< 0 & \text{for } x_1 < r_c, \\
> 0 & \text{for } x_1 > r_c.
\end{cases}
\end{equation}

In particular, in the region $x_1<r_c$ the RSET is negative implying  a violation of the null energy condition. At fixed $x_1$, the early-time approximation above behaves as $\langle T_{u_2u_2}^{\rm (2D)}\rangle/f^2\sim-1/x_1^5$. This approximation is not uniform as $x_1\to0$, because $\delta/x_1$ ceases to be small. For any fixed $dv>0$, the exact expression~\eqref{Tuu2} instead has $B(x_1)\sim-8Q^4/x_1^6$ and therefore still diverges negatively at the Cauchy horizon, i.e~at $\mathcal{C}_0$ of the left panel in Fig.~\ref{fig:spherical_penrose}. According to the discussion above, this negative flux induces a growth of the ingoing expansion $\theta_-$.

For $dv=0$, i.e.\ at $v=\tilde{v}_1$, the RSET vanishes. For any $dv>0$, however, its exact expression becomes arbitrarily large and negative as $x_1\to0$. Within the fixed-background Raychaudhuri analysis, this supports the formation of an anti-trapped region whose size increases with $v$.

\begin{figure}

    \centering
    \includegraphics[width=1\linewidth]{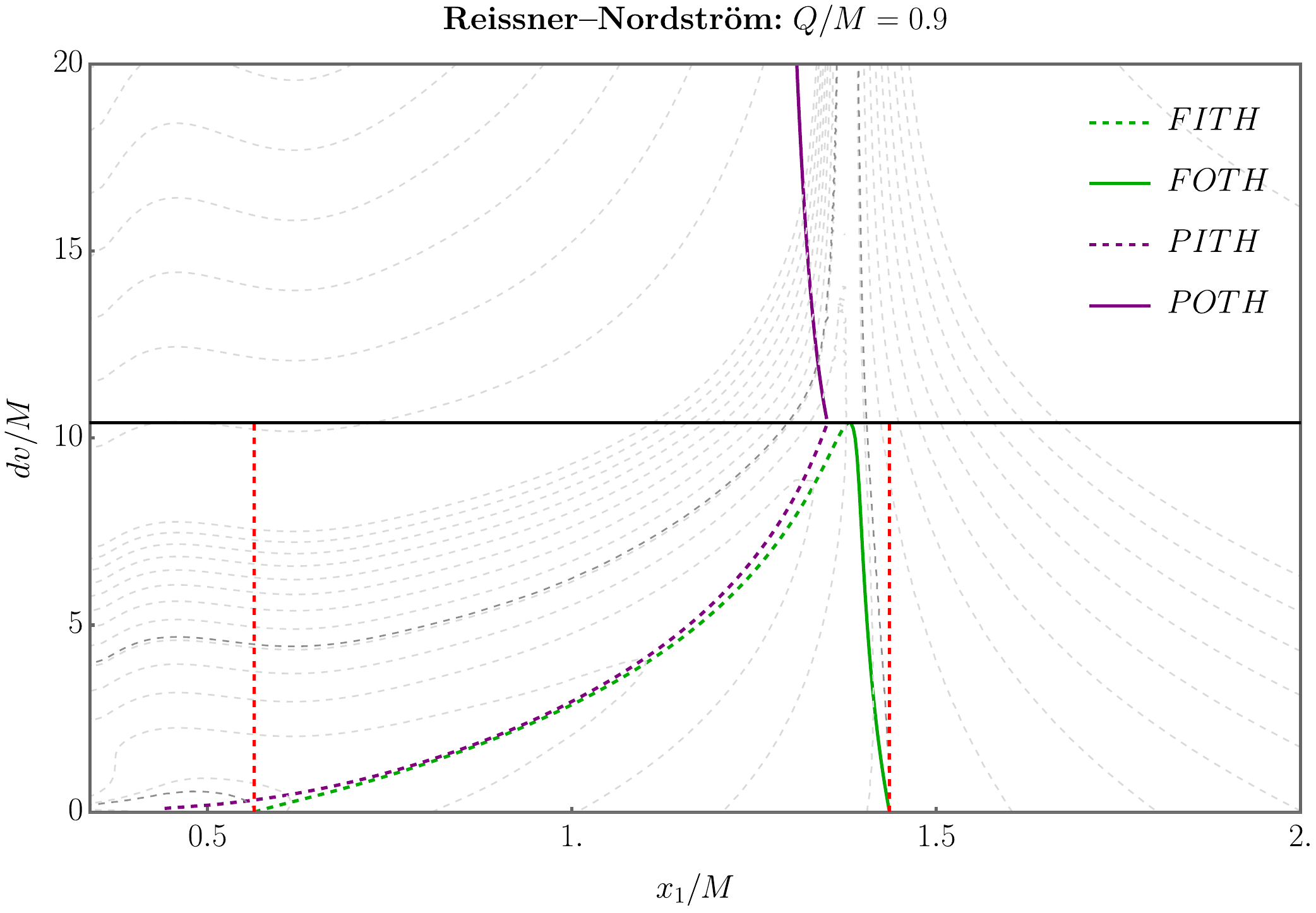}
    \caption{Plot of the location of the trapped and anti-trapped regions for the Reissner--Nordström, left panel of of Fig.~\ref{fig:spherical_penrose}. The black line represents the null shell located at $\tilde{v}_2$, which has been introduced by hand at $v_{\rm extr}$. The red dashed lines represent the static horizons $r_+$ and $r_-$ that would be present in the classical case. The green and purple  curves denote the trapped and anti-trapped regions, respectively, updated according to the fluxes computed in Sec.~\ref{section: evanescentBH}. The gray dashed lines represent the constant $r$ lines corrected with the presence of the RSET, Let us notice how the anti-trapped region grows indefinitely in terms of the radius in $u$-time (moving towards smaller $x_1$). }
    \label{fig:dynamical_RN}
\end{figure}

\begin{figure}
\centering
\includegraphics[width=0.9\linewidth]{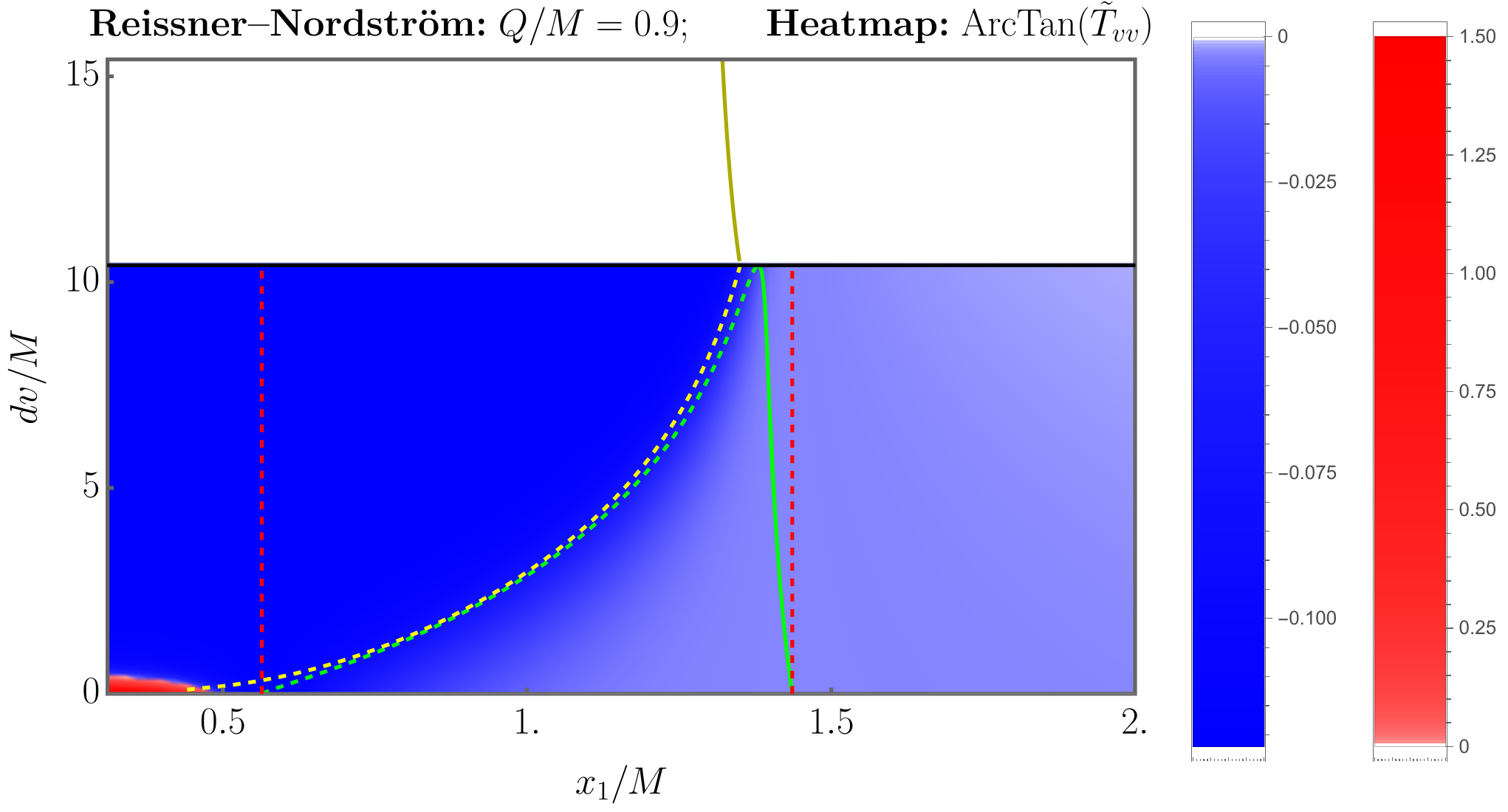}
\vspace{5mm}
\includegraphics[width=0.9\linewidth]{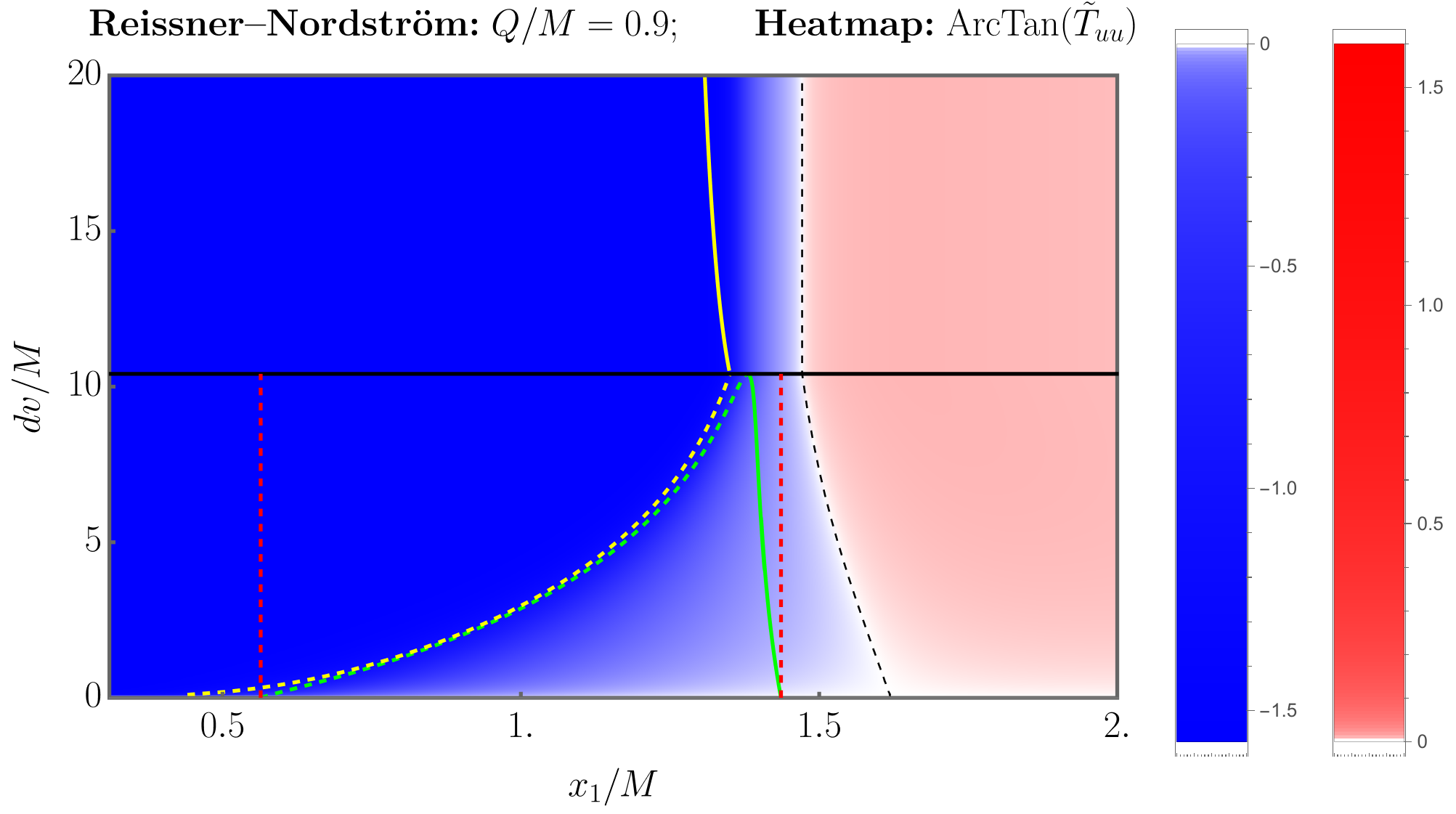}

    \caption{Heatmaps of the ingoing (top) components of the RSET, Eqs. \eqref{Tvv2} and \eqref{Tvv3}, and the outgoing (bottom) components, Eqs. \eqref{Tuu2} and \eqref{Tuu3}, as functions of $v$ and $x_1$ for the Reissner-Nordstr\"om spacetime. Red denotes positive values, whereas blue denotes negative values. The heatmaps are overlaid with the marginal surfaces (trapped in green and anti-trapped in yellow, continuos lines are outer horizons, dashed ones are inner horizons) determined in Fig. \ref{fig:dynamical_RN}. Notice that, in the bottom panel, the anti-trapped region expands indefinitely (towards smaller values of $x_1$) since the outgoing RSET is always negative and divergent in $x_1=0$. In the top panel instead, the trapped region continuously shrinks because the ingoing RSET is negative throughout the entire region shown. }
    \label{fig:heatmap_RN}
\end{figure}

The same argument can be done in region III, with the flux eventually suppressed by the factor $1/r^2$, moving the anti-trapped region closer to the Cauchy horizon at $x_1=0$.  We show this behaviour in Figures \ref{fig:dynamical_RN} and \ref{fig:heatmap_RN} (where the dynamics of the trapped and anti-trapped surfaces is compared with the ingoing/outgoing fluxes ).

\subsubsection{Bardeen black hole}
\begin{figure}

    \centering
    \includegraphics[width=1\linewidth]{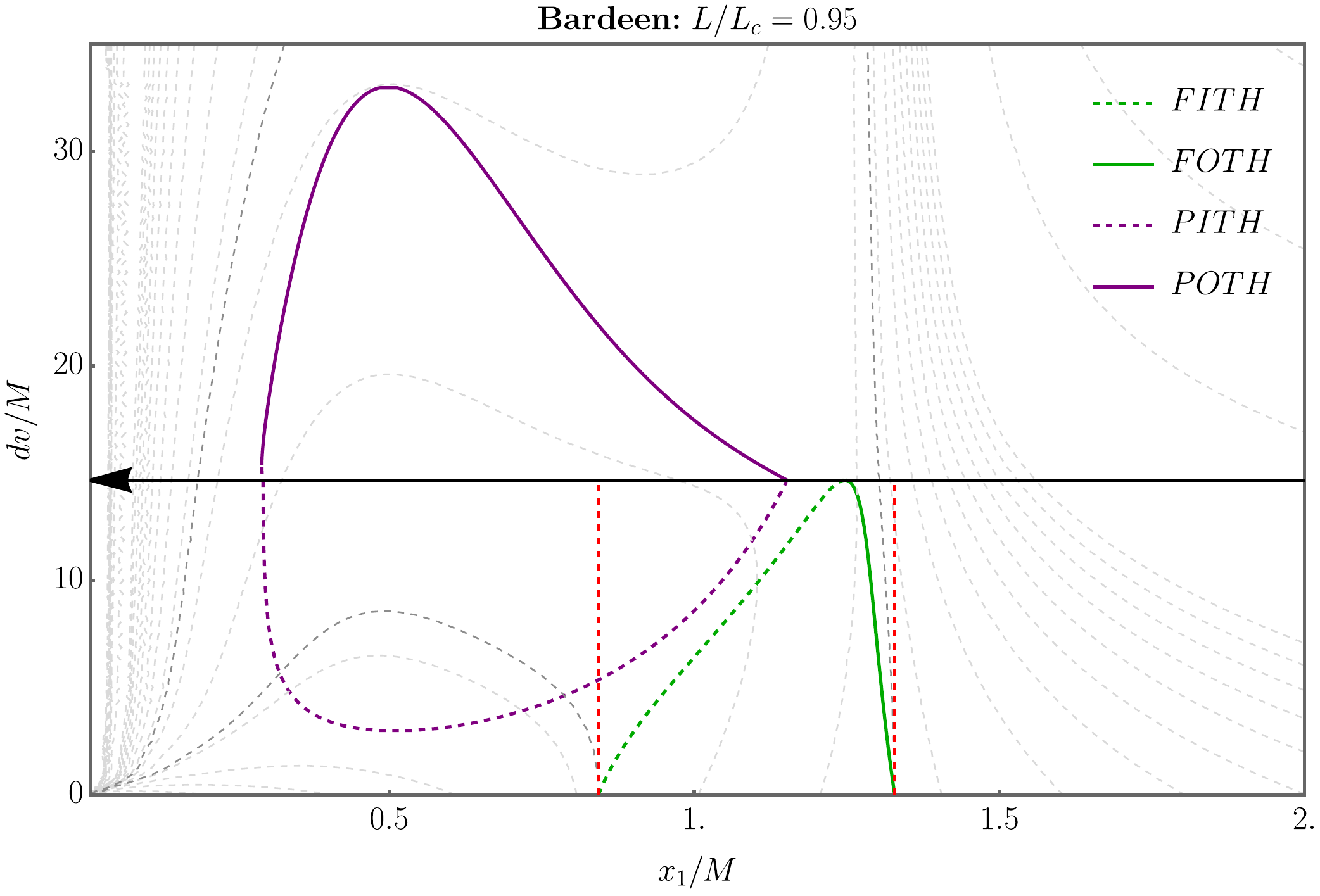}
    \caption{Plot of the location of the trapped and anti-trapped regions for the Bardeen regular black hole, right panel of Fig.~\ref{fig:spherical_penrose}. The black line represents the null shell located at $\tilde{v}_2$, which has been introduced by hand in order to close the trapped region at $v_{\rm extr}$. The red dashed lines represent the static horizons $r_+$ and $r_-$ that would be present in the classical case. The green and purple curves denote the trapped and anti-trapped regions, respectively, updated according to the fluxes computed in Sec.~\ref{section: evanescentBH}. The gray dashed lines represents the constant $r$ lines corrected with the presence of the RSET, the dark lines represent the level curves at $r_+$ and $r_-$. Let us notice that for Bardeen the anti-trapped region reaches a size smaller than $r_+$, which is the one of the initial Black Hole.
}
    \label{fig:dynamical_bardeen}
\end{figure}

\begin{figure}
\centering
\includegraphics[width=0.96\linewidth]{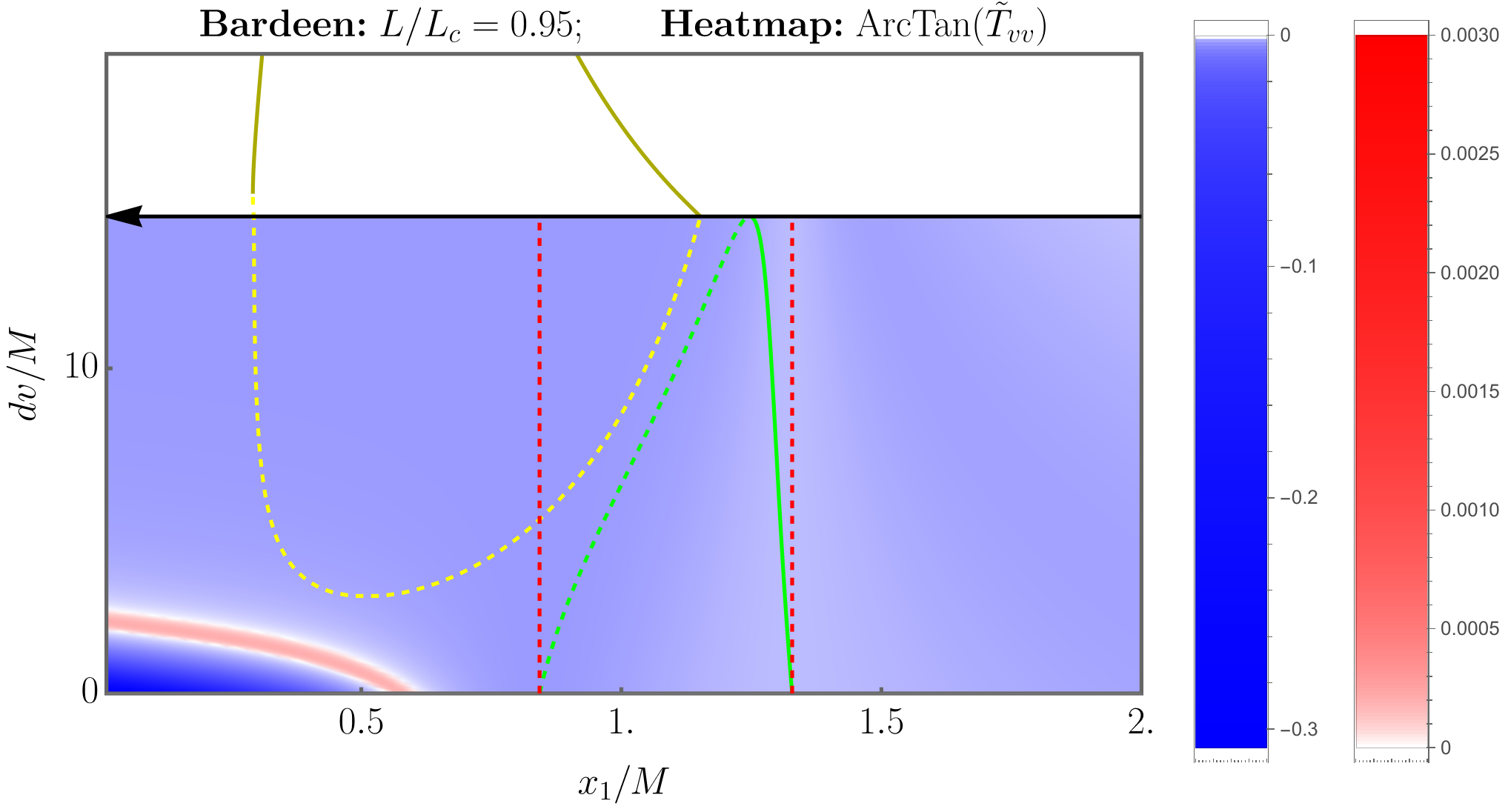}
\includegraphics[width=0.95\linewidth]{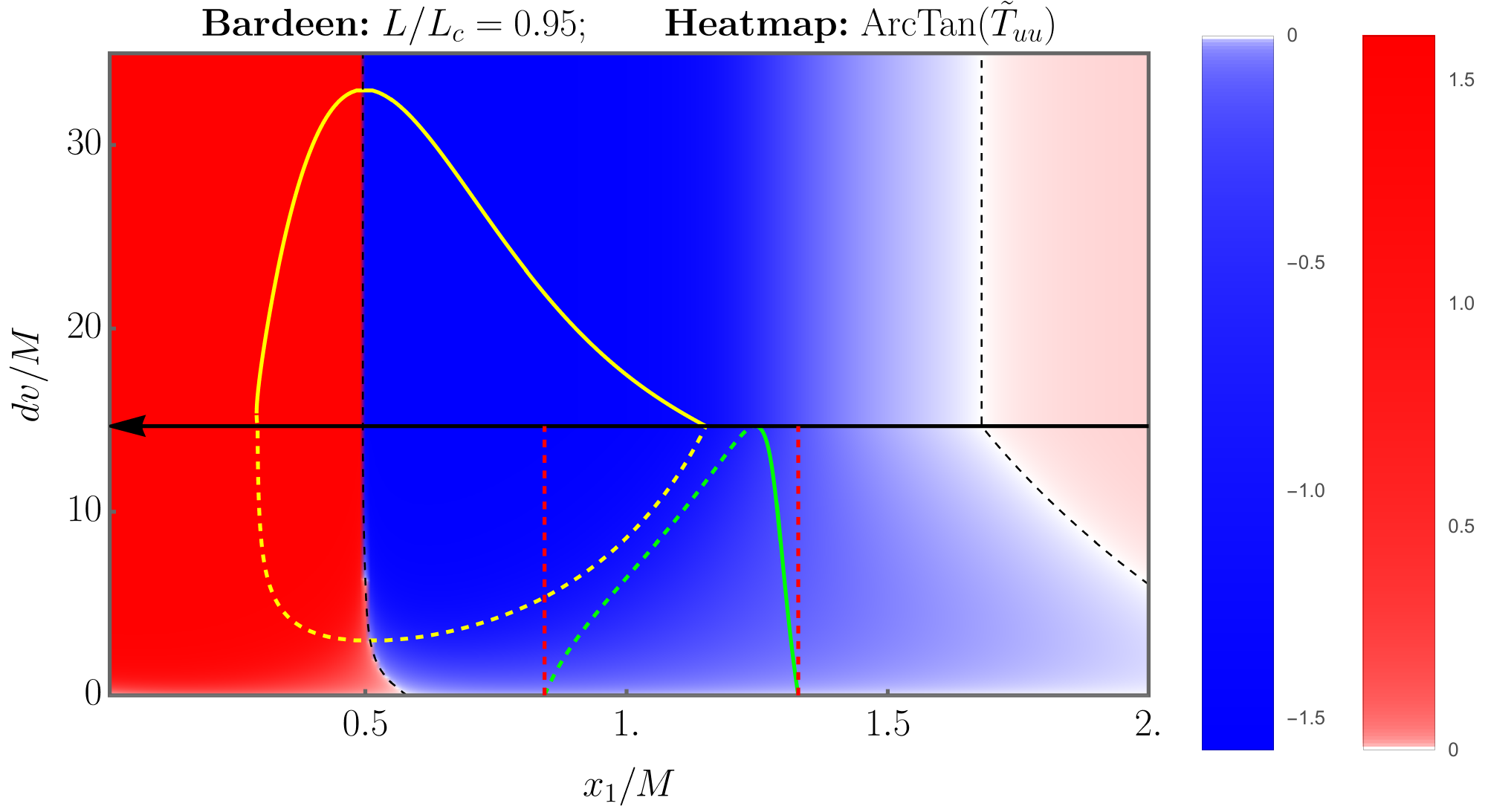}

    \caption{Heatmaps of the ingoing (top) components of the RSET, Eqs. \eqref{Tvv2} and \eqref{Tvv3}, and the outgoing (bottom) components, Eqs. \eqref{Tuu2} and \eqref{Tuu3}, as functions of $v$ and $x_1$ for the Bardeen spacetime. Red denotes positive values, whereas blue denotes negative values. The heatmaps are overlaid with the marginal surfaces (trapped in green and anti-trapped in yellow, continuos lines are outer horizons, dashed ones are inner horizons) determined in Fig. \ref{fig:dynamical_bardeen}. Notice that, in the bottom panel, the anti-trapped region expands (towards smaller values of $x_1$) when the outgoing RSET is negative and contracts when it is positive. By contrast, in the top panel, the trapped region continuously shrinks because the ingoing flux is negative throughout the (almost) entire region shown. The dashed black line (below) represents the zeros of the outgoing RSET.}
    \label{fig:heatmap-Bardeen}
\end{figure}

We will not go into the details of the RSET calculations as done previously, since this simply amounts to substituting the function $f(r)$ corresponding to the black hole under consideration. Instead, we directly present the plots of the expansions, as was done for the Reissner--Nordström case, and elaborate on their implications.
In Figures~\ref{fig:BardeenTuu} and \ref{fig:BardeenTuuLong} we show that the outgoing component of the RSET, moving from $\tilde{x}_1 \gg 1$ toward $\tilde{x}_1 \sim 0$, first presents a region where it is negative, immediately followed by a region of positive flux of larger magnitude. This indicates that, for a fixed $dv>0$, the expansion $\theta_-$ possesses two (maximum) marginally anti-trapped surfaces. As discussed previously, we expect the dissipation of $\langle T_{u_3u_3}\rangle^{\textsc{bh}}_{\rm phys}/r^2$ as $v \rightarrow \infty$ to lead to the disappearance of the anti-trapped region. This behavior is clearly observed in Fig.~\ref{fig:dynamical_bardeen} and Fig.~\ref{fig:heatmap-Bardeen}

\subsubsection{Regular black holes with AdS and Minkowski cores}
\begin{figure}
\centering
\includegraphics[width=0.45\linewidth]{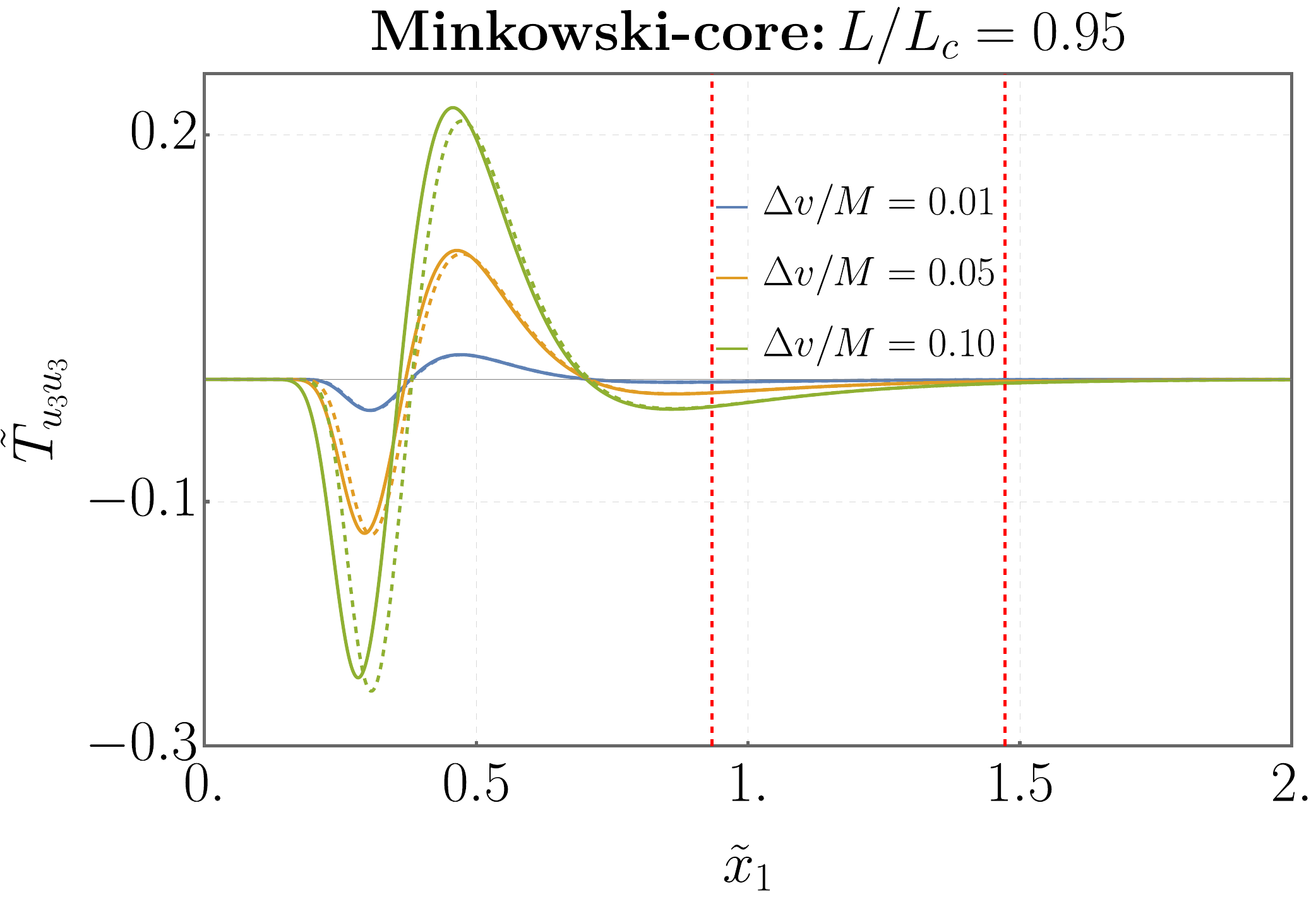}
\includegraphics[width=0.45\linewidth]{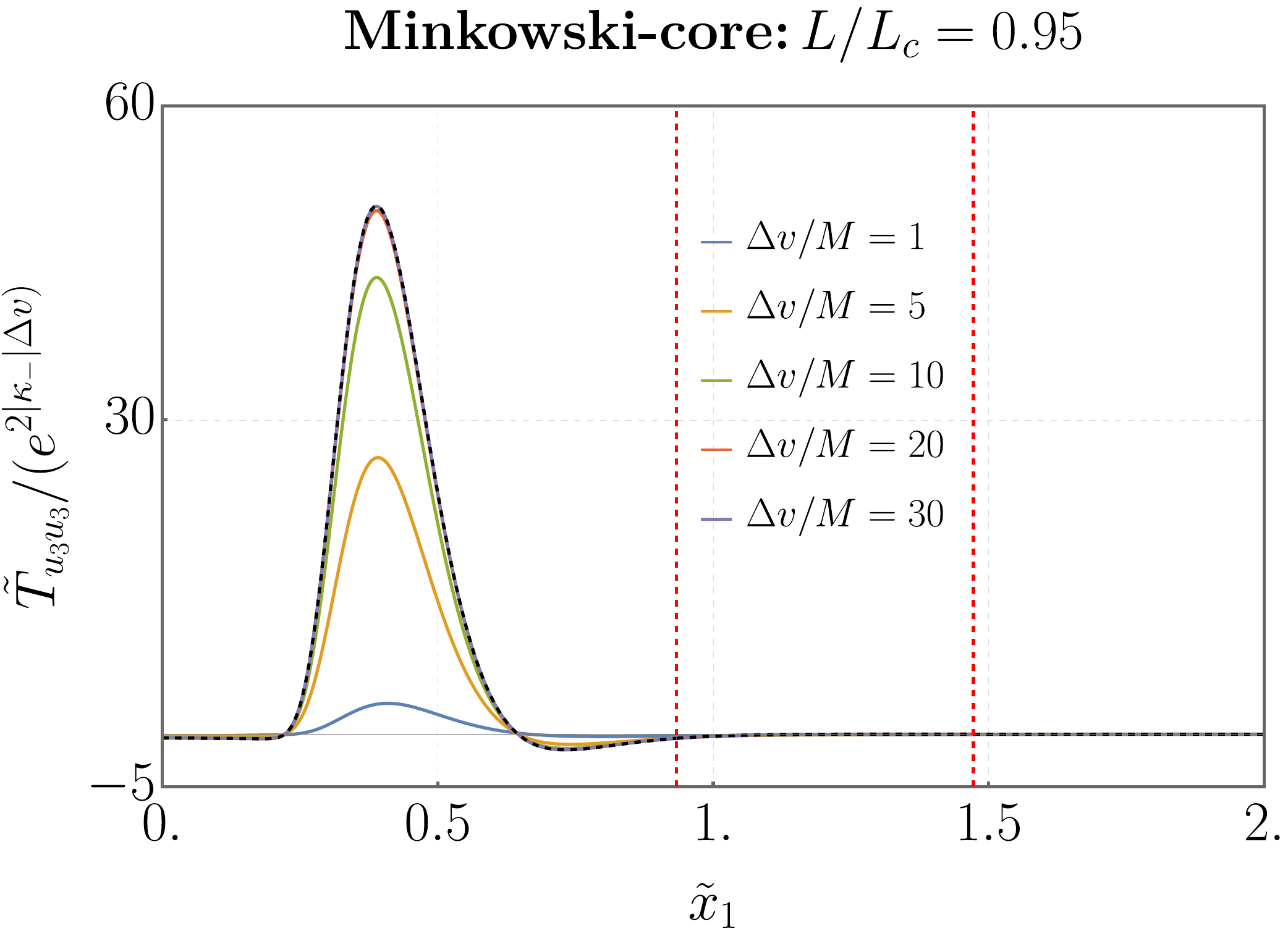}

        \centering
\includegraphics[width=\linewidth]{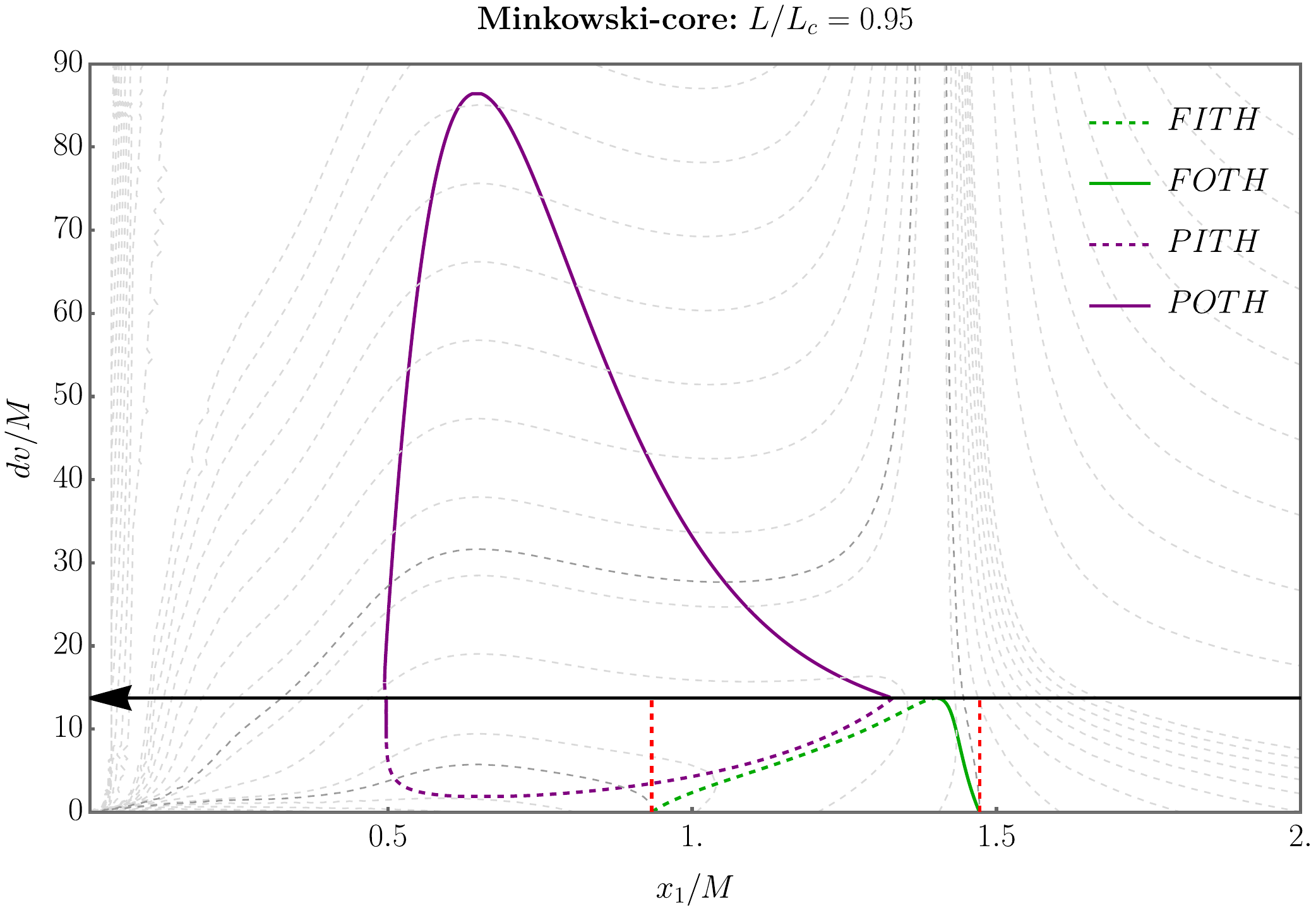}

    \caption{Plot of the location of the trapped and anti-trapped regions for the regular-black-hole with a Minkowski core as given in Eq.~\eqref{eq:Mink}. The black line represents the null shell located at $\tilde{v}_2$, which has been introduced by hand in order to close the trapped region at $v_{\rm extr}$. The red dashed lines represent the static horizons $r_+$ and $r_-$ that would be present in the classical case. The green and purple curves denote the trapped and anti-trapped regions, respectively, updated according to the fluxes computed in Sec.~\ref{section: evanescentBH}. The gray dashed lines represents the constant $r$ lines corrected with the presence of the RSET, the dark lines represent the level curves at $r_+$ and $r_-$.}

    \label{fig:Minkowski_dynamics}

\end{figure}

\begin{figure}

\centering
\includegraphics[width=0.45\linewidth]{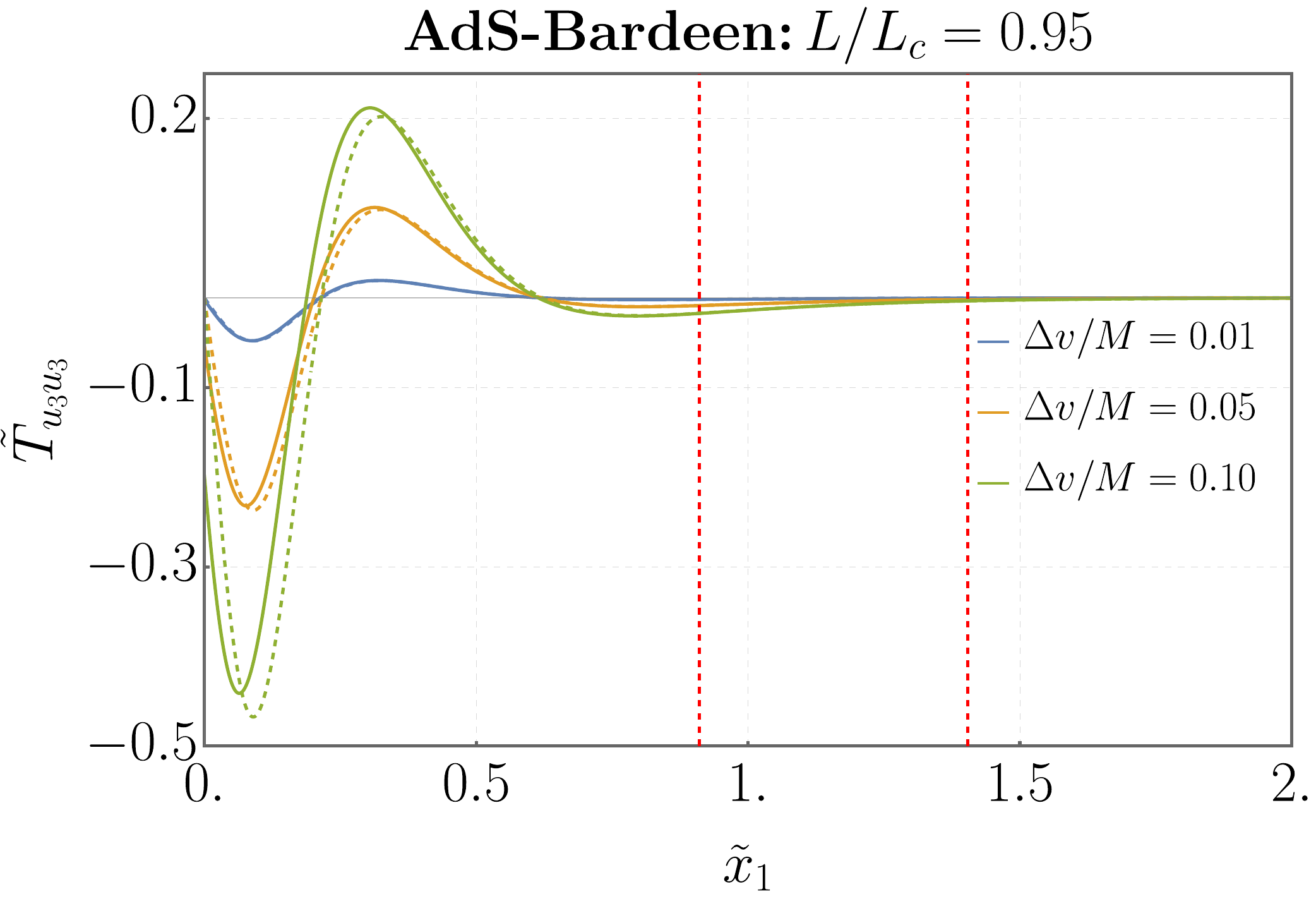}
\includegraphics[width=0.45\linewidth]{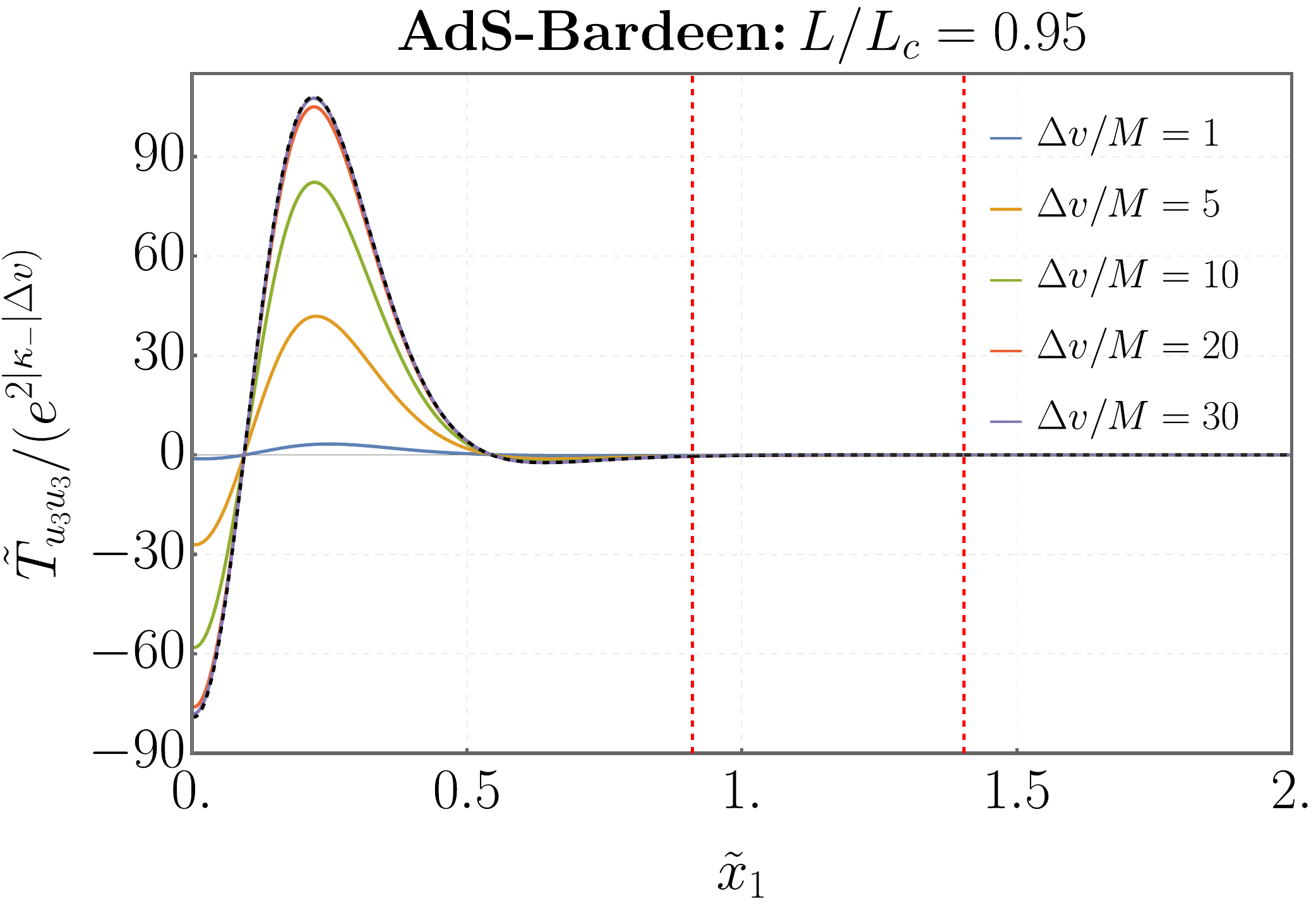}

  \centering

\includegraphics[width=\linewidth]{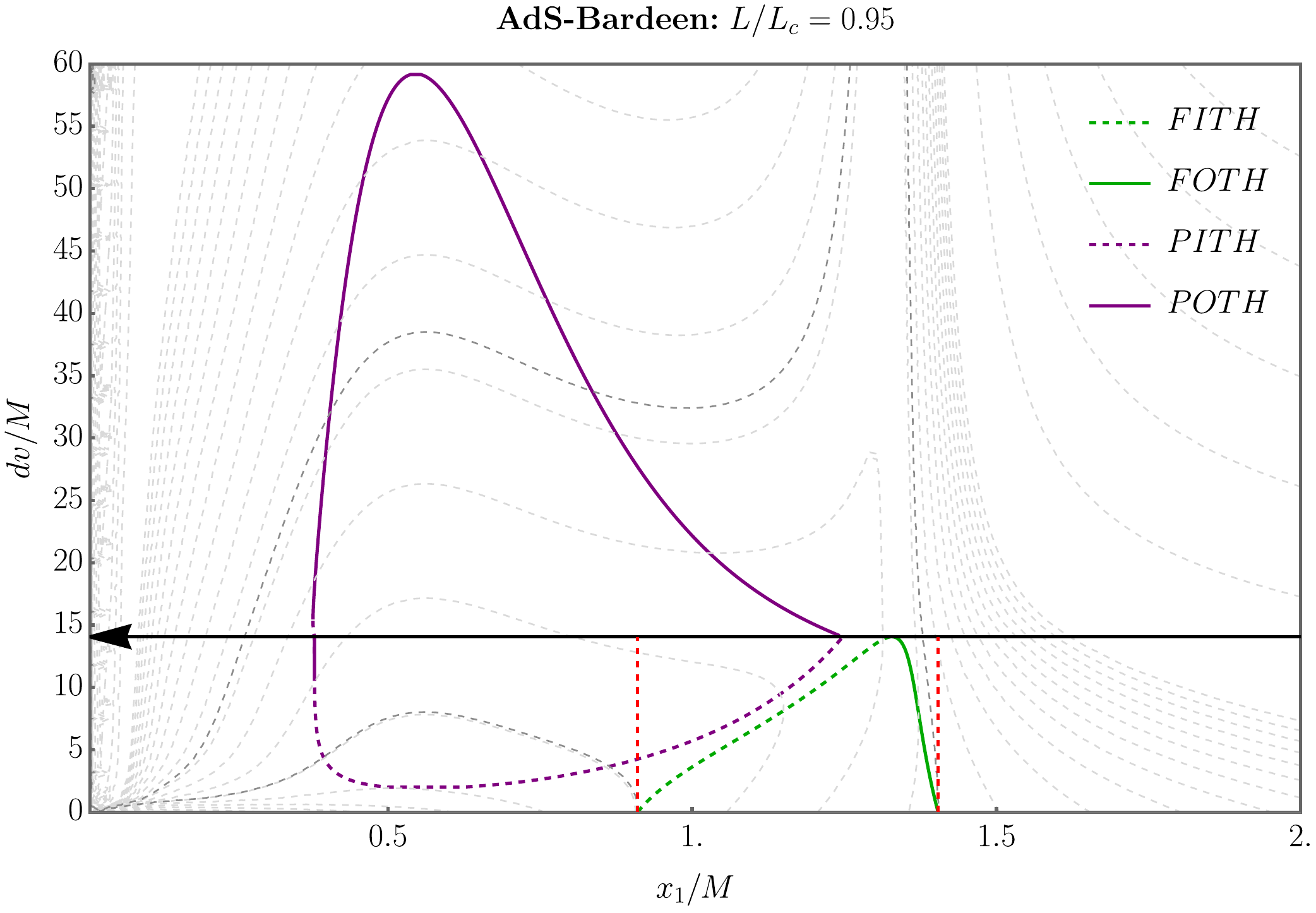}

    \caption{Plot of the location of the trapped and anti-trapped regions for the regular-black-hole with a Bardeen AdS core as given in Eq.~\eqref{eq:AdS}. The black line represents the null shell located at $\tilde{v}_2$, which has been introduced by hand in order to close the trapped region at $v_{\rm extr}$. The red dashed lines represent the static horizons $r_+$ and $r_-$ that would be present in the classical case. The green and purple curves denote the trapped and anti-trapped regions, respectively, updated according to the fluxes computed in Sec.~\ref{section: evanescentBH}. The gray dashed lines represents the constant $r$ lines corrected with the presence of the RSET, the dark lines represent the level curves at $r_+$ and $r_-$.}

    \label{fig:AdSBardeen-dynamics}

\end{figure}

For completeness, we also present the corresponding results for regular black holes with a {Minkowski core~\cite{Simpson:2019mud} (Fig. \ref{fig:Minkowski_dynamics}) and an AdS core~\cite{Arrechea:2025nlq} (Fig. \ref{fig:AdSBardeen-dynamics}), see Sec.~\ref{subsubsec:RBHs} for their respective line elements.} We observe that, although the outgoing component of the RSET differs significantly in shape in the two cases, the dynamics of the anti-trapped region remain qualitatively the same.
The reason is that all the cases studied  still exhibit an initial negative region (moving inward in $x_1$), which favors the formation of the anti-trapped region, followed by a positive region, which drives its evaporation.

This is a noteworthy result, as it suggests that the regularity of a spacetime endowed with two trapping horizons naturally favours the formation and subsequent evaporation (with respect to the coordinate $u$) of an anti-trapped region in a universal way.

\section{Discussion and conclusions}
\label{Sec:Discussion}
In this work we have developed an analytic treatment of semiclassical effects in
dynamical trapped and anti-trapped regions, with the aim of understanding the
black-hole-to-white-hole transitions recently observed in numerical simulations.
Our analysis was carried out in a deliberately simple setting: two-dimensional
collapse models, obtained by matching flat regions to static black-hole or
white-hole geometries through null shells, and the renormalized stress-energy
tensor of a massless scalar field in the $| \textit{in}\rangle$ state. Despite the
minimal nature of this construction, it captures the essential ingredients of
the semiclassical dynamics.

The first main result is that, for an evanescent black hole endowed with an
outer and an inner horizon, the outgoing component of the RSET is exponentially
amplified during the lifetime of the trapped region. This amplification is a
direct consequence of the focusing of outgoing null rays near the inner horizon.
More precisely, while the flux crossing the trapping horizons controls the
evaporation of the trapped region, the flux parallel to them controls the
evolution of the complementary null expansion. It is this latter component that
can drive the ingoing expansion through zero and thereby generate an
anti-trapped region. In this sense, the formation of a white-hole-like region is a natural consequence of the $|in\rangle$
vacuum once the semiclassical fluxes generated by collapse are followed inside
the trapped region.

\begin{figure}
    \centering    \includegraphics[width=1\linewidth]{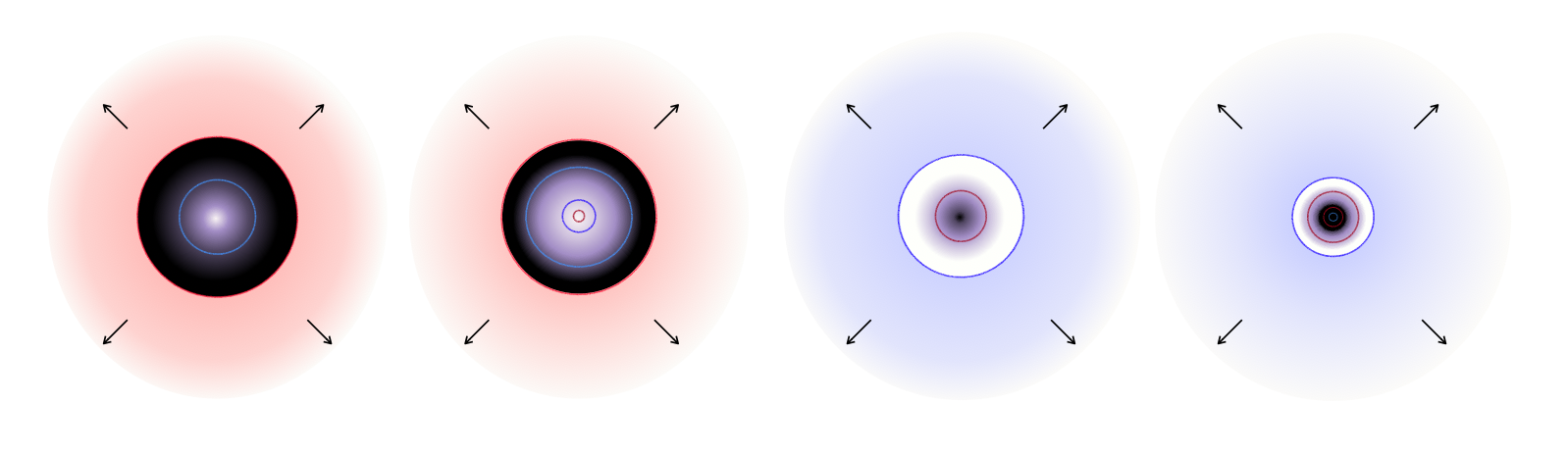}
    \caption{Pictorial representation of the cascade process described above. The initial trapped region generates fluxes that lead to the formation of an anti-trapped region inside the inner horizon. Moving to the right (i.e.\ forward in time $t$), the trapped region eventually evaporates, while the anti-trapped region generates fluxes that give rise to a new trapped region (last picture). The anti-trapped region then evaporates, and the process repeats itself, emitting positive (in red) and negative (in blue) fluxes to future null infinity. {This description has been deliberately simplified by assuming the radius of spheres to be single-valued. It is known that semiclassical backreaction may also produce wormhole necks~\cite{barenboim2025evaporationregularblackholes,boyanov2025semiclassicalevolutiondynamicallyformed}.}}
    \label{fig:BH-WH-transition}
\end{figure}

The second main result is the time-reversed counterpart of this mechanism. In an
evanescent white-hole geometry, the ingoing component of the RSET is amplified
near the outer horizon of the anti-trapped region. If sufficiently large, this
flux can in turn drive the formation of a new trapped region. The picture that
emerges therefore allows for a sequence in which a trapped region generates
fluxes that favour the formation of an anti-trapped region, while an
anti-trapped region generates fluxes that can seed a new trapped region. A qualitative picture of this possibility is shown in Fig.~\ref{fig:BH-WH-transition}.
Quantum
vacuum effects thus provide a dynamical link between black-hole and white-hole
geometries without any need for true quantum gravitational effects, {aside from those that might be responsible for the regularization of the spacetime in the first place}.

A striking aspect of our results is that such a simple toy model reproduces the
salient qualitative features of the full numerical simulations. In particular,
our analytic expressions identify a mechanism consistent with the inside-out disappearance of the trapped region and the subsequent appearance of an anti-trapped region, while suggesting how further black-to-white-hole transitions might be seeded. The mechanism responsible for these
features is transparent in the present treatment: the horizon-transverse fluxes
control the motion and eventual annihilation of the trapping horizons, whereas
the horizon-parallel fluxes --- being exponentially amplified at the FITH and POTH ---
can change the sign of the other null expansion and drive the opening of new trapped or anti-trapped regions. 
This separation of roles is one of the main lessons of this analytic approach.

The simplicity of the model also allows us to distinguish universal features
from model-dependent ones. The exponential amplification itself is universal
whenever a sufficiently long-lived region with two trapping horizons is present:
for black holes it is controlled by the inner horizon, while for white holes it
is controlled by the outer horizon. By contrast, the detailed sign and shape of
the amplified flux depend on the background geometry through derivatives of the
metric function $f(r)$. 

This dependence determines whether the anti-trapped
region is compact or unbounded, how long it persists, and how efficiently it can
seed the next trapped region. In the regular black-hole geometries considered
here, including Bardeen, Minkowski-core and AdS-core examples, the flux profiles
contain the sequence of signs required for the formation and
subsequent evaporation of a finite anti-trapped region. This suggests that the
regularity of the core, together with the presence of two horizons, strongly
favours a complete black-hole-to-white-hole transition and is hence able to affect the fluxes reaching null infinity. The core structure is no longer completely unobservable in semiclassical physics.

Regarding the timescale of this regular evaporation process, we can use the Raychaudhuri equation to set upper and lower bounds to the individual evaporation of each region. For the trapped region, for example, we know that the surface gravity of the outer and inner horizons reduces as they approach each other. This implies that our predicted timescale (as deducible from the provided plots), which assumes these fluxes are constant and finite, sets an upper bound to how fast this process can unfold. Regarding the anti-trapped region, on the contrary, our calculation sets a lower bound to its evaporation timescale. This is so because, as the anti-trapped region grows in size, the exponential growth in $v$ of the outgoing flux is cancelled. 

It is important to stress also the limitations of our analysis. We have computed the RSET on fixed backgrounds and then inferred the qualitative backreaction through the Raychaudhuri equation. The metrics used in the intermediate regions are not self-consistent solutions of the semiclassical Einstein equations. Moreover, we have worked within a two-dimensional Polyakov-type approximation, neglecting backscattering, higher angular modes, and the possible competition with classical instabilities. These simplifications prevent us from extracting precise
timescales or final states.\footnote{More so, the black-to-white hole transitions observed here invite to consider definitions of evaporation timescale relevant for astrophysics, such as those based on the time measured by very distant timelike observers.} Nevertheless, the agreement with the qualitative behaviour of numerical simulations indicates that the mechanism identified here is not tied to a particular numerical scheme, although establishing its robustness would require a self-consistent backreaction analysis.

Our results therefore support the following physical interpretation. In
semiclassical gravity, the disappearance of a trapped region need not mark the end of the evolution. The same quantum fluxes responsible for its evaporation
can reorganize the causal structure of the interior and produce an anti-trapped region. That region can then radiate and, under suitable conditions, trigger the
formation of a new trapped region. The resulting cascade is naturally damped by radiation escaping to infinity, and may eventually terminate in a horizon-free
geometry once the amplified fluxes are no longer sufficient to create the next trapped or anti-trapped region. Establishing whether this endpoint is generic, and determining its dependence on the matter content and on the underlying regular geometry, requires solving the full semiclassical backreaction problem. Still, we think that the present investigation, even if preliminary, strongly suggest that, in the absence of unbounded fluxes associated to timelike singularities, 
semiclassical backreaction may generically tend to reduce or erase trapped and anti-trapped regions under the conditions identified here.

Several directions remain open. A first priority is to go beyond the fixed background approximation and incorporate the RSET into a self-consistent evolution, which could potentially be implemented by smearing the ingoing shells into continuous flux distributions. A second one is to assess the robustness of the mechanism beyond the two-dimensional approximation, including greybody effects and the contribution of higher multipoles~\cite{Ori:2025zhe}. Finally, it will be important to understand how the semiclassical instability described here interacts with classical matter and its associated inner-horizon \cite{Carballo-Rubio:2024dca} and white-hole instabilities~\cite{PhysRevLett.33.442}. The analytic framework developed in this work provides a useful starting point for these investigations, because it isolates the simple physical mechanism behind the numerical black-hole-to-white-hole
transition: the exponential amplification of vacuum fluxes in regions bounded by inner and outer horizons. If this mechanism survives a fully self-consistent treatment of backreaction, semiclassical gravity may do more than evaporate black holes: it may turn their end into the beginning of a new phase of spacetime.

\acknowledgments

The authors want to thank Valentin Boyanov, Francesco di Filippo, Raul Carballo-Rubio, Matt Visser, Carlo Rovelli and Francesca Vidotto for insightful discussions.

\appendix

\section{Appendix: Physical RSET}
\label{app:physRSET}

Let us consider a spherically symmetric spacetime with the metric \eqref{double-null_sperical_metric}. If horizons ($r_h$ such that $f(r_h)=0$) are present, then the metric expressed in the set of coordinates $(u,v)$ is singular on them.

This, as well known, is just an artifact of the coordinate frame. In section \ref{sec:2D_RSET} we used Ingoing Eddington-Finkelstein (IEF), which is related to ingoing null-rays. A more physical framework is given by the Painlevè-Gullstrand (PG) coordinates, which are related to free-falling observers. 

In this appendix we compute in detail the RSET components with respect to these coordinates and then compute the energy density observed by a free-falling observer and by a static observer. In this way we can show that the exponential growth found in Sec.~\ref{section: evanescentBH} is physical and not merely an artifact of the reference frame.

The PG time is defined by
\begin{equation}
  T = t + \int\frac{\sqrt{1-f}}{f}\,dr \equiv t + h(r),
  \qquad h'(r) = \frac{\alpha}{f},\quad \alpha\equiv\sqrt{1-f}.
\end{equation}
The metric becomes
\begin{equation}
  ds^{2} = -f\,dT^{2} + 2\alpha\,dT\,dr + dr^{2},
\end{equation}
with $\det g = -1$, giving the inverse metric
\begin{equation}
  g^{TT} = -1, \qquad g^{Tr} = \alpha, \qquad g^{rr} = f.
\end{equation}

From $u = T - h(r) - r^{*}(r)$ and $v = T - h(r) + r^{*}(r)$ one reads off
the Jacobian

\begin{align}
  \frac{\partial u}{\partial T}\bigg|_{r} &= 1, &
  \frac{\partial u}{\partial r}\bigg|_{T} &= -h' - \frac{1}{f}
    = -\frac{\alpha}{f} - \frac{1}{f} = -\frac{1+\alpha}{f},
  \label{eq:jac_u}\\
  \frac{\partial v}{\partial T}\bigg|_{r} &= 1, &
  \frac{\partial v}{\partial r}\bigg|_{T} &= -h' + \frac{1}{f}
    = -\frac{\alpha}{f} + \frac{1}{f} = \frac{1-\alpha}{f}.
  \label{eq:jac_v}
\end{align}
From this is trivial to determine the PG components of the RSET
\begin{align}
  T_{TT}^{\mathrm{PG}} &= T_{uu} + 2\,T_{uv} + T_{vv},
  \label{eq:PG_TT}\\
  T_{Tr}^{\mathrm{PG}}
    &= \frac{1}{f}\left[{-(1+\alpha)\,T_{uu} - 2\alpha\,T_{uv} + (1-\alpha)\,T_{vv}}\right],\\
    T_{rr}^{\mathrm{PG}}
    &= \frac{(1+\alpha)^{2}}{f^{2}}\,T_{uu}
      - \frac{2}{f}\,T_{uv}
      + \frac{(1-\alpha)^{2}}{f^{2}}\,T_{vv}.
  \label{eq:PG_rr}
\end{align}

\subsection{Energy Density for a Free-Falling Observer via PG Coordinates}
 
A free-falling observer (FFIO) released from rest at spatial infinity has
conserved energy $E = f\dot{t} = 1$.  The normalization
$g_{\mu\nu}u^{\mu}u^{\nu}=-1$ in the static metric gives
\begin{equation}
  \dot{r} = -\alpha = -\sqrt{1-f}.
\end{equation}
In PG coordinates the four-velocity components are
$u^{T}=1$ and $u^{r}=-\alpha$.
The energy density measured by the FFIO is
\begin{equation}
  \rho_{\mathrm{obs}} = T_{\mu\nu}u^{\mu}u^{\nu}
  = T_{TT}(u^{T})^{2} + 2T_{Tr}\,u^{T}u^{r} + T_{rr}(u^{r})^{2}.
\end{equation}
Substituting $u^{T}=1$ and $u^{r}=-\alpha$:
\begin{align}
  \rho_{\mathrm{obs}} &= T_{TT} - 2\alpha\,T_{Tr} + \alpha^{2}\,T_{rr},\\
  &=\frac{(1+\alpha)^{2}}{f^{2}}\,T_{uu}
    + \frac{2}{f}\,T_{uv}
    + \frac{(1-\alpha)^{2}}{f^{2}}\,T_{vv},
    \notag\\
    &=(1+\alpha)^{2}\,(T_{uu})_{\rm phys}^{\textsc{bh}}
    + 2\,(T_{uv})_{\rm phys}^{\textsc{bh}}
    + \frac{(1-\alpha)^{2}}{f^{2}}\,(T_{vv})_{\rm phys}^{\textsc{bh}}, 
    \qquad \alpha = \sqrt{1-f(r)}.\notag
  \label{eq:rho_PG_start}
\end{align}

\subsection{Energy Density for a Static Observer}

A static observer (fixed $r$) has $\dot{r}=0$.  From the normalization
$g_{\mu\nu}u^{\mu}u^{\nu}=-1$ in the static metric:
\begin{equation}
  -f\,(u^{T})^{2} = -1
  \qquad\Longrightarrow\qquad
  u^{T} = \frac{1}{\sqrt{f}}.
\end{equation}
In PG coordinates, since $T=t+h(r)$ and $\dot{r}=0$ implies $\dot{T}=\dot{t}$,
the four-velocity components are
\begin{equation}
  u^{T} = \frac{1}{\sqrt{f}}, \qquad u^{r} = 0.
\end{equation}

Note that the static observer exists only where $f>0$, i.e.\ outside the
horizon; as $r\to r_{H}$ the required proper acceleration diverges.
Since $u^{r}=0$ only $T_{TT}$ contributes:
\begin{align}
  \rho_{\mathrm{obs}}^{\mathrm{static}}
  &= \frac{T_{TT}}{f},\\
  &= \frac{1}{f}\!\left(T_{uu} + 2\,T_{uv} + T_{vv}\right),\notag\\
  &= f (r) (T_{uu})_{\rm phys}^{\textsc{bh}} + 2\,(T_{uv})_{\rm phys}^{\textsc{bh}} + \frac{1}{f}(T_{vv})_{\rm phys}^{\textsc{bh}},\notag
  \label{eq:rho_static}
\end{align}

\section{Appendix: The Schwarzian derivative}
Definition:
\begin{equation}
    \{f,x\} = \frac{f'''}{f'} - \frac{3}{2}\left(\frac{f''}{f'}\right)^2
    \label{schwarzian}
\end{equation}
where $f' = \frac{df}{dx}$.

\vspace{1cm}
Properties:
\begin{enumerate}
    \item Chain rule:
    $\{ f(g(x)), x\} = g'(x)^2\{f,g\} + \{g,x\}$
    \item Inverse Schwarzian: 
    $\{x,f\} =- \frac{1}{f'(x)^2}\{f,x\}$
    \item Implicit Schwarzian $(f(x) = g(y)$ ):
    $\{y,x\} = f'(x)^2  \left[\{y,g\}-\{x,f\}  \right] $
\end{enumerate}

In this section we explicitly derive the Schwarzian derivatives appearing in the computation of the RSET, starting from the matching conditions introduced in Sec.~\ref{section: evanescentBH}.

\subsection{Schwarzian derivative from matching conditions: Black hole scenario}

From the matching at $v=\tilde v_1$ and $v=\tilde v_2$, Eqs.~\eqref{v1_match} and \eqref{v2_match}, we have
\begin{equation}
\frac{du_2}{du_1} = \frac{1}{f(x_1)}, 
\qquad
\frac{du_2}{du_3} = \frac{1}{f(x_2)},
\end{equation}
where $x_1 = \frac{\tilde v_1 - u_1}{2}$ and $x_2 = \frac{\tilde v_2 - u_3}{2}$.
From which we derive immediately 
\begin{equation}
\frac{dx_1}{du_1} = -\frac{1}{2}, 
\qquad
\Rightarrow
\qquad
\begin{cases}
    \frac{dx_1}{du_2} = \frac{dx_1}{du_1}\frac{du_1}{du_2} = -\frac{1}{2} f(x_1)\\
    \frac{d^2 x_1}{du_2^2} = -\frac{1}{2} f'(x_1)\frac{dx_1}{du_2}
= \frac{1}{4} f'(x_1) f(x_1).
\end{cases}
\end{equation}

We first compute the Schwarzian derivative $\{u_1,u_2\}$. Using the definition
\begin{equation}
\{u_1,u_2\} = \frac{u_1'''(u_2)}{u_1'(u_2)} - \frac{3}{2} \left( \frac{u_1''(u_2)}{u_1'(u_2)} \right)^2,
\end{equation}
it is convenient to express derivatives with respect to $u_2$ in terms of $u_1$. From the matching condition,
\begin{align}
\frac{du_1}{du_2} &= f(x_1)\\
\frac{d^2 u_1}{du_2^2} &= \frac{d}{du_2} f(x_1)
= f'(x_1)\frac{dx_1}{du_2}, \\
\frac{d^3 u_1}{du_2^3} &= f''(x_1)\left(\frac{dx_1}{du_2}\right)^2 + f'(x_1)\frac{d^2 x_1}{du_2^2}.
\end{align}

Substituting into the Schwarzian definition, after straightforward algebra, one obtains
\begin{equation}
\boxed{\{u_1,u_2\} = -\frac{1}{8}\left[ f'(x_1)^2 - 2 f(x_1) f''(x_1) \right]
= -\frac{1}{8} B(x_1)},
\end{equation}
where we used the definition of $B(r)$ given in Eq. \eqref{B_function}.

An analogous computation yields
\begin{equation}
\{u_3,u_2\} = -\frac{1}{8} B(x_2).
\end{equation}

Finally, the Schwarzian $\{u_1,u_3\}$ can be obtained using the chain rule:
\begin{equation}
\{u_1,u_3\} = \left( \frac{du_2}{du_3} \right)^2 \{u_1,u_2\} + \{u_2,u_3\}.
\end{equation}

Using
\begin{equation}
\frac{du_2}{du_3} = \frac{1}{f(x_2)},
\qquad
\{u_2,u_3\} = - \left( \frac{du_2}{du_3} \right)^2 \{u_3,u_2\},
\end{equation}
we obtain
\begin{equation}
\boxed{\{u_1,u_3\}
= -\frac{1}{8} \frac{B(x_1) - B(x_2)}{f(x_2)^2}}.
\end{equation}

\subsection{Schwarzian derivative from matching conditions: White hole scenario}

For the evanescent white hole, the computation proceeds analogously, but the non-trivial coordinate transformations involve the $v$ coordinate.

From the matching conditions \eqref{u1_match} and \eqref{u2_match}, we have
\begin{equation}
\frac{dv_2}{dv_1} = \frac{1}{f(y_1)},
\qquad
\frac{dv_2}{dv_3} = \frac{1}{f(y_2)},
\end{equation}
with
\begin{equation}
y_1 = \frac{v_1 - \tilde u_1}{2},
\qquad
y_2 = \frac{v_3 - \tilde u_2}{2}.
\end{equation}

Repeating the same steps as above, one finds
\begin{equation}
    \boxed{\{v_1, v_2\} = -\frac{1}{8} B(y_1)},
\end{equation}

\begin{equation}
    \boxed{\{v_1,v_3\}= -\frac{1}{8} \frac{B(y_1) - B(y_2)}{f(y_2)^2}}.
\end{equation}

These expressions play a crucial role in determining the RSET in the different regions of the spacetime.

\section{Appendix: Check for Regularity}
\label{app:reg_check}

Let us try to understand how the coordinates $u_1, u_2, u_3 $ work when the argument of the function $B(r)$ approaches the horizon $r \rightarrow r_h$.
So, Let us analyze the region II, also, from now on we recognize the surface gravity of the horizon $f'(r_h) = 2 \kappa_h$. The definition of $u_2$ is:
\begin{equation}
    u_2 = v - 2 \int_{r_0}^{r}\frac{dx}{f(x)},
\end{equation}
now, $r_0$ is arbitrary, so without loss of generality, we can fix $r_0 = r_h + \epsilon$ with $\epsilon>0$. From the Penrose diagram \ref{fig:spherical_penrose}, we see that $r = r_h$ is a null curve in region II; by the matching conditions we expect that $x_1 \rightarrow r_h \iff x_2 \rightarrow r_h$. If $r \rightarrow r_h$ , in region II we can approximate $f(r) \sim 2 \kappa_h(r - r_h)$, and then:
\begin{align} 
    u_2& = v - 2 \int_{r_0}^{r}\frac{dx}{f(x)} \notag \\
    &\sim v - \frac{1}{\kappa_h} \int_{r_0}^{r}\frac{dx}{x - r_h} \notag
    \\
    & \sim v - \frac{1}{\kappa_h} \log|\kappa_h ( r - r_h)| + C,
    \label{u_2-NH}
\end{align}
where $C = \frac{ \log|\kappa_h(r_0 - r_h)|}{\kappa_h}$.
The matching applies substituting $(v,r) \rightarrow (v_\alpha,r_\alpha)$ where $\alpha= \{1,2\}$ and $x_1 = \frac{\tilde{v}_1 - u_1}{2}$ and $x_2 = \frac{\tilde{v}_2 - u_3}{2}$ as usual.
We got that $u_2$ diverges when $r \rightarrow r_h$(with a sign dependent on $\kappa_h$).
\newline
It is interesting (and it will be very useful) to understand how $u_1$, $u_2$, $u_3$ (or equivalently $x_1$, $r$, $x_2$)are related; in order to do that we can use the result \eqref{u_2-NH}, for a general $v \in [\tilde{v}_1, \tilde{v}_2]$ and for $v= \tilde{v}_1$

\begin{equation}
    \tilde{v}_1 - \frac{1}{\kappa_h} \log| \kappa_h(x_1 - r_h)| = v- \frac{1}{\kappa_h} \log|\kappa_h( r - r_h)|.
\end{equation}
From which we derive
\begin{equation}
    x_1 - r_h = (r - r_h)e^{-\kappa_h (v - \tilde{v}_1)}.
\end{equation}
With $r=x_2$ for $v=\tilde{v}_2$.
It means, from \eqref{v1_match}, \eqref{v2_match}, that as long as $u_3 \rightarrow \tilde{v}_2-2r_h$ also $u_1 \rightarrow \tilde{v}_1 - 2r_h$.
So, in the limit $r,x_1, x_2 \rightarrow r_h$ we have:
\begin{align}
    B(r) &\sim 4 \kappa_h^2,
    \label{eq:B_horizon}\\
    B(x_1)-B(r)
    & \sim 2\kappa_hf^{(3)}(r_h) \left[(r - r_h)^2 - (x_1 - r_h)^2 \right]  \\& \sim 2\kappa_hf^{(3)}(r_h) (r- r_h)^2\left[1 -e^{-2\kappa_h(v-\tilde{v}_1)}\right]\notag .
\end{align}

Let us now make explicit the connection between these expressions and the RSET. Inserting
\eqref{eq:B_horizon} into \eqref{eq:RSET-regionII}, and $f''(r)\to f''(r_h)$ into \eqref{Tuv2}, one obtains at
once the horizon values \eqref{Tvv2_horizon} and \eqref{Tuv2_horizon}. The outgoing component is
the non-trivial one: both $B(x_1)-B(r)$ and $f(r)^2\simeq 4\kappa_h^2(r-r_h)^2$ vanish
quadratically, so that the ratio appearing in \eqref{Tuu2} is finite,
\begin{align}
    \langle T_{u_2u_2} \rangle_{\rm phys}^{\textsc{bh}}\Big|_{r_h}
    &= \frac{1}{192\pi}\,
      \frac{2\kappa_h f^{(3)}(r_h)(r-r_h)^2\left[1-e^{-2\kappa_h(v-\tilde{v}_1)}\right]}
           {4\kappa_h^2 (r-r_h)^2}\\
    &= \frac{1}{384\pi}\frac{f^{(3)}(r_h)}{\kappa_h}\left[1-e^{-2\kappa_h(v-\tilde{v}_1)}\right]\notag,
\end{align}
which is Eq.~\eqref{Tuu2_horizon}; the same manipulation with $r\to x_2$ and
$v-\tilde{v}_1\to\Delta v$ gives the region III result \eqref{Tuu3_horizon}.

\section{Evanescent Schwarzschild black hole}
\label{app:Schwarzschild}

The purpose of this appendix is twofold: to recover the classic results of
Hiscock~\cite{Hiscock:1980ze,Hiscock:1981xb} as a particular case of the general formulae of
Sec.~\ref{section: evanescentBH}, and to extend that analysis to the question addressed in this
work, namely the formation of an anti-trapped region. We shall see that the outgoing flux is
positive throughout the spacetime, so that no anti-trapped region can form: the single-horizon
case does not transition.

\subsection{Setup}

For the Schwarzschild metric $f(r)=1-2M/r$ one has
\begin{equation}
    f'(r)=\frac{2M}{r^2},\quad
    f''(r)=-\frac{4M}{r^3},\quad
    f^{(3)}(r)=\frac{12M}{r^4},\quad
    B(r)=\frac{4M}{r^4}\left(2r-3M\right),
    \label{eq:Schw_f}
\end{equation}
with a single non-degenerate horizon at $r_h=2M$, of FOTH type, with
$\kappa_h=f'(r_h)/2=1/(4M)>0$ and $B(r_h)=4\kappa_h^2$. The matching
conditions~\eqref{v1_match} and~\eqref{v2_match} become
\begin{equation}
    \frac{du_2}{du_\alpha}=\frac{1}{f(x_\alpha)}=\frac{v_\alpha-u_\alpha}{v_\alpha-u_\alpha-4M},
    \qquad \alpha=1,3,
    \label{eq:Schw_matching}
\end{equation}
and the ray-tracing relation~\eqref{r1_r2_relation} integrates to
\begin{equation}
    x_2-x_1+2M\ln\left|\frac{x_2-2M}{x_1-2M}\right|=\frac{\Delta v}{2},
    \label{eq:Schw_raytracing}
\end{equation}
which reproduces~\eqref{u_2-NH} in the limit $x_1,x_2\to r_h$.

\subsection{The RSET in the $|in\rangle$ state}

Substituting~\eqref{eq:Schw_f} into the general
expressions~\eqref{eq:RSET-regionII} and~\eqref{eq:RSET-regionIII} one finds, region by region:

\begin{enumerate}
    \item Region I: $\langle T_{ab}\rangle^{\textsc{bh}}_{\rm phys}=0$.

    \item Region II:
    \begin{subequations}
    \label{eq:Schw_RSET_II}
    \begin{align}
        \langle T_{u_2u_2}\rangle^{\textsc{bh}}_{\rm phys}
        &= \frac{M}{48\pi f(r)^2}\left[\frac{2x_1-3M}{x_1^4}-\frac{2r-3M}{r^4}\right],
        \label{eq:Schw_Tuu_II}\\
        \langle T_{vv}\rangle^{\textsc{bh}}_{\rm phys}
        &= \frac{M}{48\pi}\,\frac{3M-2r}{r^4},
        \label{eq:Schw_Tvv_II}\\
        \langle T_{u_2v}\rangle^{\textsc{bh}}_{\rm phys}
        &= -\frac{M}{24\pi r^3}.
        \label{eq:Schw_Tuv_II}
    \end{align}
    \end{subequations}

    \item Region III:
    \begin{subequations}
    \label{eq:Schw_RSET_III}
    \begin{align}
        \langle T_{u_3u_3}\rangle^{\textsc{bh}}_{\rm phys}
        &= \frac{M}{48\pi}\,\frac{x_2^2}{(x_2-2M)^2}
           \left[\frac{2x_1-3M}{x_1^4}-\frac{2x_2-3M}{x_2^4}\right],
        \label{eq:Schw_Tuu_III}\\
        \langle T_{vv}\rangle^{\textsc{bh}}_{\rm phys}
        &= \langle T_{u_3v}\rangle^{\textsc{bh}}_{\rm phys}=0 .
    \end{align}
    \end{subequations}
\end{enumerate}
Here $r=r(u_2,v)$ is the radius reached at advanced time $v$ by the ray that entered region II at
$x_1$, so that $x_1$ and $x_2$ are to be regarded as functions of $u_2$ and $u_3$ through
Eqs.~\eqref{eq:Schw_matching}--\eqref{eq:Schw_raytracing}. Using $u_1=\tilde{v}_1-2x_1$ and
$u_3=\tilde{v}_2-2x_2$, and setting $\tilde{v}_1=0$, $\tilde{v}_2=v_0$,
Eqs.~\eqref{eq:Schw_RSET_II}--\eqref{eq:Schw_RSET_III} take the form
\begin{align}
    \langle in|T_{u_2u_2}|in\rangle
    &= \frac{1}{48\pi}\left[\frac{M}{r^4}\left(3M-2r\right)
       -\frac{16M}{u_1^4}\left(3M+u_1\right)\right],\\
    \langle in|T_{u_3u_3}|in\rangle
    &= \frac{1}{3\pi}\frac{(v_0-u_3)^2}{(v_0-u_3-4M)^2}
       \left[\frac{M}{(v_0-u_3)^4}\left(3M-v_0+u_3\right)-\frac{M}{u_1^4}\left(3M+u_1\right)\right],
\end{align}
in which the results of~\cite{Hiscock:1980ze,Hiscock:1981xb} are immediately recognised.

Two features of~\eqref{eq:Schw_RSET_II} are worth pointing out already at this stage. First, the
ingoing flux $\langle T_{vv}\rangle$ is negative for $r>3M/2$ and positive for $r<3M/2$; in
particular it is negative on the horizon, in agreement with the general statement following
Eq.~\eqref{Tvv2_horizon}. Second, $\langle T_{u_2v}\rangle$ is negative everywhere, being
proportional to $f''(r)$ and hence to the 2D Ricci scalar.

\subsection{Regularity on the horizon}

The conditions~\eqref{regularBH_RSET} are readily checked. The transverse and mixed components
are finite,
\begin{equation}
    \langle T_{vv}\rangle^{\textsc{bh}}_{\rm phys}\Big|_{r_h}=-\frac{1}{768\pi M^2}
    =-\frac{\kappa_h^2}{48\pi},
    \qquad
    \langle T_{u_2v}\rangle^{\textsc{bh}}_{\rm phys}\Big|_{r_h}=-\frac{1}{192\pi M^2}
    =\frac{f''(r_h)}{96\pi}.
\end{equation}
Using the general formula \eqref{Tuu2_horizon} we get immediately
\begin{equation}
    \langle T_{u_2u_2}\rangle^{\textsc{bh}}_{\rm phys}\Big|_{r_h}
    = \frac{1}{128\pi M^2}\left[1-e^{-(v-\tilde{v}_1)/2M}\right].
    \label{eq:Schw_Tuu_horizon}
\end{equation}
 The RSET is therefore regular on the horizon and assumes the same value shown in \cite{Hiscock:1980ze}.
Note that, $r_h$ being a FOTH ($\kappa_h>0$), the bracket saturates to unity: no exponential
amplification takes place, in contrast with what happens at a FITH.

\subsection{Positivity of the outgoing flux}

Writing the outgoing flux as the accumulation integral~\eqref{Tuu2_integral},
\begin{equation}
    \langle in|T_{u_2u_2}|in\rangle
    =\frac{1}{192\pi}\int_{\tilde{v}_1}^{v}
     f^2\!\left[r(u_2,v')\right]f^{(3)}\!\left[r(u_2,v')\right]dv',
\end{equation}
and observing that $f^{(3)}(r)=12M/r^4>0$ for every $r>0$, the integrand is non-negative along
any outgoing ray, so that
\begin{equation}
    \langle T_{u_2u_2}\rangle^{\textsc{bh}}_{\rm phys}\ \ge\ 0
    \qquad\text{everywhere in regions II and III}.
\end{equation}
This is to be contrasted with the Reissner--Nordstr\"om case, where
$f^{(3)}(r)=12(Mr-2Q^2)/r^5$ changes sign at $r_c=2Q^2/M$, and with the regular black holes of
Sec.~\ref{subsubsec:RBHs}, where a negative region is always present.

By the discussion of Sec.~\ref{sec:BH-WH_transition}, the absence of any negative outgoing flux
prevents $\theta_-$ from ever crossing zero: no anti-trapped region can form through this
mechanism.




\bibliographystyle{JHEP}
\bibliography{biblio}

\end{document}